\documentclass[onecolumn]{aastex702}
\usepackage{natbib}
\usepackage{soulutf8}
\usepackage{makecell} 
\usepackage{graphicx}
\usepackage{bm}
\usepackage{booktabs}
\usepackage[position=top]{subfig}

\graphicspath{{Figures/}{sed\_combined/}{sed\_plot\_one\_less\_excess/}}  
\usepackage{rotating}

\setcitestyle{citesep={,}}

\begin{document}

\title{The Nearest, Brightest Debris Disk Host Stars: Updated Photosphere Models and Dust Properties for Stars with Saturated 2MASS Photometry}%

\author{Inbok Yea, Sarah Dodson-Robinson, Cesar Julian Bustos Sanchez, John E. Gizis}%
\email{}
\affiliation{University of Delaware}
\date{August 2026}%

\begin{abstract}
	To discover and characterize debris disks, one must obtain high-quality photometry in order to accurately fit photosphere models for the host stars. Since the $K$ band always falls on the Rayleigh-Jeans side of a star's spectral energy distribution (SED), its reliability is particularly important for uncovering mid-infrared excess emission from dust. While Two Micron All Sky Survey (2MASS) photometry is the near-infrared benchmark for most stars that host debris disks discovered by Spitzer and the Wide Field Infrared Survey Explorer (WISE), the brightest stars ($K_s < 5$) saturated the 2MASS detectors. To accurately model debris disks that orbit the nearest, brightest stars, we construct new photosphere models by replacing saturated 2MASS data with $JHK$ photometry from the NASA Catalog of Infrared Observations, archival $R_C I_C$ photometry, and saturation-corrected WISE W1 photometry. We estimate the dust temperature and minimum distance from the star by fitting blackbody and modified blackbody models to infrared photometry from WISE, Spitzer, and Herschel. Our models allow us to identify disks that subtend large angles on the sky; these disks are candidates for follow-up with ground-based angular differential imaging, the Space Telescope Imaging Spectrograph (STIS) on the Hubble Space Telescope (HST), and NIRCam on JWST. Finally, we find that the disk surrounding $\gamma$~Tri likely has warm dust that could be resolvable with JWST's MIRI instrument at $23$ or $25.5 \micron$. Multiwavelength observations provide benchmarks for collisional cascade models and may reveal planet--disk interactions.
\end{abstract}




\section{Introduction}
\label{sec:intro}

	Resolved images of circumstellar dust reveal the dynamical influence of neighboring planets, making debris disks high-value targets for the next generation of space missions \citep{kalas95,kalas05,su09,apai15,debes19}. Indeed, debris disk imaging is one of the science drivers behind the designs of JWST's NIRCam and MIRI coronagraphs \citep{gardner04} and the Nancy Grace Roman Space Telescope Coronagraph \citep{bailey23, anche24}. The simplest way of discovering debris disks, pioneered during the InfraRed Astronomical Satellite (IRAS) mission, is to identify excess infrared emission above what a star photosphere should produce \citep{aumann84, smith84, aumann85}. The infrared excess is then attributed to reprocessed starlight emitted by dust grains, which are produced by a collisional cascade in a planetesimal belt such as our own asteroid belt or the Edgeworth--Kuiper belt. When infrared excess is detected in two or more photometric bands, one can roughly estimate the temperature and location of the orbiting dust, paving the way for a direct imaging campaign.
	
	The main goals of direct imaging are to map out debris disk morphology \citep[e.g.,][]{gaspar23,matthews23,boccaletti24,rebollido24} and probe grain size and composition \citep[e.g.,][]{chen20,singh21,arnold22}.
Grains with radii of $\sim 5 \micron$, a size expected to be present in debris disks
orbiting stars of all spectral types \citep{pawellek15}, 
can be seen both in thermal emission with MIRI and in scattered light with NIRCam, providing a multiwavelength view of dust warmer than $\sim 120$~K \citep{su13}. Multiwavelength imaging allows observers to simultaneously detect planet-generated dust perturbations near the parent planetesimal ring \citep{strubbe06} and characterize the smaller grains with orbits that are evolving due to radiative forces, as seen in detections of dust being blown out of Fomalhaut's Kuiper Belt analog \citep{adams18, gaspar23, thebault23}. High-resolution images may also reveal disk-interstellar medium (ISM) interactions \citep[e.g.,][]{schneider18}.
	
	Accurate photosphere models provide the foundation for debris disk discovery and characterization. Detecting infrared excess from asteroid/Kuiper Belt analogs requires correctly extrapolating the star photosphere's Rayleigh-Jeans tail out to $\lambda > 20 \micron$. Near-infrared photometry is particularly important for photosphere modeling because for all stars with $T_{\rm eff} > 3000$~K, the $JHK$ bands fall on the Rayleigh-Jeans side of the blackbody (BB) function. Most debris disk searches use data from the Two Micron All Sky Survey (2MASS) \citep{cutri03,skrutskie19} to constrain models of the star's mid-infrared emission \citep[e.g.,][]{beichman06,bryden06, su06,  trilling08, bryden09, morales09, dodsonrobinson11, chen12}. Unfortunately, the nearest, brightest stars---whose debris disks would make the best JWST targets because they subtend large angles on the sky---saturated the 2MASS detectors.
	
	\begin{figure}[h]
		\centering
		\includegraphics[width=0.7\textwidth]{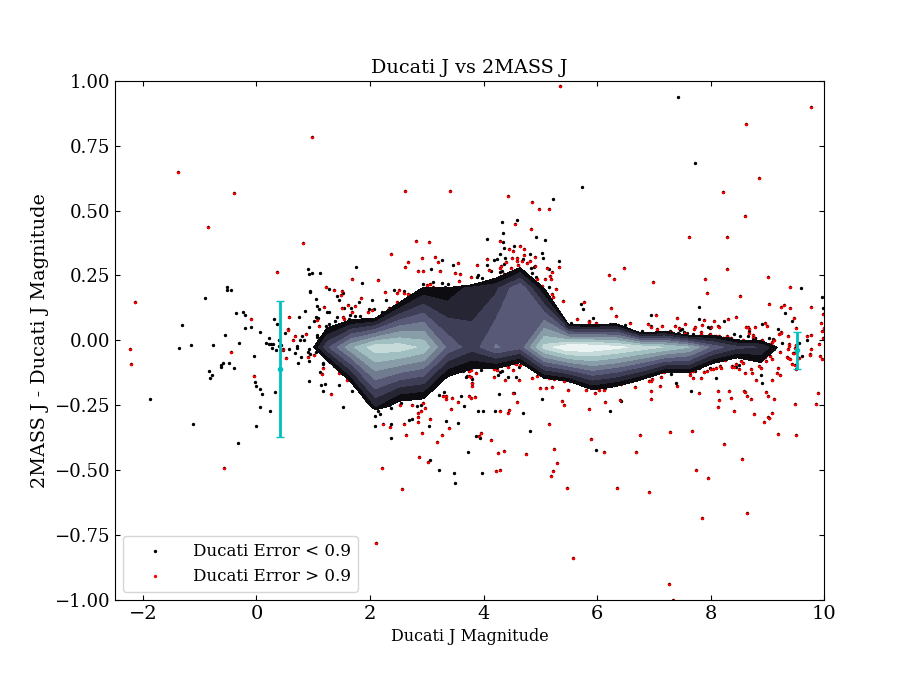}
		\caption{Scatter plot of \citet{ducati02} $J$ magnitude vs (2MASS $J$ -- Ducati $J$ magnitude). All the stars indicated as variable by Simbad have been excluded. Red points indicate observations with a Ducati $J$ magnitude error greater than $0.9$, while black points have a Ducati $J$ magnitude error less than $0.9$. A density plot has been overlaid on the most populous part of the diagram to show the distribution of data points. The sudden increase in scatter as the Ducati $J$ magnitude drops below 5 is indicative of 2MASS saturation. Two representative error bars have been placed on the plot. On the right is the average error in (2MASS $J$ -- Ducati $J$ magnitude) for stars with Ducati $J > 5.5$ (light blue), and on the left is the representative error in (2MASS $J$ -- Ducati $J$ magnitude) for stars with Ducati $J < 5.5$.}
		\label{fig:ducvs2mass}
	\end{figure}
	
	The 2MASS photometric system is reliably and consistently calibrated for fainter stars observed in the $1.3$ s exposure frames and for brighter stars observed in the $51$ ms frames (``Read 1''). However, the cores of the very brightest stars (magnitude $< 4$--$5$) are saturated ===even in the shortest exposures \citep{cutri06}. For these stars, the 2MASS project team estimated positions and magnitudes by fitting the nonsaturated wings of the stellar image. This procedure was primarily intended to allow bright star artifacts to be identified; the magnitudes appear in the 2MASS catalog but are flagged, not intended for astrophysical use. Errors are typically $0.1$--$0.2$ mag, up to 20\% of the star flux at that wavelength. Figure 1, which compares $J$ magnitudes from the \citet{ducati02} catalog and 2MASS, shows the effect of saturation on the 2MASS $J$ band. At $J \sim 5.5$, the scatter in Ducati $J$ vs (2MASS $J$ -- Ducati $J$) increases, as evidenced by the widening density overlay. Previous investigations have recognized the importance of avoiding saturated 2MASS photometry when characterizing debris disks: in their study of the $\eta$~Cru system, \citet{hengst17} used legacy $JHK$ photometry from \citet{morel78}, while \citet{ishihara17} used the IRSF telescope to collect new near-IR photometry for bright stars in the AKARI catalog.

	 Here we present new photosphere models for 62 known debris disk host stars with saturated 2MASS photometry. Then, using the new photosphere models, and using different statistical techniques  (see \S 2.3) from previous studies, we aim to reevaluate disk models to obtain new temperatures and sizes. For SED modeling, instead of 2MASS data, we use near-IR photometry from the NASA Catalog of Infrared Observations (CIO) \citep{gezari93, ducati02}, archival $R_c$ and $I_c$ photometry from \citet{cousins76}, \citet{bessell90}, \citet{mermiliod97}, and \citet{mallama14}; and saturation-corrected WISE W1 photometry \citep{cutri12data, patel14}. For each star, we calculate the excess fluxes in WISE, Spitzer, and Herschel mid- and far-IR passbands, then use BB and modified blackbody (MBB) models of the dust emission to predict the disk angular size. We discuss our target selection and data analysis procedures in \S \ref{sec:targets}. In \S \ref{sec:photospheres}, we present our new stellar photosphere models. In \S \ref{sec:diskmodels}, we re-analyze the dust properties for systems that have statistically significant infrared excess and identify the disks that should be resolvable with JWST's NIRCam and/or MIRI coronagraphs. We present our conclusions in \S \ref{sec:conclusion}.

\section{Target selection and photosphere modeling}
\label{sec:targets}

	\subsection{Target selection}
	
	This study includes 62 main-sequence stars with spectral types ranging from A to G. We chose stars that (1) are within 100 parsecs or less, so that reddening corrections to the optical photometry are not required \citep{redfield04}; (2) have saturated 2MASS photometry ($K_s < 5$); and (3) have infrared excess detections reported in the literature. Targets were mainly drawn from the \citet{chen14} meta-analysis of debris disks previously identified by the Spitzer Infrared Spectrograph (IRS) and MIPS surveys \citep{rieke05,beichman05,beichman06,bryden06,su06}. In addition, we added targets from catalogs by \citet{gaspar13}, \citet{patel14}, and \citet{ishihara17}. We removed 38 stars without $R_C$, $I_C$, or non-2MASS $JHK$ photometry from our sample (table \ref{tab:starsexclude}). Well-known and characterized disks such as Vega and Fomalhaut were also removed, as was $\eta$~Cru, which was the subject of a recent paper that examined the SED and resolved Atacama Large Millimeter/submillimiter Array (ALMA) image \citep{yang24}. 
	
	\begin{deluxetable*}{c}
		\tablecaption{Stars excluded due to a lack of either $R_c$, $I_c$, or non-2MASS $JHK$. \label{tab:starsexclude}}
		\tablewidth{6in}
		\tabletypesize{\normalsize}
		\tablehead{
			\colhead{Hipparcos Number} 
		} 
		\startdata
		1292\\
		1499\\
		7978\\
		8241\\
		11477\\
		12484\\
		13141\\
		21010\\
		25486\\
		30503\\
		30593\\		
		33719\\
		37891\\		
		40693\\
		40843\\
		43797\\
		49809\\
		53699\\
		56830\\		
		61174\\
		62904\\
		64924\\
		65530\\
		69989\\
		72848\\
		82587\\
		85537\\
		92024\\
		93815\\
		99188\\
		100412\\
		101507\\
		105312\\
		107350\\
		107649\\
		114924\\
		114948\\
		118121\\	
		\enddata	
		
	\end{deluxetable*}

	\subsection{Stellar Photosphere Models}
	
	To construct our photosphere models, we collected $B_T$ and $V_T$ photometry from the Hipparcos/Tycho catalogs \citep{perryman97, hog00} and $R_C$ and $I_C$ photometry from \citet{cousins76}, \citet{bessell90}, \citet{mermiliod97}, and \citet{mallama14}. For near-IR photometry, we used the catalog compiled by \citet{ducati02}, which in turn draws from the NASA CIO \citep{gezari93}. The CIO includes hundreds of thousands of individual observations of about 20,000 sources. Near-IR observations were taken between 1963 and 1990. The CIO is the most comprehensive source of near-IR photometry other than 2MASS. We also added W1 photometry from the WISE All-Sky catalog \citep{cutri12,wise20} to our collection of near-infrared photometry. Since stars with saturated $K_s$ are also saturated in W1, we adopted the saturation correction developed by \citet{patel14}, which is benchmarked against the systematic W1 vs.\ $K_s$ -- W1 trend for nonsaturated stars.
	
	 We note that for the stars with spectral types A, F, and G that we study here, W1 ($3.4 \micron$) is unambiguously on the Rayleigh-Jeans side of the star SED. In principle, the saturation-corrected W1 photometry alone could be a sound basis for synthetic flux predictions at mid- and far-IR wavelengths, with no $J$, $K$, $R_c$ or $I_c$ data required. However, the profile fitting method used to estimate W1 fluxes for saturated stars is less reliable for W1 $\lesssim 4.5$~mag \citep{cutri12}, so the additional information provided by ground-based photometry is important for this study, which includes 43 targets with W1 $< 4.5$. Table \ref{tab:optical} lists the optical and near-IR photometry used for each SED fit. Many targets do not have $JK$ photometry, but most have $R_C$ and $I_C$, and all have W1, which means all photosphere models have constraints on the Rayleigh-Jeans side of the SED. 
	
\begin{figure}[h]
		\centering
		\includegraphics[width=0.8\textwidth]{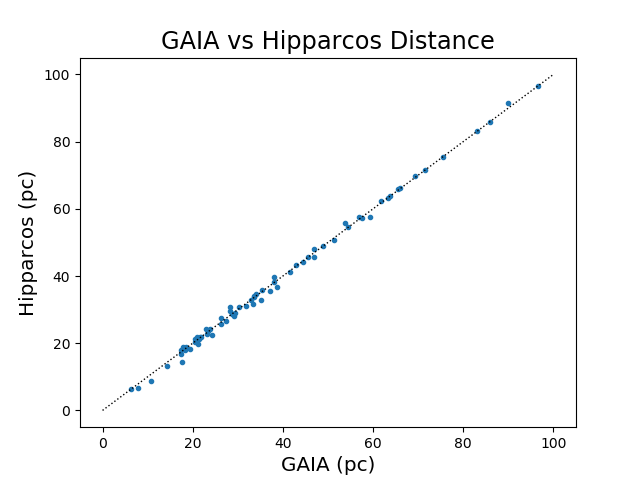}
		\caption{Comparison of Hipparcos and Gaia distance measurements. The black dotted line denotes exact matches.}
		\label{fig:distance}
	\end{figure}

	For the $B_T$, $V_T$, $R_C$, $I_C$, $J$, and $K$ bands, we converted magnitudes to flux in Janskys using the zero-point fluxes from \citet{mann15}. The zero point and color correction for W1 were given by \citet{wright10}. Star distances, which are necessary to scale the observed flux to the emitted flux, were obtained from the Gaia Data Release 2 \citep[GR2,][]{brown18,gaia20}. However, the Gaia DR2 is known to have systematic errors for very bright stars \citep{drimmel19}. 
Since previously published SED fits to our target stars were created using Hipparcos photometry, we compared the distances measured by Gaia and Hipparcos for the stars in our sample (Figure \ref{fig:distance}). For all targets, the two measurements agree within a factor of 20\% or better. Distance errors at the 20\% level primarily affect the fitted flux distance factor $\mathcal{F} = (R_* / d)^2$, where $R_*$ is the star radius and $d$ is the distance to the star. The factor takes into account the fact that the intensity of a point light source is proportional to the inverse square of the distance. Distance errors at the 50\% level or more can change the best-fit value of $\log g$.

	
	For SED fitting, we used the VOSA online tool hosted by the Spanish Virtual Observatory (SVO)\footnote{http://svo2.cab.inta-csic.es/theory/vosa/} \citep{bayo08}. VOSA uses a $\chi^2$ minimization to fit user-uploaded photometry with a model SED. VOSA features an extensive suite of model atmospheres. 
	We used the Castelli--Kurucz ATLAS9 atmosphere library for SED modeling \citep{castelli04}. The models have three parameters: effective temperature $T_{\rm eff}$, $\log \; g$, and metallicity [Fe/H].  VOSA also provides Bayesian posterior probability calculations for each parameter, which we used to calculate the uncertainties on our fitted temperatures (see \S \ref{subsec:excessidentification}). As for the calculation of synthetic flux, SVO contains a library of filter profiles and zero-points from the literature \citep{rodrigo1, rodrigo2,rodrigo24}, which includes the work of \citet{mann15}. We will discuss the calculation procedure over the course of the next section.



	\subsection{Infrared Excess Identification}
	\label{subsec:excessidentification}
	
	To detect and characterize orbiting dust, we used archival mid-IR and far-IR photometry from the following instruments and filters: WISE W3 and W4 \citep{cutri12}; Spitzer MIPS $24$ and $70 \micron$, Herschel PACS $70$, $100$, and $160 \micron$, and Herschel SPIRE $250$, $350$, and $500 \micron$ \citep{eiroa13, gaspar13, chen14, cotten16, pawellek21,marton24}. For Hip 41307 (30 Mon), we added photometry from T-ReCS on the Gemini South telescope at $10.4$ and $18.3 \micron$ \citep{moerchen10}. For Hip 10670 and Hip 87108, we added the $850 \micron$ observations from the SONS survey \citep{holland17}. Finally, for Hip 22263, we added the SCUBA photometry from \citet{greaves09}. Most debris disk surveys with MIPS conducted observations at both $24$ and $70 \micron$, though Hip 113044 was observed at $70 \micron$ but not $24 \micron$. We excluded MIPS $70 \micron$ observations in which the star--disk system was not detected and only an upper limit on flux was available.
 Although the Akari $L18W$ data set would have helped characterize warm dust, we did not use it because the Akari fluxes were always at least $2 \sigma$ higher than the photosphere models at $18 \micron$, even for stars that had no excess in the neighboring W4 or MIPS $24 \micron$. We suspect there is a zero-point offset between Akari and W4/MIPS.
	
	In addition to the WISE, MIPS, and Herschel photometry, we used calibrated Spitzer IRS staring-mode spectra from the CASSIS archive \citep{lebouteiller11}.
We based our analysis on data from the LL1, and LL2, and SL1 modules, which cover $19.91$--$39.90$, $13.90$--$21.27$, and $7.46$--$14.29 \micron$ respectively. The 1D spectra in the CASSIS archive were extracted from the 2D basic calibrated data using the SMART-AdOpt technique of \citet{lebouteiller10}, which includes both an oversampled point-spread function (PSF) and a relative spectral response function (RSRF) constructed by comparing observed spectra of calibrator stars with theoretical models. Use of the RSRF greatly reduces instrumental artifacts such as the commonly encountered $24 \micron$ flux deficit.\footnote{A description of instrumental artifacts is available at https://irsa.ipac.caltech.edu/data/SPITZER/docs/irs/features/.} We integrated the IRS spectra to calculate spectrophotometry in three wavelength bands: $28$--$32$, $15$--$19$, and $11$--$15 \micron$. The $28$--$32 \micron$ band, which traces warm dust with $T_{\rm dust} \ga 115$~K, was defined by \citet{lawler09}. The $15$--$19 \micron$ band traces temperatures of $T_{\rm dust} \ga 220$ K and provides similar information to what one might obtain from Akari $18 \micron$. Finally, the $11$--$15 \micron$ band is sensitive to hot dust with $T_{\rm dust} \ga 280$~K. In the subsequent text, we will refer to the IRS spectrophotometry using each band's central wavelength: $30$, $17$, and $13 \micron$.

For all IRS staring-mode targets that also have MIPS $24 \micron$ photometry, 
we normalized the LL1 spectrum by integrating across the MIPS $24 \micron$ bandpass, then scaling the result to the observed MIPS $24 \micron$ flux. We then scaled the LL2 spectrum to match the LL1 spectrum in the overlap region around $20 \micron$. Hip 2505, Hip 20635, Hip 66234, and Hip 95347 have IRS spectra but no MIPS $24 \micron$ observations, so the normalization procedure was performed using W4 photometry instead. For Hip 113044, which has a calibrated IRS spectrum in the CASSIS archive but no MIPS $24 \micron$ or W4 photometry, we assumed that the 10 shortest-wavelength LL2 resolution elements traced the photosphere, then matched the LL1 data to the LL2 in the overlap region. Since Hip 10670 was observed in IRS mapping mode instead of staring mode and therefore does not have a calibrated spectrum in the CASSIS archive, we used the $13$ and $31 \micron$ spectrophotometry tabulated by \citet{chen14} instead of deriving our own. Finally, for Hip 23693, we excluded the $13 \micron$ data since the SL1 spectra for the object have an unphysical positive $dF_{\nu}/d \lambda$ trend and noticeable slit-loss effect. See Tables \ref{tab:IR1} and \ref{tab:IR2} for the full list of mid- and far-IR photometry.

\begin{figure}[h]
	
	\centering
	\begin{tabular}{cc}

	\setlength{\fboxsep}{20pt} 
    \setlength{\fboxrule}{0pt}
    \fbox{\includegraphics[width=.43\textwidth,height=.40\textwidth]{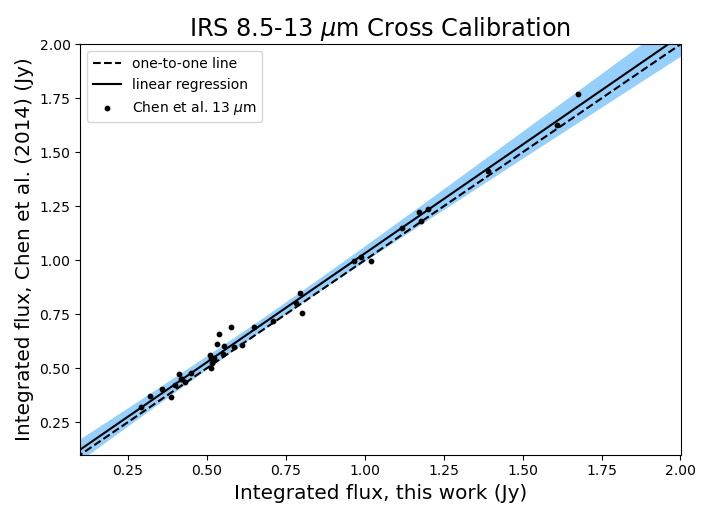}} &
		\setlength{\fboxsep}{0pt} 
    \setlength{\fboxrule}{0pt}
    \fbox{\includegraphics[width=.43\textwidth,height=.40\textwidth]{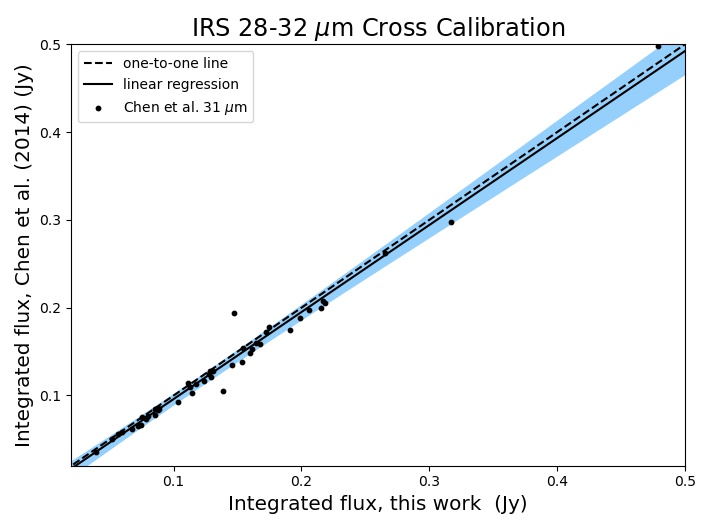}} \\
	\end{tabular}
	\caption{Comparison of IRS flux density measurements between this work and \citet{chen14} at $8.5$--$13$ (left) and $28$--$32 \micron$ (right). The solid line indicates the best-fit linear regression, while the dashed line indicates a perfect one-to-one match. The blue shaded area represents the 95\% confidence interval on the linear regression. We find good agreement between our results and those of \citet{chen14}.
	}
	\label{fig:irscomparison}
\end{figure}

To verify the reliability of our IRS spectrophotometry, we perform a cross-calibration with \citet{chen14} by measuring integrated fluxes in the two pseudofilters they used: $8.5$--$13 \micron$, which was designed to probe silicate emission features, and $28$--$32 \micron$. (Note that we do not use the $8.5$--$13 \micron$ pseudofilter for science, as our focus is on BB continuum emission rather than mineralogy; we instead use the longer-wavelength bands centered at $13$ and $17 \micron$.) Figure \ref{fig:irscomparison} shows the results of the cross-calibration. In each panel, the slope of a linear regression (solid line) is consistent with 1, and all flux measurements agree within the error bars (omitted from the plot in order to improve readability).
	
\begin{figure}[h]
		\centering
		\includegraphics[scale=0.9]{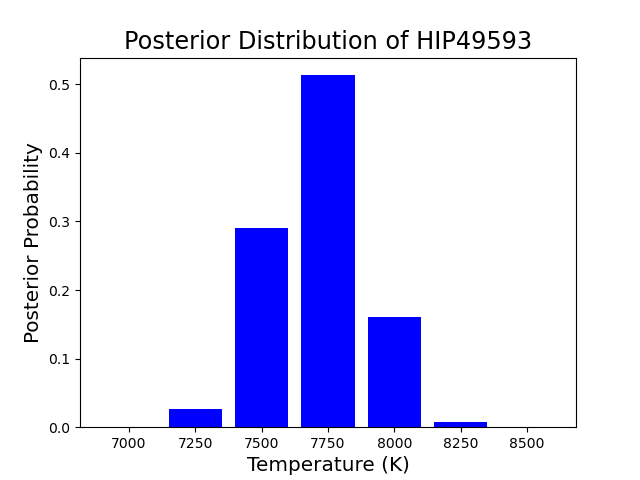}
		\caption{Posterior probability distribution of Hip 49593 $T_{\rm eff}$. The accompanying Zenodo archive \citep{zenodo} shows $T_{\rm eff}$ posterior probability distributions for all target stars.}
		\label{fig:posterior_example}
	\end{figure}

To detect orbiting dust, we searched all mid- and far-IR wavelengths for excess emission. 
The significance of an infrared excess is determined in part by the error on the synthetic flux prediction. When fitting SEDs, VOSA performs Bayesian inference calculations, allowing the user to examine the posterior probability distributions for the free parameters $T_{\rm eff}$, $\log g$, and [Fe/H]. In turn, the user can calculate the posterior probability that an SED model corresponding to a specific combination of atmospheric parameters is the correct model. 
We calculated the synthetic-filter fluxes associated with each model contributing to the $T_{\rm eff}$, $\log g$, and [Fe/H] posterior probability distributions by integrating the model across the filter transmission profile,
	\begin{equation}
		F_{x,i} = \frac{\int f_i(\lambda) T_x(\lambda) d\lambda}{\int T_x(\lambda) d\lambda},
	\end{equation}
where $F_{x,i}$ is the synthetic flux for model $i$ and filter $x$, $f_i$ is the flux density of the SED model, and $T_x(\lambda)$ is the transmission as a function of wavelength $\lambda$ for filter $x$ \citep{mann15}. We then found the error bar $s_x$ on each synthetic-filter flux prediction $F_x$ by calculating the root mean square error of all values of $F_{x,i}$ weighted based on the posterior probability of their corresponding models,
	\begin{equation}
	s_x = \sqrt{\sum_i w_i (F_{x,i} - \overline{F_x})^2},
	\label{eq:modelstd}
	\end{equation}
where $\overline{F_x}$ is the average predicted flux in filter $x$ from all models and the weighting factor $w_i$ is the posterior probability of model $i$.
Figure \ref{fig:posterior_example} shows an example posterior probability distribution for the effective temperature of the A7V star Hip 49593. The accompanying Zenodo archive \citep{zenodo} at doi:10.5281/zenodo.21327122 shows temperature posterior distributions for the rest of the stars. Since the model atmosphere grid has a coarse resolution of 250~K, there are some stars for which only one model $T_{\rm eff}$ has significant probability in the posterior distribution. In such cases, the SED model uncertainty is reflected in the $\log g$ posterior probability distribution.

	
	As in \citet{sibthorpe18}, we designate an infrared excess as when the observed flux in filter $x$, $F_{o,x}$, is more than three standard deviations above the synthetic-filter flux calculated from the model photosphere:
	\begin{equation}
	F_{o,x} - F_x > 3 \sigma_x
	\end{equation}
The error $\sigma_x$ for excess calculation is the quadrature sum of the observed photometric error $\sigma_{o,x}$ and the predicted synthetic-filter flux error:
	\begin{equation}
	\sigma_x = \sqrt{s_x^2 + \sigma_{o,x}^2}.
	\label{eq:sigma}
	\end{equation}
Published Spitzer MIPS and Herschel uncertainties include contributions from the sky background, detector/photon noise, and calibration uncertainty (see \citet{bryden06} for MIPS and \citet{sibthorpe18} for Herschel). Uncertainties listed in the WISE All-Sky Catalog do not include the flux calibration contribution. For W3 and W4, we take $\sigma_{o,x}$ to be the quadrature sum of the published error bars and the calibration uncertainties, which are 4.5\% and 5.7\%, respectively \citep{jarrett11}.
	
	\subsection{Disk SEDs}
	\label{subsec:diskseds}
	
	Our basic model for each debris disk SED is 
the BB function
	\begin{equation}
		F_{\lambda} = B_{\lambda}(T) = A \frac{2 \pi h c^2}{\lambda^5} \frac{1}{e^{h c/ \lambda k_b T_{\rm dust}} - 1},
		\label{eq:blackbody}
	\end{equation}
where $h$ is Planck's constant and $k_B$ is Boltzmann's constant. The two parameters are the dust temperature $T_{\rm dust}$ and the multiplicative constant $A$. In reality, debris disks are not perfect blackbodies, but there are good reasons for making this simplification. Most star--disk systems in our sample do not have infrared excess detections in enough wavelengths to warrant a more complex model, which would have at least four free parameters in the most simplistic single-disk radiative transfer scenario: dust mass, inner disk radius, outer disk radius, and minimum grain size. The grain size number density distributions would have to be prescribed.
Blackbody models incorporate the assumption that all dust grains have the same temperature regardless of size. Such models provide a lower limit to the disk's distance from the star, as small grains ($1$--$10 \micron$) have to superheat above the BB temperature in order to emit light in the mid- and far-infrared \citep{kruegel03}.
	
	We use different BB fitting procedures for targets with infrared excess detections in different numbers of filters. For stars with infrared excess detected in only a single filter, we assume that the wavelength $\lambda_{\rm max}$ of the peak BB flux density $F_{\nu}$ coincides with the central wavelength of the filter \citep[e.g.][]{bryden09, dodsonrobinson11}. Wien's law then yields a dust temperature and luminosity estimate.
The approximation that the excess comes from near the peak of the BB function is somewhat justifiable given that our sample's smallest fractional infrared excesses belong to disks detected in a single filter. In other words, if we see infrared excess at wavelengths far from $\lambda_{\rm max}$, we expect to find excess at $\lambda_{\rm max}$ as well, yielding detections on at least two wavelengths.

For most systems with infrared excess at two wavelengths (seven exceptions will be discussed below), the two-parameter dust BB model is uniquely specified. We use a root-finding algorithm to numerically solve for $T_{\rm dust}$ and $A$ in Equation \ref{eq:blackbody}. We estimate error bars on $T_{\rm dust}$ and $A$ as follows. For each wavelength at which infrared excess is detected, we create a new realization of the excess flux by randomly sampling a normal distribution with the observed flux $F_{o,x}$ as the mean and the reported uncertainty $\sigma_{o,x}$ as the standard deviation. We then use the root-finding algorithm to find the values of $T_{\rm dust}$ and $A$ that
match the simulated observations. By repeating the random-sampling process 2000 times, we create a probability distribution for the dust temperature based on the observational uncertainties. The best-fit dust temperature is the median of the distribution, and the error on the dust temperature is determined by the 84th and 16th percentiles. Note that we do not fit two-parameter BB models to targets with infrared excess in W4 and MIPS $24 \micron$ only, as the two filters have overlapping wavelength coverage. Instead, we treat such targets as having a single-filter excess detection in MIPS $24 \micron$.

\begin{figure}[h]
	\centering
	\includegraphics[width=0.7\textwidth]{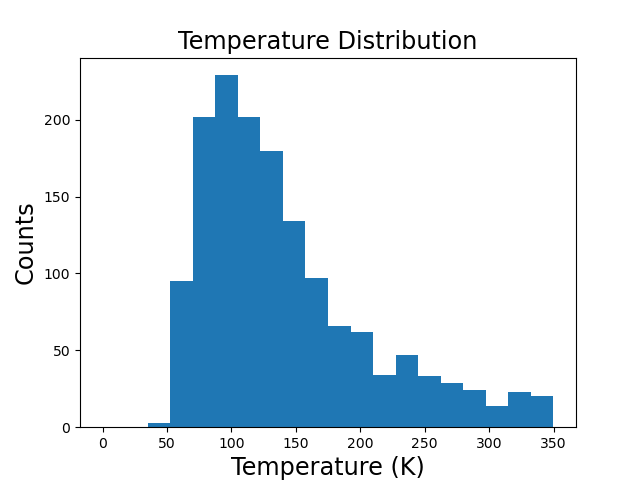}
	
	\caption{The Hip 23693 dust temperature posterior distribution. Hip 1473, Hip 42438, Hip 43726, Hip 65721, Hip 103389, and Hip 111188 show similarly wide dust temperature posterior distributions.}
	\label{fig:2pthist}
\end{figure}

Hip 1473, Hip 23693, Hip 42438, Hip 43726, Hip 65721, Hip 103389, and Hip 111188 have excess detections at $70$ and $100 \micron$, but the observed $70/100 \micron$ flux ratio is nearly Rayleigh-Jeans. In such cases, the excess fluxes do not specify a unique solution to Equation \ref{eq:blackbody}; instead, parameters $T_{\rm dust}$ and $A$ are heavily degenerate, and a range of solutions is possible (see histogram in Figure \ref{fig:2pthist}). Since none of the seven stars have an infrared excess detection in MIPS $24 \micron$, we incorporate the upper limits on the $24 \micron$ dust flux to find a BB model. Following \citet{sawicki12}, we minimize a modified $\chi^2$ that includes contributions from upper limits:
\begin{equation}
    \chi^2 = \sum_i \frac{(\mathcal{D}_{o,i} - \mathcal{D}_i)^2}{\sigma_i^2} - 2 \sum_j \ln \left \{ \sqrt{\frac{\pi}{2}} \: \sigma_j \left[ 1 + {\rm erf} \left( \frac{\mathcal{D}_{{\rm lim},j} - \mathcal{D}_j}{\sqrt{2} \sigma_j} \right) \right] \right \}.
    \label{eq:chi2limits}
\end{equation}
In Equation \ref{eq:chi2limits}, erf is the error function, subscript $i$ denotes filters with excess detections, and subscript $j$ indicates filters without $3 \sigma_x$ excesses (Equation \ref{eq:sigma} with $x = i$ or $x = j$). $\mathcal{D}_{o,i}$ is the observed dust flux (i.e. the infrared excess), $\mathcal{D}_i$ is the dust flux predicted by the BB model, and $\sigma_i$ and $\sigma_j$ are the dust flux uncertainties in filters $i$ and $j$. $\mathcal{D}_{{\rm lim},j} = 3 \sigma_j$ is the upper limit to the dust flux in filter $j$. The values of $T_{\rm dust}$ and $A$ found by modified $\chi^2$ minimization are physically reasonable but not necessarily unique. We do not attempt to calculate uncertainties on either parameter.

Finally, we note that Hip~1473 has significant excess in three filters, but we discarded the PACS $100 \micron$ photometry because of obvious background emission, which is visible in the Herschel Science Archive image, and followed the BB model fitting procedure for targets with infrared excess in two filters only.
	
	For the 13 stars with infrared excess at three or more wavelengths, the BB dust model is overconstrained (i.e., more data points than free parameters). Here we find the best-fit BB model by using \texttt{emcee} \citep{foreman13} to perform a Markov Chain Monte Carlo (MCMC) exploration of the $(T_{\rm dust}, A)$ parameter space. We choose uniform priors for the model parameters: $10$~K $< T_{\rm dust} < 500$~K, $1 < A < 10,000$. We assume a Gaussian distribution for the log-likelihood function, 
	\begin{equation}
		\log \mathcal{L} = -\sum_x \frac{(F_{o,x} - F_x)^2}{2 \sigma_x^2} - n \: \log{\sigma_x} - \frac{n}{2}\log(2 \pi),
	\end{equation}
where $\sigma_x$ is the infrared excess error given by Equation \ref{eq:sigma} and $n$ is the number of infrared excess measurements. We determine the best-fit BB model parameters by finding the 16th, 50th, and 84th percentiles of the $T_{\rm dust}$ and $A$ posterior distributions from each MCMC simulation.

For 11 stars with infrared excess at four or more wavelengths, we explore more complex models in addition to the BB. We use \texttt{emcee} to fit (1) the MBB model developed by \citet{backman93}, (2) the two-component (2BB) disk model \citep[e.g.,][]{chen14, su14, ballering17}, which combines emission from two blackbodies that represent spatially separated dust rings such as our own asteroid belt and the Kuiper belt, and (3) a 2BB disk model consisting of a cold MBB ($< 120$~K) and a warm BB (BB+MBB). The MBB is given by
\begin{equation}
F_{\lambda} = \left[ H(\lambda_0 - \lambda) + H(\lambda - \lambda_0) \left( \frac{\lambda}{\lambda_0} \right)^{-\beta} \right] B_{\lambda}(T),
\label{eq:mbb}
\end{equation}
where $H()$ is the Heaviside function and the free parameters are $T_{\rm dust}, A, \lambda_0, \beta$. The MBB is especially useful for targets with infrared excess detected by Herschel since it accounts for the fact that dust grains have poor emissivity at wavelengths much longer than their circumference \citep{booth13, pawellek14}.

Although the MBB, 2BB, and BB+MBB often yield SED fits with higher likelihoods than the BB model, the fit improvement is not always enough to justify increasing the number of free parameters from two for the BB to four for the MBB and 2BB or six free parameters for the BB+MBB. We choose between the BB and the more complex models using the likelihood ratio test. We calculate the test statistic
\begin{equation}
\mathcal{R} = -2 (\log \mathcal{L}_{bb} - \log \mathcal{L}_{\rm complex}),
\label{eq:lrtest}
\end{equation}
where $\mathcal{L}_{bb}$ is the likelihood of the best-fit BB model and $\mathcal{L}_{\rm complex}$ is the likelihood of the best-fit MBB or two-component model (2BB or BB--MBB). $\mathcal{R}$ follows a $\chi^2_K$ distribution, where $K$ is the number of free parameters that must be added to the BB model to form the complex model. For a risk level of $P(\mathcal{R} > \mathcal{R}_c) = \int_{\mathcal{R}_c}^{\infty} \chi^2_K(\mathcal{R}) \: d\mathcal{R} = 0.01$, we select a critical value of $\mathcal{R}_c = 9.21$ if $K = 2$ and $\mathcal{R}_c = 13.28$ if $K = 4$ (i.e., the BB model is being compared to the BB--MBB model). If $\mathcal{R} > \mathcal{R}_c$, we reject the null hypothesis that the BB is an adequate model and select the more complex model. When the likelihood ratio test indicates that both the MBB and the 2BB model are preferred over the simple BB, we choose the model with the highest likelihood; when the BB--MBB is preferred over the simple BB model, we do an additional likelihood test between the BB--MBB model and the single MBB.

\section{Results: Star Photospheres}
\label{sec:photospheres}

Here we examine the effects of replacing saturated 2MASS photometry with older, unsaturated NASA CIO and/or ground-based $R_C I_C$ photometry along with saturation-corrected W1. The best-fit values of $T_{\rm eff}$ and $\log g$ are listed in Table \ref{tab:IR1}. In \S \ref{subsec:literaturecompare} we compare our model photospheres with high-resolution spectroscopy and previous SED fits from the literature. We assess the degree to which binarity might affect our infrared excess measurements in \S \ref{subsec:binary}.

\subsection{Comparison with literature results}	
\label{subsec:literaturecompare}

\begin{figure}[ht]
		\centering
		\includegraphics[width=0.9\textwidth]{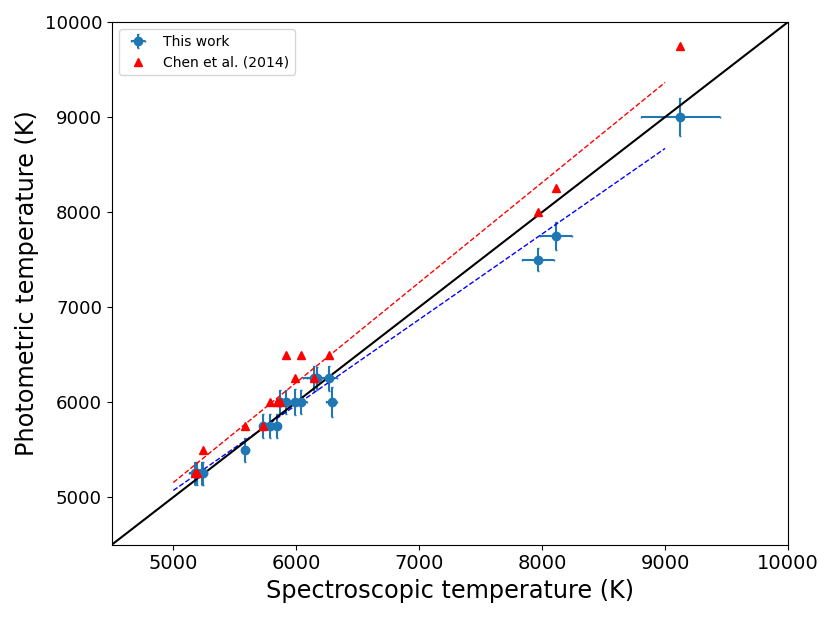}
		\caption{A comparison between spectroscopic temperature and photometric temperature for the stars in this study (blue) and \citet{chen14} (red). A linear fit was performed for both datasets. In general, this study generates temperatures closer to spectroscopic observations.}
		\label{fig:tempcompare}
	\end{figure}

For stars with ground-based, high-resolution spectroscopy, we test the performance of our photosphere models by comparing $T_{\rm eff}$ derived in this work with $T_{\rm eff}$ derived from high-resolution spectroscopy. We note that some of our target stars have not been observed with high-resolution spectrographs. The FG stars are more likely to have spectroscopic $T_{\rm eff}$ measurements because they have been surveyed by planet searches. Sources of spectroscopic data are listed in Table \ref{tab:IR1}. Figure \ref{fig:tempcompare} compares $T_{\rm eff}$ measurements from our new photosphere models vs.\ high-resolution spectroscopy for 19 stars in our study (blue). Since 16 of those stars had photosphere models from \citet{chen14}, we also plot the Chen et al.\ model $T_{\rm eff}$ as a function of spectroscopic $T_{\rm eff}$ (red). In Figure \ref{fig:tempcompare}, 15/19 targets have NASA CIO-based model SED temperatures within one error bar of the spectroscopic $T_{\rm eff}$ measurements, where the error bar on the model SED $T_{\rm eff}$ is measured from the VOSA posterior probability distribution. The blue dashed line fitted to the targets with CIO-based photosphere models closely follows the black line indicating equal photometric and spectroscopic $T_{\rm eff}$. The red dashed line is fitted to the 16 targets from \citet{chen14}. The Chen et al.\ photosphere models examined here have a modest temperature offset from the spectroscopic data, though this small sample may not be representative of their entire catalog.

		\begin{figure}[ht]
		\centering
		\includegraphics[width=0.7\textwidth]{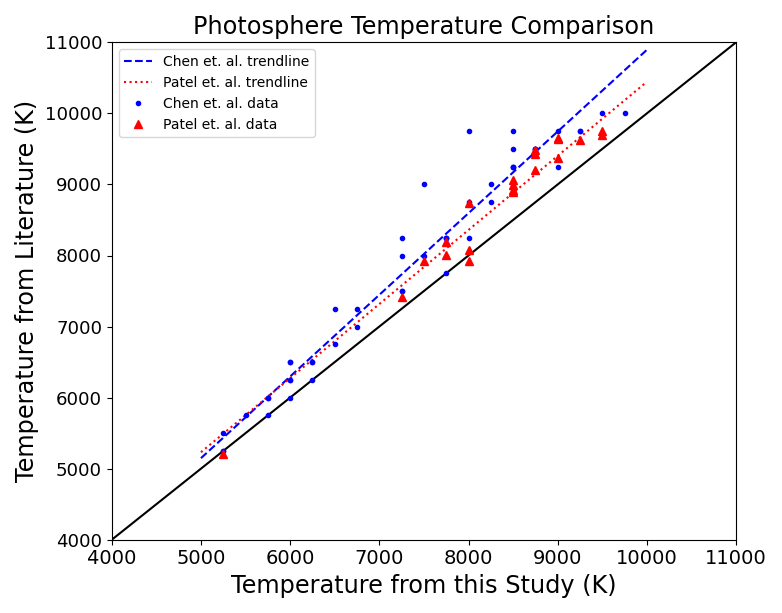}
		\caption{A comparison of photosphere model temperatures from this study (Set 1), \citet{patel14}, and \citet{chen14}. }
		\label{fig:litcompare}
	\end{figure}
	
	
	In addition to benchmarking our results against high-resolution spectroscopic data, we also compared our photosphere temperatures directly with those of \citet{chen14} and \citet{patel14} (Figure \ref{fig:litcompare}). Combined, those two studies have 56 stars in common with this work. NASA CIO and ground-based $R_C I_C$ photometry tend to record higher fluxes in the near-IR than saturated 2MASS data (Figure \ref{fig:ducvs2mass}), which means our model photospheres have to produce redder colors than 2MASS-based models. Therefore, our new photosphere model temperatures tend to be slightly lower than those of \citet{chen14}, as recorded by the blue dashed line fitted to the data in common to both studies. Our results are more consistent with those of \citet{patel14}, who used 2MASS $JHK_s$ and saturation-corrected WISE near-IR photometry in order to derive star temperatures.
	

	Although the motivation behind our new set of SED fits (Set 2) was to remove the reliance on saturated 2MASS photometry, there are many factors that might cause a photosphere model in this study to differ from a previously published model: filter profiles and zero-points, model atmospheres chosen, functionality of the SED fitting software, star distances, and more. In an attempt to isolate the effects of changing the SED data sources from 2MASS to CIO and saturation-corrected W1, we computed a second set of SED models using 2MASS near-IR photometry and excluding CIO/W1 (Set 1). While we cannot replicate SED models from the literature without actually using the authors' software, Set 2 is designed to mimic previous work from \citet{beichman06}, \citet{bryden09}, \citet{dodsonrobinson11}, and \citet{chen14}. 
	
\begin{figure}[h]
	\centering
	\includegraphics[width=\textwidth]{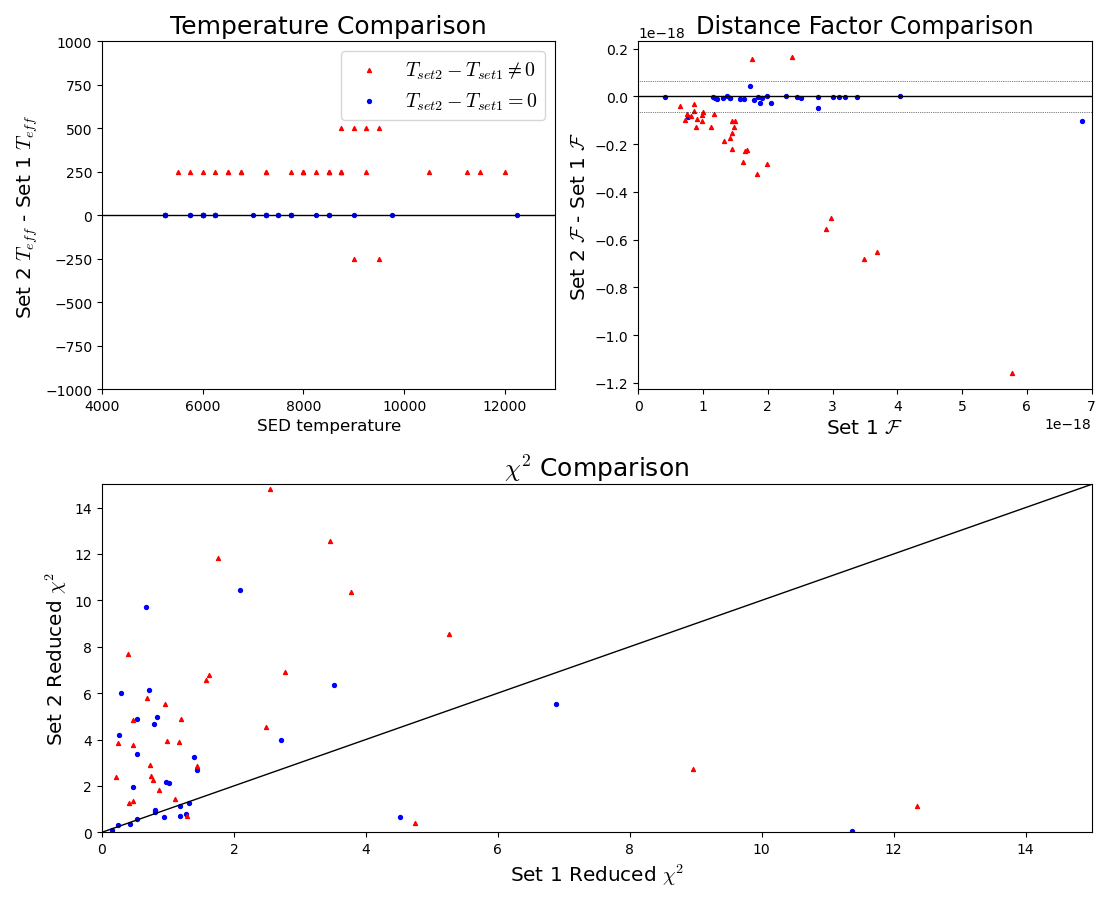}
	\caption{A comparison of photosphere temperature $T_{\rm eff}$, the distance factor $\mathcal{F}$, and reduced $\chi^2$ of SED fits obtained using archival CIO photometry plus saturation-corrected W1 (Set 1) and 2MASS photometry (Set 2). In the upper left panel, we see that the two sets of models predict different temperatures for 32 stars (shown as red triangles), and the same temperatures for 30 stars (shown as blue circles). Some of these temperature values overlap, since model parameter values come in discrete steps. On the upper right, our comparison of flux distance factor shows that 29 of the 32 stars with temperature disagreements also have statistically significant differences in $\mathcal{F}$. The dashed lines represent $\pm 3 \sqrt{s_{\mathcal{F}_1}^2 + s_{\mathcal{F}_2}^2}$. The bottom panel compares reduced $\chi^2$ from the two sets of photosphere models, with a black line representing equal values of reduced $\chi^2$. In general, fits in Set 1 have lower reduced $\chi^2$ than fits in Set 2.}
	\label{fig:sedmodelscompare}
\end{figure}

	In Figure \ref{fig:sedmodelscompare}, we compare the model $T_{\rm eff}$, reduced $\chi^2$, and the distance factor $\mathcal{F}$ from the two sets of SED models. The 2MASS (Set 2) and CIO-based (Set 1) SEDs yield temperatures different by 250~K--500~K for 32 of our target stars, which are plotted in red triangles in all panels of Figure \ref{fig:sedmodelscompare}. Out of these thirty two targets, twenty nine have factors that differ by more than $3 \sqrt{s_{\mathcal{F}_1}^2 + s_{\mathcal{F}_f2}^2}$, where $s_{\mathcal{F}_1}$ and $s_{\mathcal{F}_2}$ are the error bars on the fitted CIO and 2MASS-based distance factors computed as in Equation \ref{eq:modelstd}, with synthetic-filter flux replaced with the distance factor $\mathcal{F}$. These are all the red triangles outside the two horizontal dotted lines in the upper right plot. As for the 30 targets with the same temperature between Set 1 and Set 2, represented by blue circles, aside from two stars, their distance factor differences $s_{\mathcal{F}_1}$ and $s_{\mathcal{F}_2}$ are within the dotted lines representing $\pm 3 \sqrt{s_{\mathcal{F}_1}^2 + s_{\mathcal{F}_f2}^2}$. In other words, for 28 out of 30 objects with the same Set 1 and Set 2 temperature, the values of the distance factor between Set 1 and Set 2 are within the margin of error. It is also worth mentioning, for one of the blue data points, the value of the factor is larger than $10^{-17}$. Since adding the point would make the plot hard to see, we opted not to show it in the plot. Most Set 2 models have higher reduced $\chi^2$ values than their corresponding CIO-based models in Set 1 regardless of whether the model temperature differs (bottom panel of Figure \ref{fig:sedmodelscompare}). Note that 14 objects have fit reduced~$\chi^2 < 1$ in at least one set of models, which might suggest overestimation of the error bars for some photometric data.

\begin{figure}[ht]
	\hspace{-1.4cm}
	\includegraphics[width=1.\textwidth] {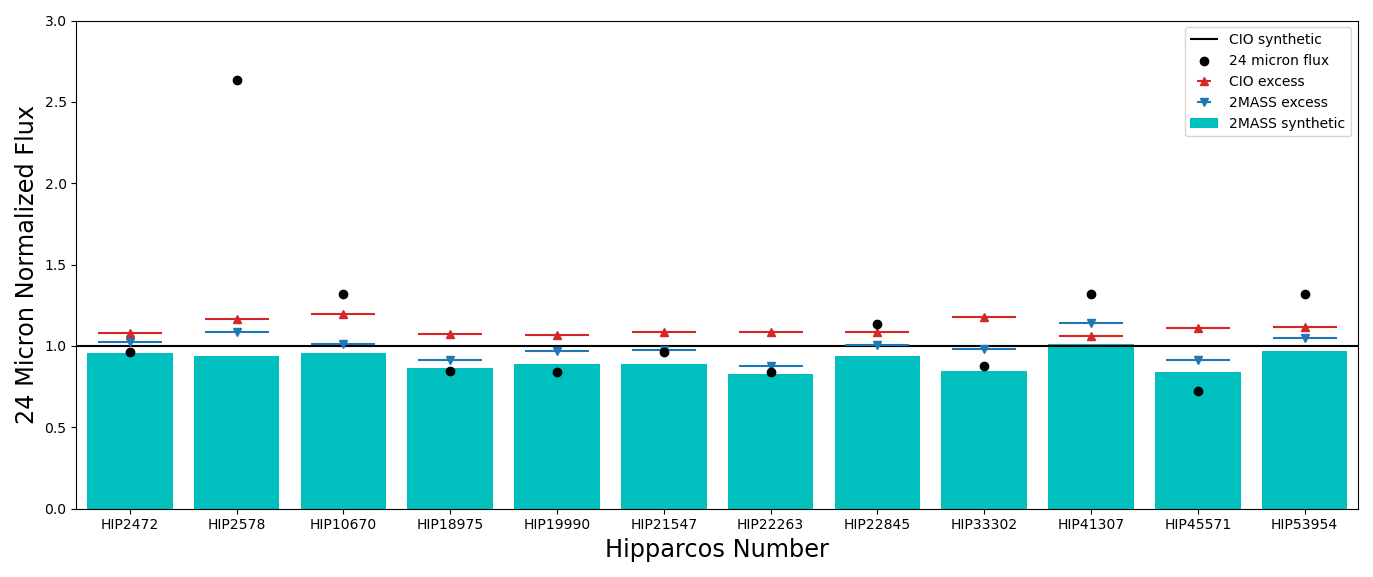}
	\includegraphics[width=1.\textwidth] {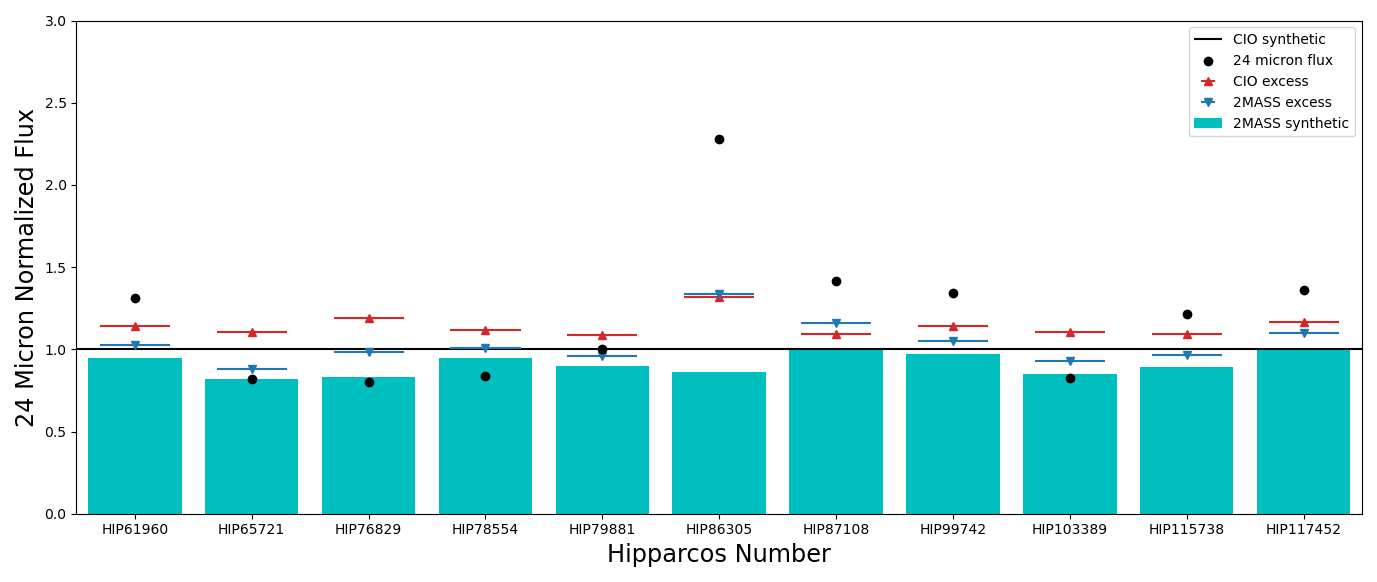}
	\caption{Comparison between predicted synthetic fluxes at $24 \micron$ for stars where the CIO-based and 2MASS-based photosphere models yielded different best-fit values of $T_{\rm eff}$. The vertical scale is normalized so that the predicted $24 \micron$ synthetic flux from CIO-based models is $1.0$. The green bar is the predicted synthetic flux from 2MASS-based models. The upward-pointing red triangles mark the infrared excess detection threshold for CIO-based models, while the downward-pointing blue triangles represent the excess detection threshold for 2MASS-based models. Black dots are the observed $24 \micron$ fluxes. Hip 21547 and Hip 79881 have $3 \sigma_x$ excess detection with the 2MASS-based photosphere models but not with CIO-based models.}
	\label{fig:diff_temp_24mic}
\end{figure}

	Figure \ref{fig:diff_temp_24mic} compares the photosphere model predictions in the MIPS $24 \micron$ band for the 23 stars with different model $T_{\rm eff}$ between CIO-based and 2MASS-based models that have available MIPS $24 \micron$ observations. Doing so, we seek to observe how fitting the two different models affects excess detection.
For the $24 \micron$ predicted flux, the difference between the two models can differ up to a factor of $18\%$. We see that the choice of photosphere model can occasionally make a difference in whether there is an infrared excess detection, as shown by Hip~79881, which has $3 \sigma$ excess in the predicted flux obtained from fitting 2MASS photometry, but not the predicted flux obtained from fitting CIO photometry. While for the most part, the effects of using different near-IR photometry for SED modeling do not have significant differences in excess detection, considering that most relative excess at $24 \micron$ tends to be less than $50\%$ in our samples, it is a factor worth considering.

 \begin{figure}
	\centering
	\begin{tabular}{cc}
		\includegraphics[width=0.50\textwidth]{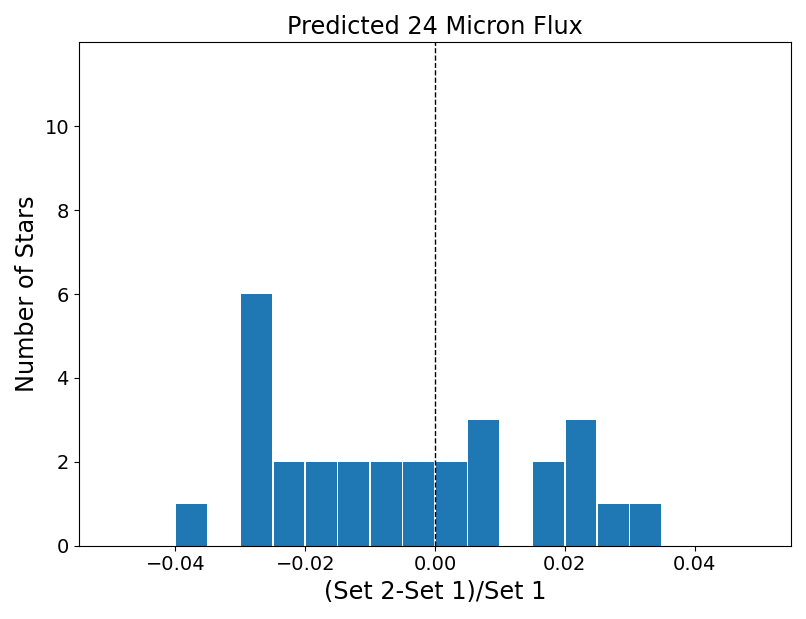} &
		\includegraphics[width=0.5\textwidth]{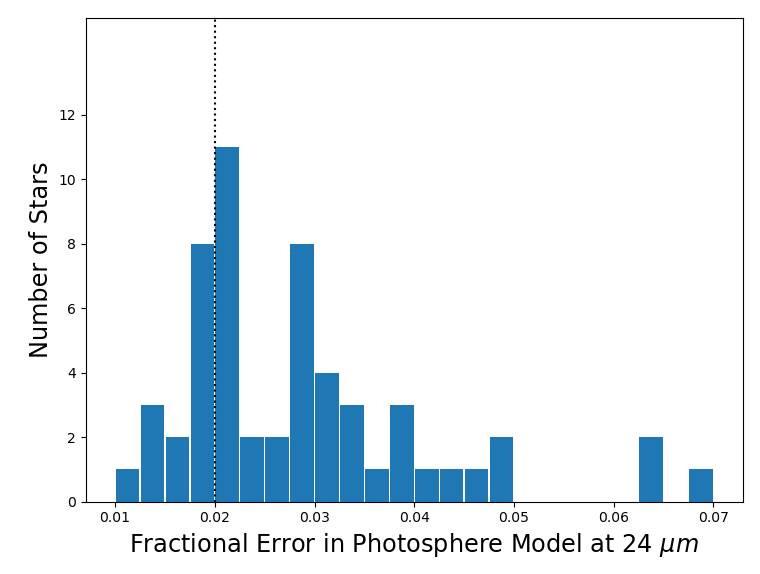} \\
	\end{tabular}
		\caption{Left: predicted $24 \micron$ synthetic flux comparison between Set 1 (archival CIO and saturation-corrected W1 photometry) and Set 2 (2MASS photometry). For seven stars, the difference in predicted flux is $> 1$\%. Right: fractional error on synthetic $24 \micron$ flux for models in Set 1. The vertical dotted line shows a fractional error of 2\%, as is commonly assumed in the literature.}
	\label{fig:24compare}
\end{figure}

 Finally, the left panel of Figure \ref{fig:24compare} compares the predicted synthetic $24 \micron$ flux from Set 1 and Set 2. As expected from the scatter in 2MASS $J$ magnitudes for bright stars with $J < 5$ in the \citet{ducati02} catalog (Figure \ref{fig:ducvs2mass}), there are seven stars for which the Set 1/Set 2 discrepancy in predicted $24 \micron$ star flux is $> 1$\%. For comparison, the fractional errors in the observed MIPS $24 \micron$ fluxes of Hip 23693 and Hip 49593 are both 1.0\%. Replacing saturated 2MASS photometry with infrared CIO, ground-based $R_C I_C$, and W1 photometry can affect the disk detection threshold, as seen in the next section. The right panel of Figure \ref{fig:24compare} shows $s_x / F_{x,i}$, the fractional error in the predicted $24 \micron$ flux from the CIO-based photosphere models. In our sample of 62 stars, 49 have $24 \micron$ synthetic flux errors above the commonly assumed value of 2\%. Our rigorous propagation of errors also affects our infrared excess detection rate.
 
 \subsection{Binary Stars}
 \label{subsec:binary}

Our sample contains 39 stars that are the primaries of binary or multiple systems. Their SEDs may have contributions from cooler companions that could affect our ability to accurately measure infrared excess. To assess the effects of multiplicity on SED fitting, we obtained the $V$-band magnitudes of each primary star and its brightest companion from the Washington Visual Double Star Catalog \citep{mason17}. Since all companions to our target stars are hot enough for the $K$ band to be firmly on the Rayleigh-Jeans side of the BB function, the secondary/primary flux ratio in the $K$ band should be similar to the flux ratio in the mid-infrared. Therefore, we predicted the $K_s$ band magnitudes of the primary star in each pair using either the \citet{mamajek13} $M_V$ vs $W1 -- K_s$ empirically calibrated relation for primary stars with saturation-corrected W1 photometry, or their $M_V$ vs $V -- K_s$ table for primary stars without WISE photometry. We also used the \citet{mamajek13} $M_V$ vs $V -- K_s$ table to find $K_s$ for all secondary stars. (Recall that we cannot simply use the 2MASS $K_s$ fluxes due to saturation.) We then computed (a) the expected secondary/primary $K_s$ band flux ratio $F_{K,2} / F_{K,1}$ and (b) the ratio of the combined $K_s$ band flux of the primary and secondary star to the predicted $K_s$ band flux from the SED model, $(F_{K,2} + F_{K,1}) / F_{K,{\rm SED}}$. Table \ref{tab:binary} lists $F_{K,2} / F_{K,1}$ and $(F_{K,2} + F_{K,1}) / F_{K,{\rm SED}}$ for the 23 binary stars with secondary stars brighter than $M_V = 12.3$. Companion stars fainter than $M_V = 12.3$ yield $F_{K,2} / F_{K,1} < 0.01$ and contribute negligibly to the combined SED \citep{yelverton19}.




Whether or not the brighter secondary stars could masquerade as debris disks depends on the angular separation of the pair. While all but two of our binary pairs have been resolved at visual wavelengths (Table \ref{tab:binary}), Spitzer MIPS and Herschel PACS and SPIRE have resolutions ranging from 5'' to 35'', depending on the instrument and observing mode \citep{rieke04,griffin10,poglitsch10}. A binary star with a high secondary-to-primary flux ratio and an angular separation in the 5''--35'' range might have an SED model that accounts for the primary star only, even though both stars are contributing to the far-IR flux. We found that 17 binary systems have $(F_{K,2} + F_{K,1}) / F_{K,{\rm SED}} > 1$, which means the secondary could trigger a false-positive dust detection (Table \ref{tab:binary}, column 2). Of those 17 systems, Hip 1473, Hip 19990, Hip 22845, Hip 23693, Hip 33302, Hip 57757, Hip 74824, Hip 75304, and Hip 111188 have $1 < (F_{K,2} + F_{K,1}) / F_{K,{\rm SED}} < 1.1$, which is not high enough for the secondary star to be mistakenly identified as a disk according to the VOSA-calculated errors on our photosphere models (\S \ref{subsec:excessidentification}). Hip 45571 has no significant infrared excess over our CIO-based photosphere models (\S \ref{subsec:nondetections}), so we can rule out false positives. Hip 8497, Hip 15197, Hip 15371, and Hip 20635 all have separations $> 35 \arcsec$, which is enough for the pair to be resolved in all mid- and far-IR photometry. Hip 2505 and Hip 23497 are targets with infrared excess comparable to the flux of the secondary star. However, since the binary separations are around $0.4 \arcsec$ for both targets, each system is unresolved in all photometry, including visual and near-IR. Furthermore, Hip 2505 and Hip 23497 are both binaries consisting of two stars with identical spectral types. A single photosphere model should be a good representation of the total star flux for both systems. Finally, Hip 55705, which has $(F_{K,2} + F_{K,1}) / F_{K,{\rm SED}} = 1.2$, has no statistically significant infrared excess. We conclude that our infrared excess detections are not contaminated by binary stars.

Our analysis of binary systems indicates that for all binary targets, either our photosphere model can adequately describe the total flux from both members of the binary or the angular separation is high enough for the two stars to be resolved at all wavelengths. Our photosphere models of close binaries and multiple stars should not be used for astrophysical characterization of the stars but only for identification of infrared excess. So long as there is good photometry at $1$--$4 \micron$, the SED fit can properly characterize the Rayleigh-Jeans tail of the composite SED of both stars. In their study of debris disks in binary systems, \citet{yelverton19} also 
argued that either the primary star will dominate the emission or the binary pair will have a similar spectral type; either way a single SED model can approximate the combined emission.

\section{Results: Disk Models}
\label{sec:diskmodels}
	
\begin{figure}
	\centering
	\captionsetup[subfloat]{labelformat=empty}
	\hspace*{-1.0cm}
	\begin{tabular}{ccc}
		\subfloat[Hip 2472]{\includegraphics[width=0.3\textwidth]{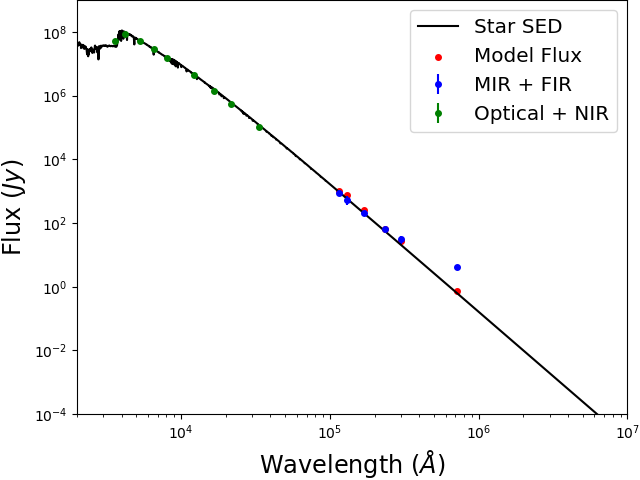}} &
		\subfloat[Hip 2505]{\includegraphics[width=0.3\textwidth]{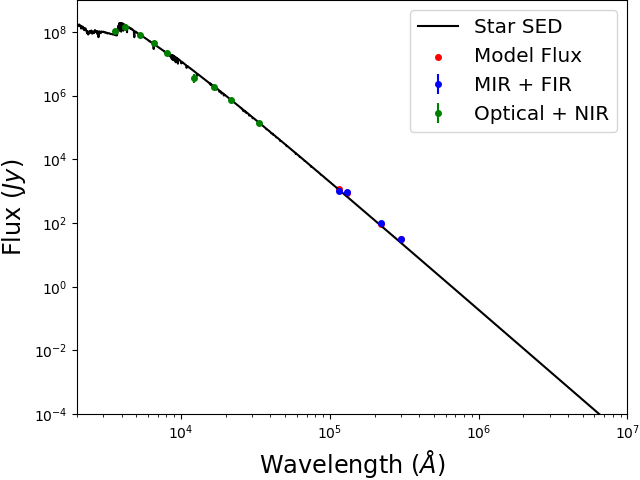}} &
		\subfloat[Hip 5862]{\includegraphics[width=0.3\textwidth]{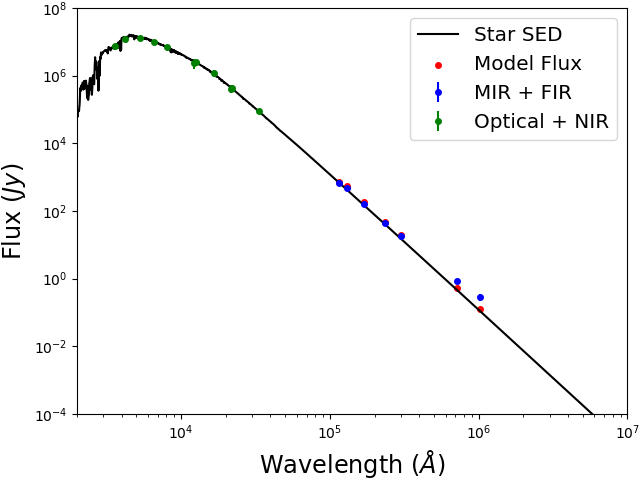}} \\
		
		\subfloat[Hip 13209]{\includegraphics[width=0.3\textwidth]{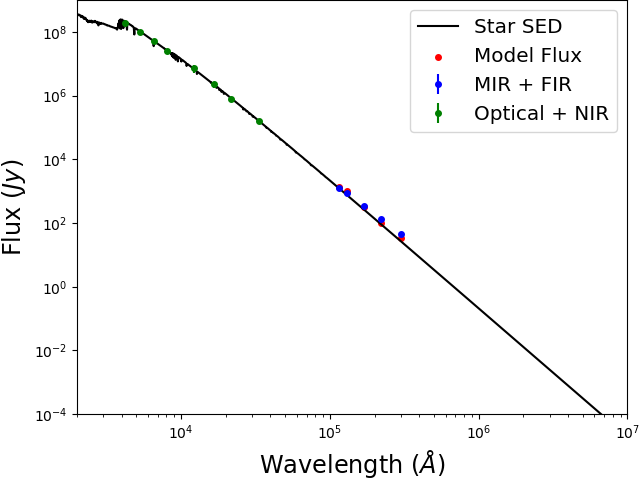}} &
		\subfloat[Hip 18975]{\includegraphics[width=0.3\textwidth]{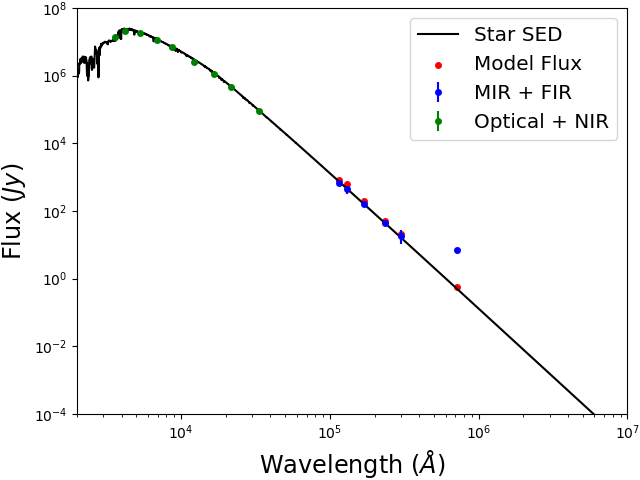}} &
		\subfloat[Hip 19990]{\includegraphics[width=0.3\textwidth]{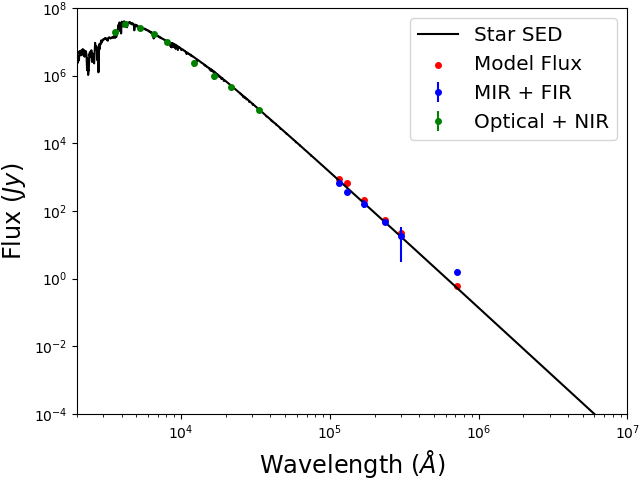}} \\
		
		\subfloat[Hip 20635]{\includegraphics[width=0.3\textwidth]{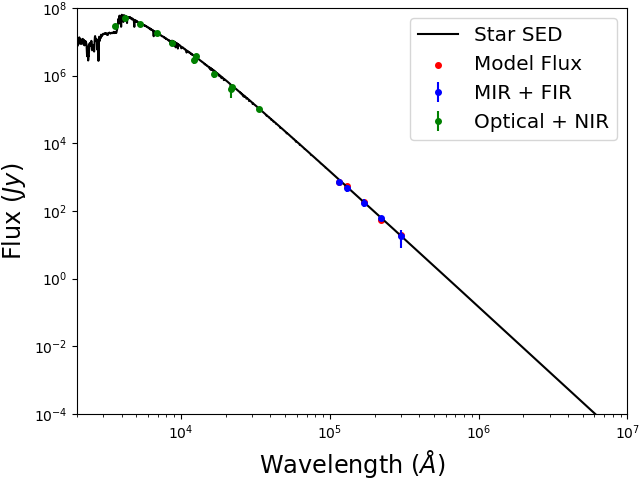}} &
		\subfloat[Hip 22509]{\includegraphics[width=0.3\textwidth]{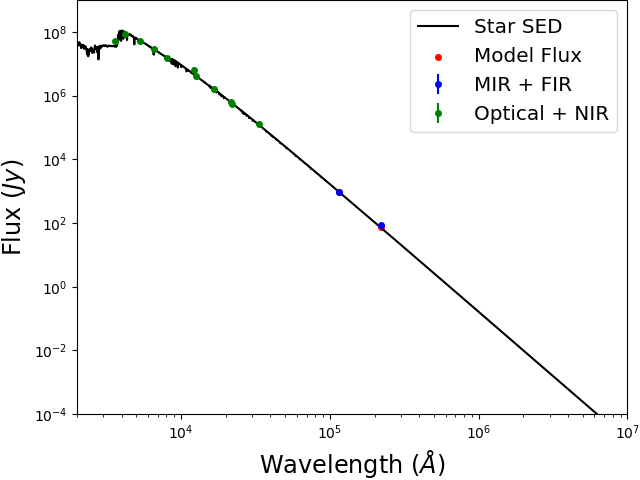}} &	
		\subfloat[Hip 23497]{\includegraphics[width=0.3\textwidth]{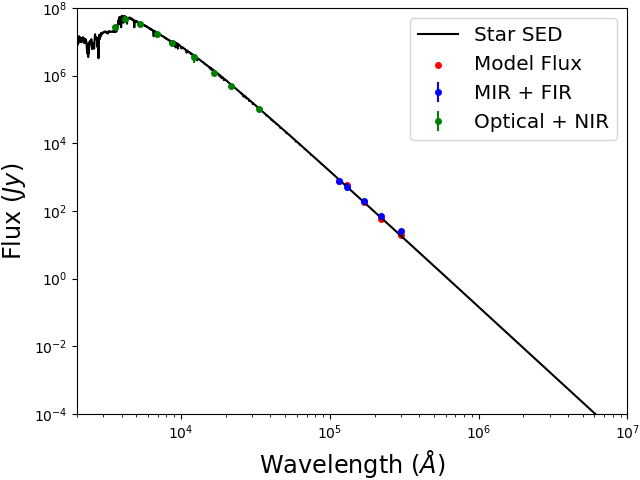}} \\
		
		\subfloat[Hip 27435]{\includegraphics[width=0.3\textwidth]{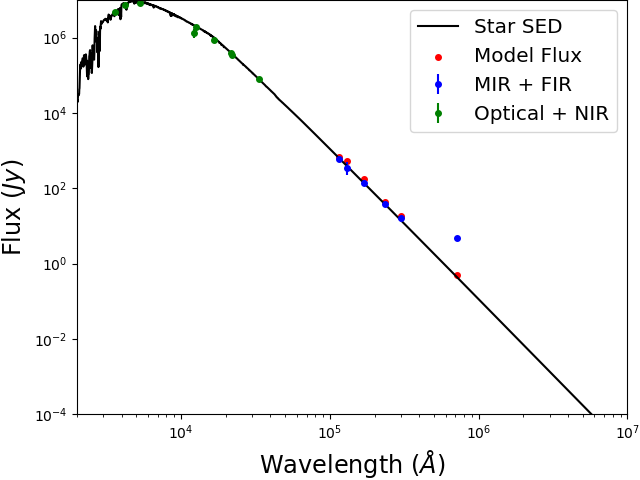}} &
		\subfloat[Hip 33302]{\includegraphics[width=0.3\textwidth]{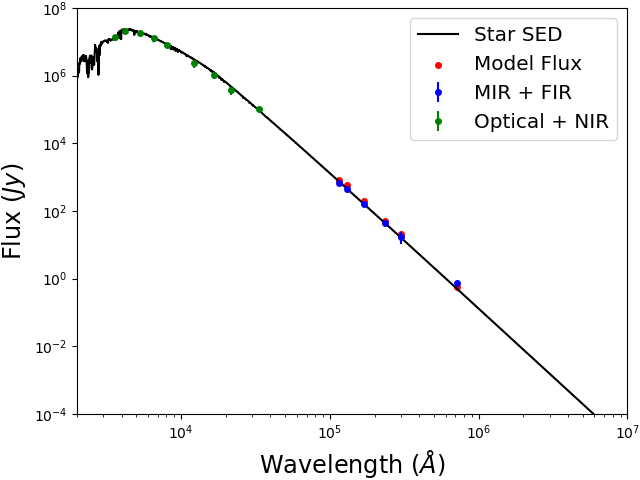}} &
		\subfloat[Hip 45571]{\includegraphics[width=0.3\textwidth]{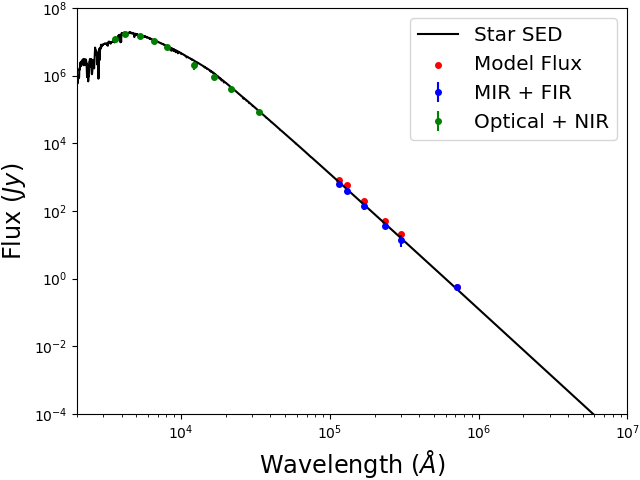}} \\
		
		\subfloat[Hip 53954]{\includegraphics[width=0.3\textwidth]{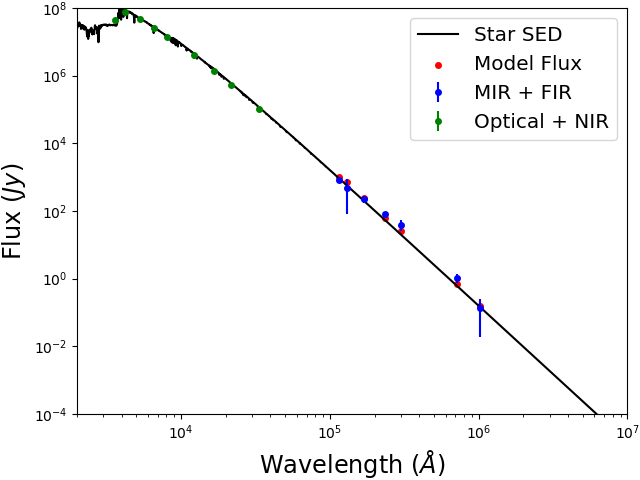}} &		
		\subfloat[Hip 55705]{\includegraphics[width=0.3\textwidth]{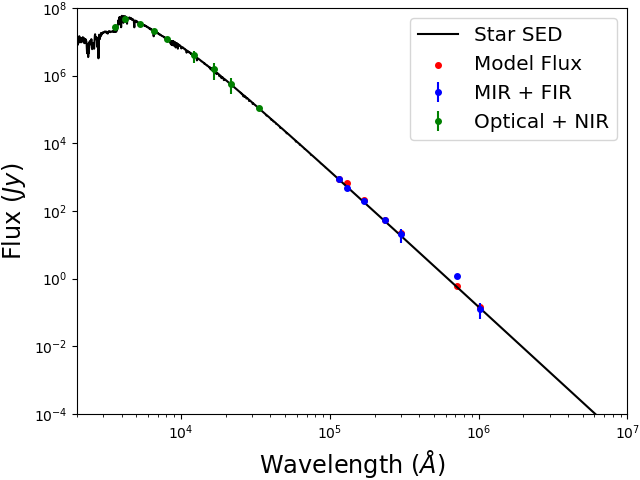}} &
		\subfloat[Hip 57757]{\includegraphics[width=0.3\textwidth]{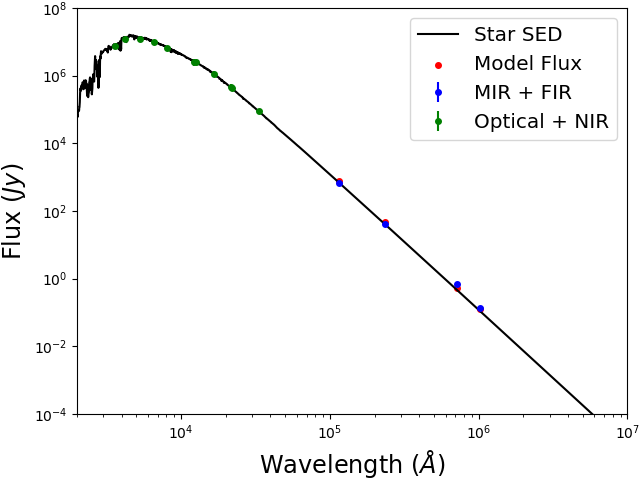}} \\ 	
	\end{tabular}
	\caption{ Best-fit SEDs for star--disk systems with one or no infrared excess detection. Note that model flux in the legends means synthetic flux.}
	\label{fig:oneorlessexcessSED1}
\end{figure}

\begin{figure}
	\centering
	\captionsetup[subfloat]{labelformat=empty}
	\hspace*{-1.0cm}
	\begin{tabular}{ccc}
	\subfloat[Hip 66234]{\includegraphics[width=0.3\textwidth]{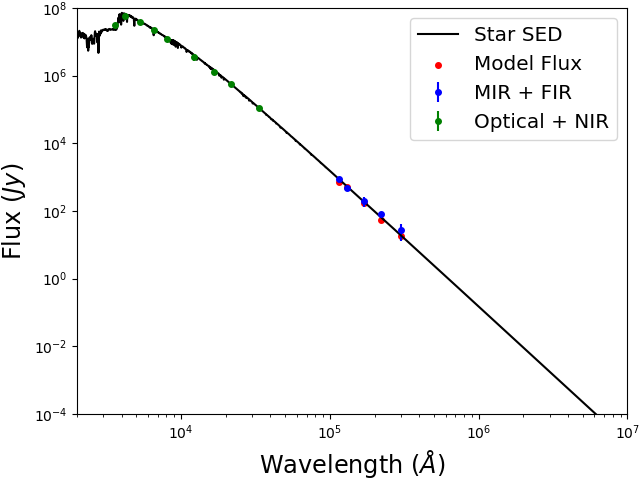}} &
	\subfloat[Hip 66765]{\includegraphics[width=0.3\textwidth]{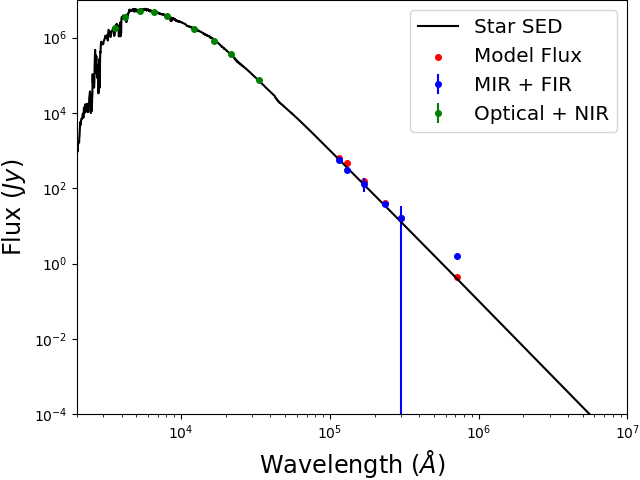}} &
	\subfloat[Hip 74702]{\includegraphics[width=0.3\textwidth]{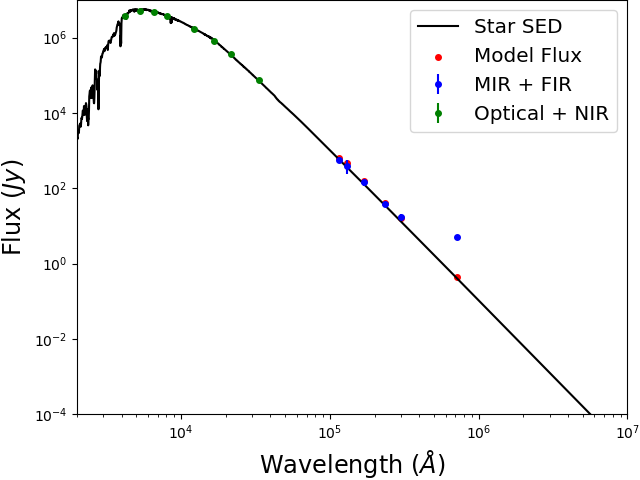}} \\ 
	
	\subfloat[Hip 74824]{\includegraphics[width=0.3\textwidth]{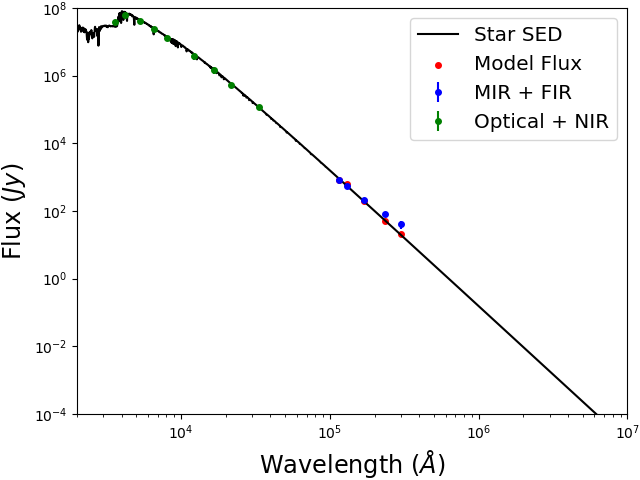}} &
	\subfloat[Hip 75304]{\includegraphics[width=0.3\textwidth]{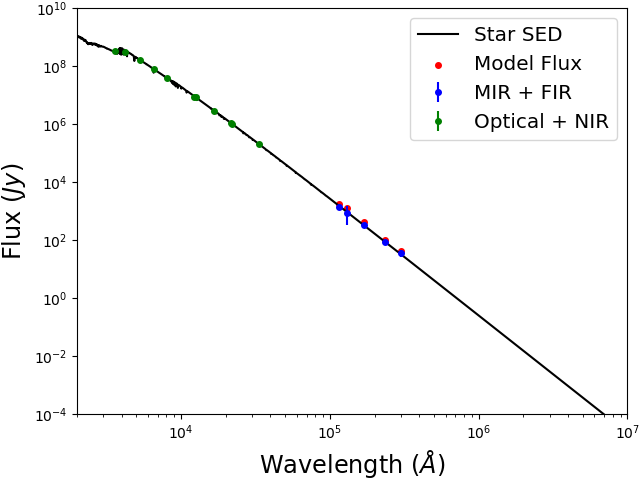}} &
	\subfloat[Hip 76829]{\includegraphics[width=0.3\textwidth]{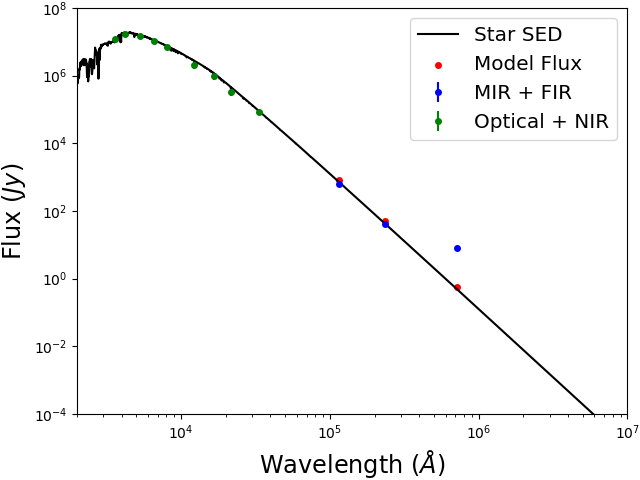}} \\ 
	
	\subfloat[Hip 78554]{\includegraphics[width=0.3\textwidth]{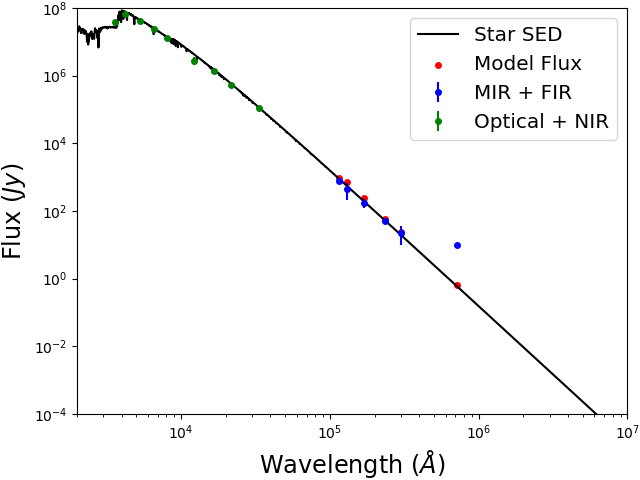}} &
	\subfloat[Hip 79881]{\includegraphics[width=0.3\textwidth]{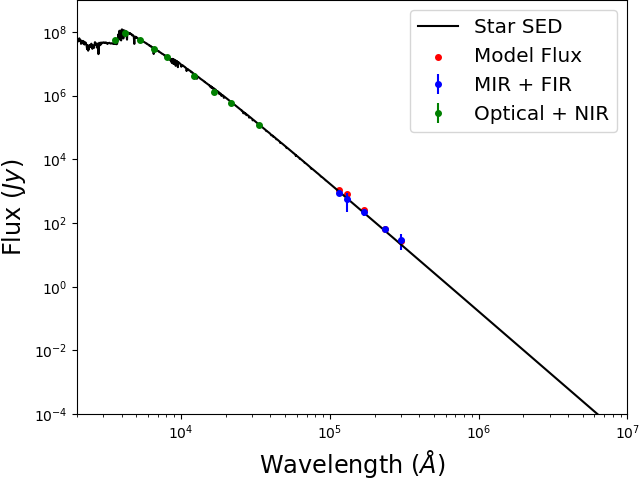}} &
	\subfloat[Hip 82673]{\includegraphics[width=0.3\textwidth]{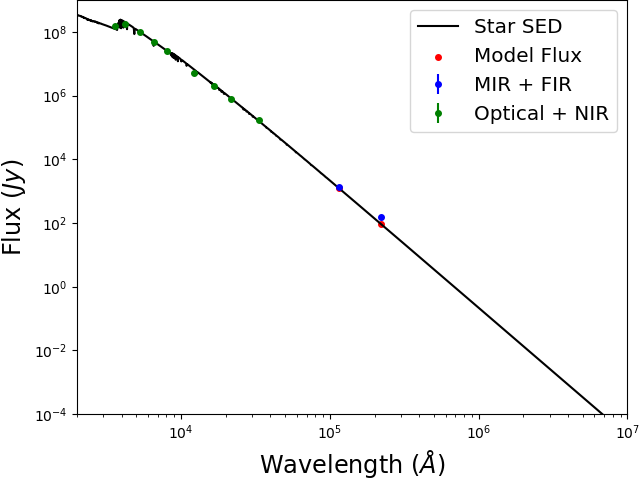}} \\ 
	
	\subfloat[Hip 93542]{\includegraphics[width=0.3\textwidth]{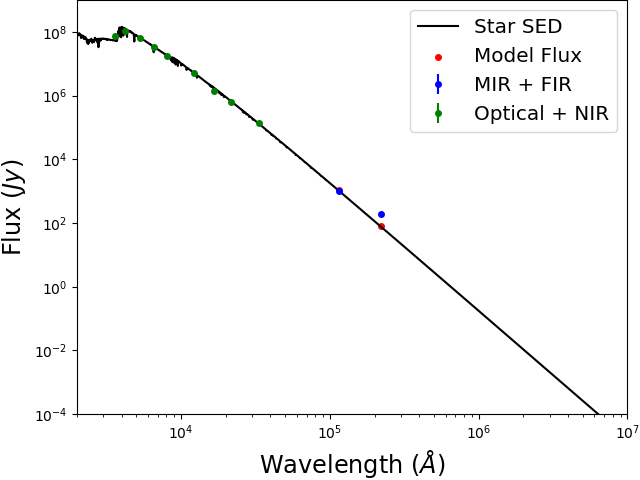}} &
	\subfloat[Hip 95347]{\includegraphics[width=0.3\textwidth]{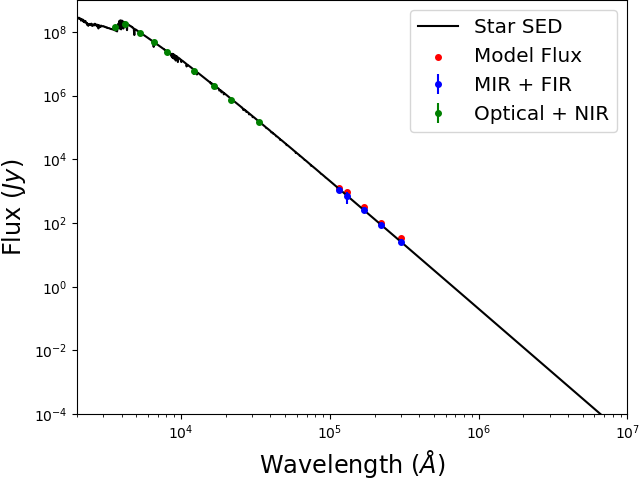}} &
	\subfloat[Hip 109857]{\includegraphics[width=0.3\textwidth]{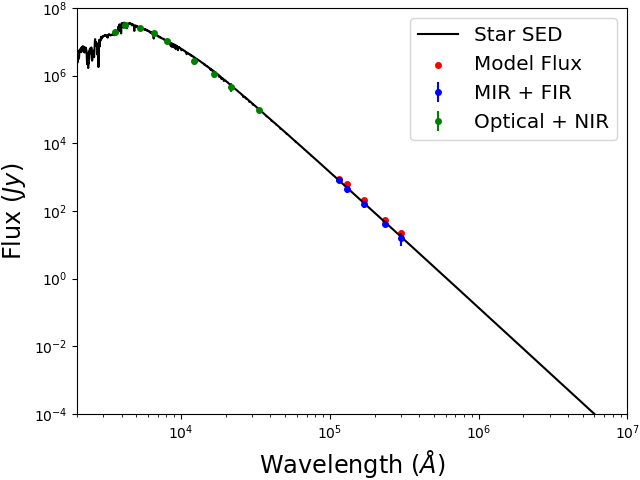}} \\ 
	
	\subfloat[Hip 113044]{\includegraphics[width=0.3\textwidth]{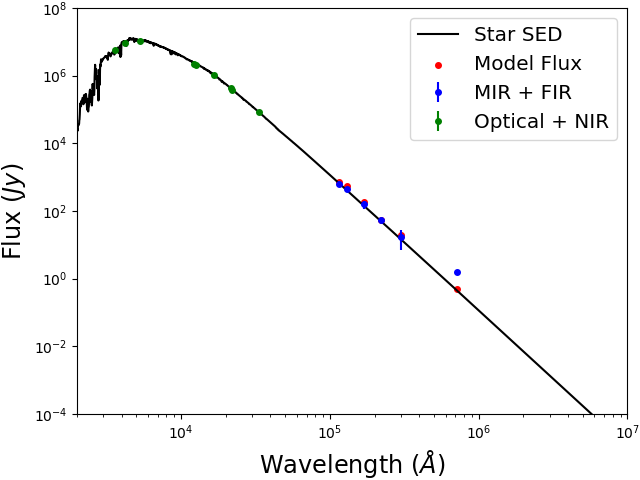}} &
	\subfloat[Hip 115738]{\includegraphics[width=0.3\textwidth]{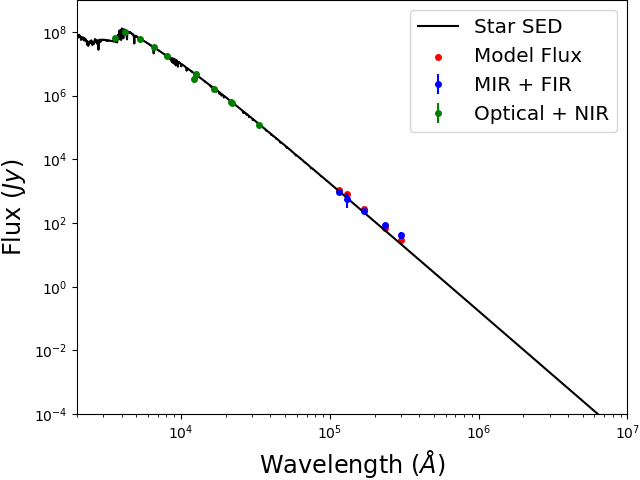}} &
	\subfloat[Hip 116631]{\includegraphics[width=0.3\textwidth]{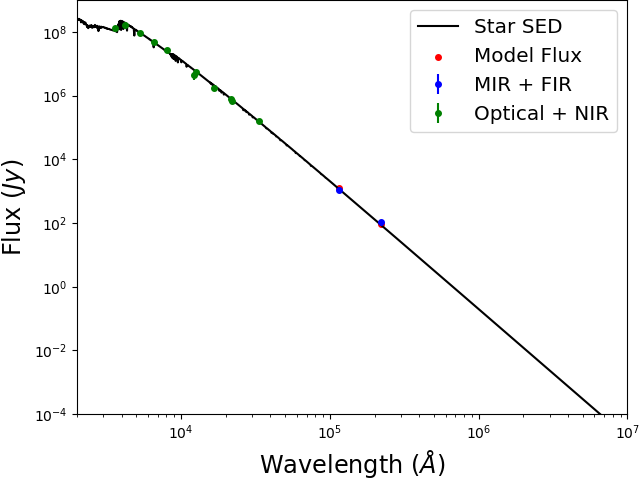}} \\
	\end{tabular}
	\caption{ Best-fit SEDs for star--disk systems with one or no infrared excess detection. Note that model flux in the legends means synthetic flux.}
	\label{fig:oneorlessexcessSED2}
\end{figure}

\begin{figure}
	\centering
	\captionsetup[subfloat]{labelformat=empty}
	\hspace*{-1.0cm}
	\begin{tabular}{ccc}
	\subfloat[Hip 1473]{\includegraphics[width=0.3\textwidth]{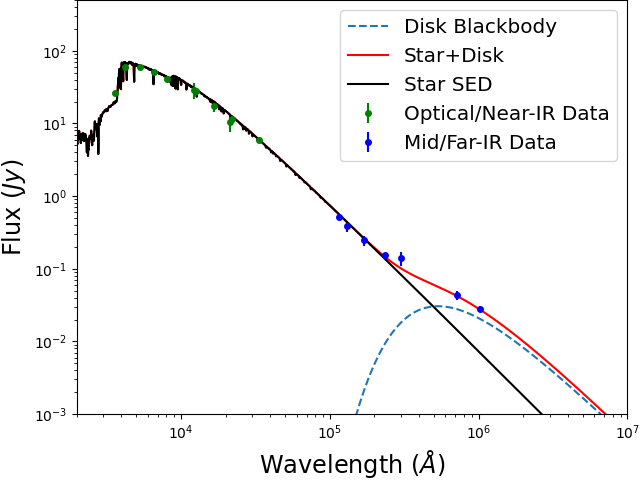}} &
	\subfloat[Hip 2578]{\includegraphics[width=0.3\textwidth]{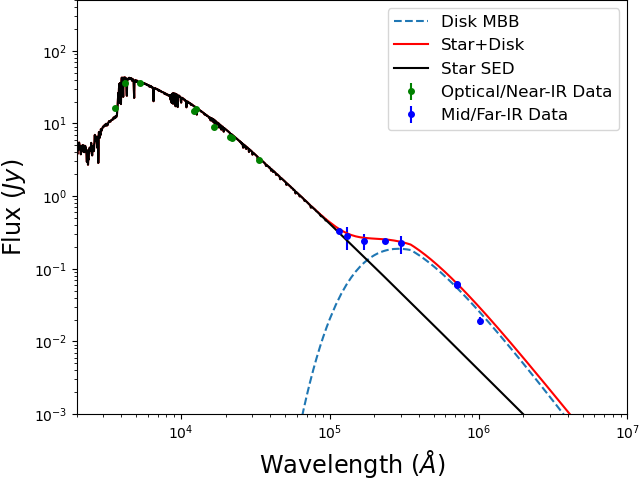}} &
	\subfloat[Hip 8497]{\includegraphics[width=0.3\textwidth]{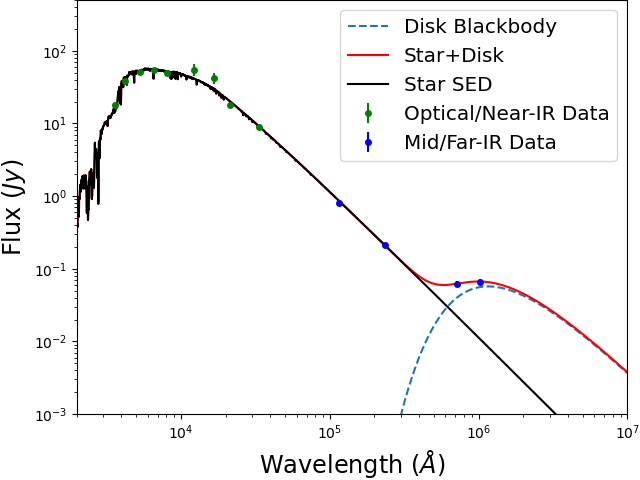}} \\
	
	\subfloat[Hip 10670]{\includegraphics[width=0.3\textwidth]{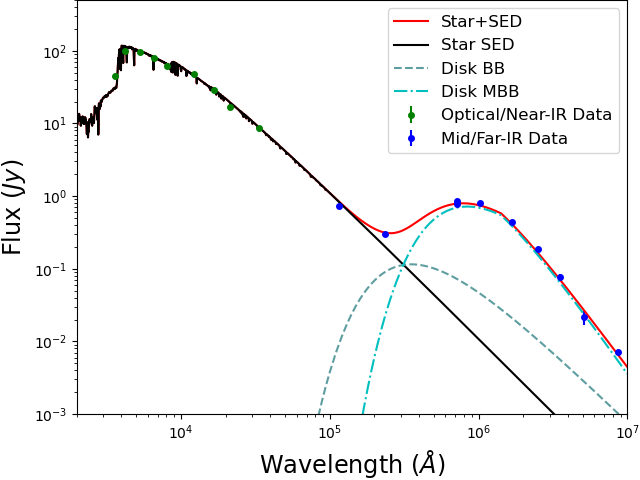}} &
	\subfloat[Hip 13402]{\includegraphics[width=0.3\textwidth]{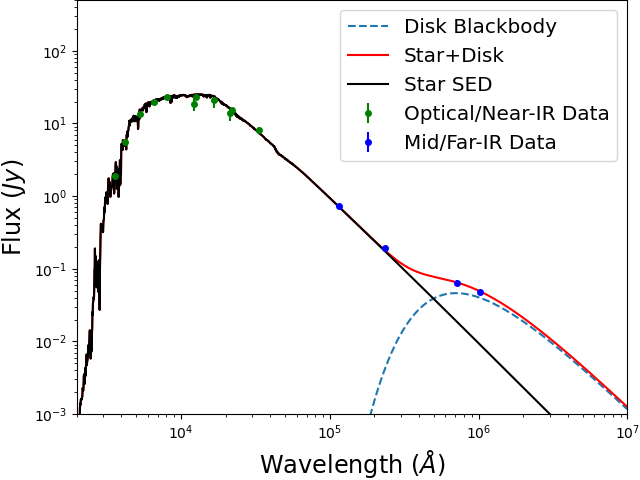}} &
	\subfloat[Hip 14954]{\includegraphics[width=0.3\textwidth]{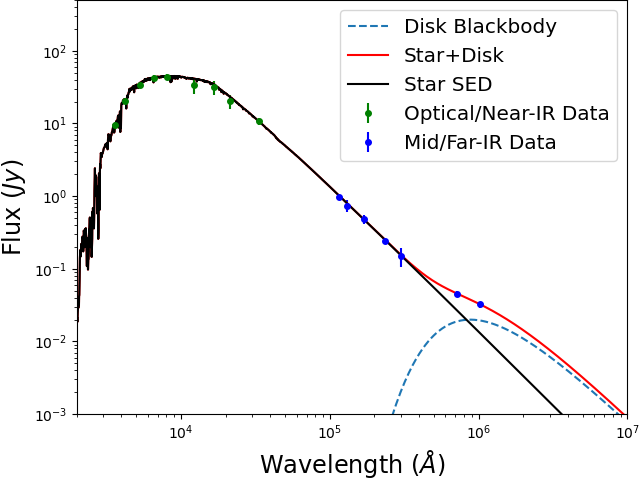}} \\
	 
	\subfloat[Hip 15197]{\includegraphics[width=0.3\textwidth]{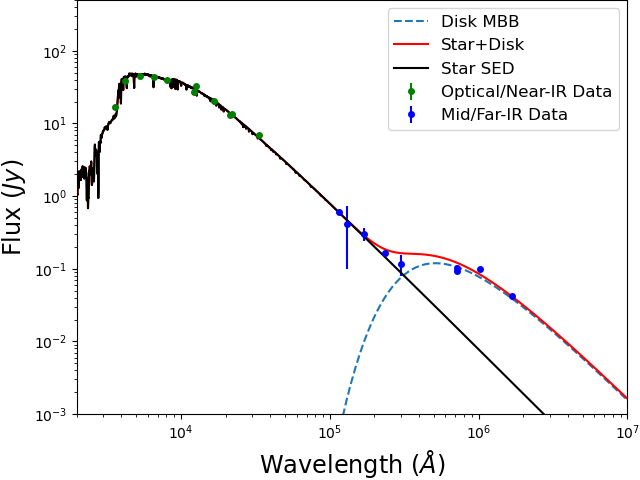}} &
	\subfloat[Hip 15371]{\includegraphics[width=0.3\textwidth]{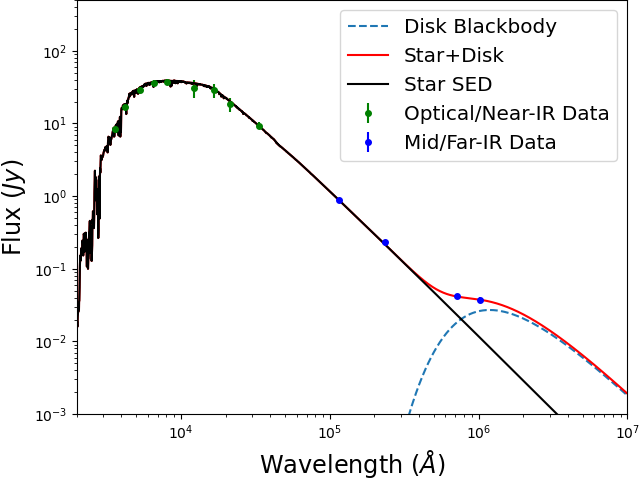}} &	
	\subfloat[Hip 16852]{\includegraphics[width=0.3\textwidth]{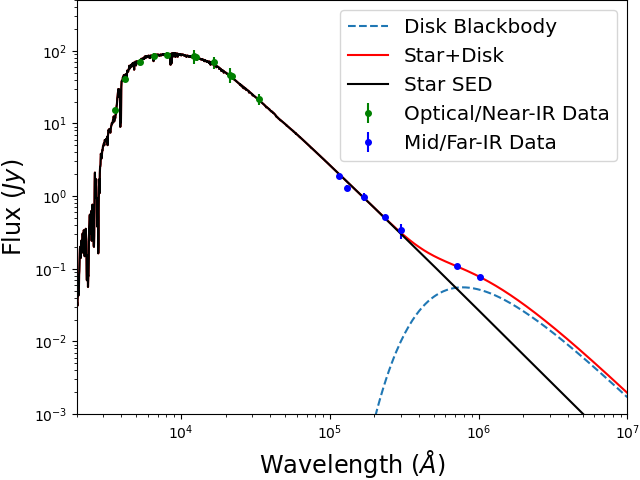}} \\
	
	\subfloat[Hip 17439]{\includegraphics[width=0.3\textwidth]{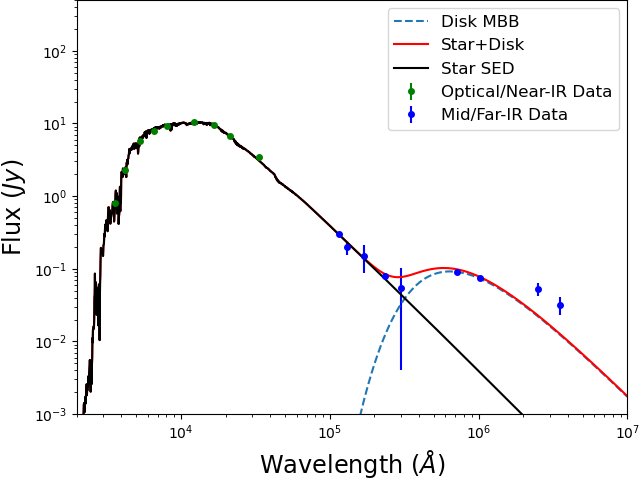}} & 
	\subfloat[Hip 19893]{\includegraphics[width=0.3\textwidth]{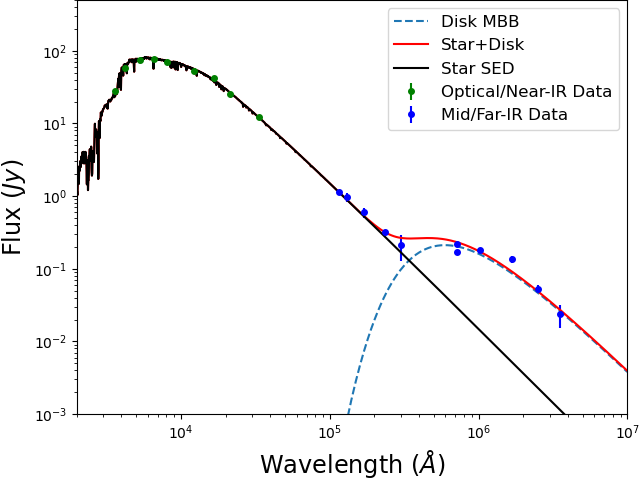}} &
	\subfloat[Hip 20901]{\includegraphics[width=0.3\textwidth]{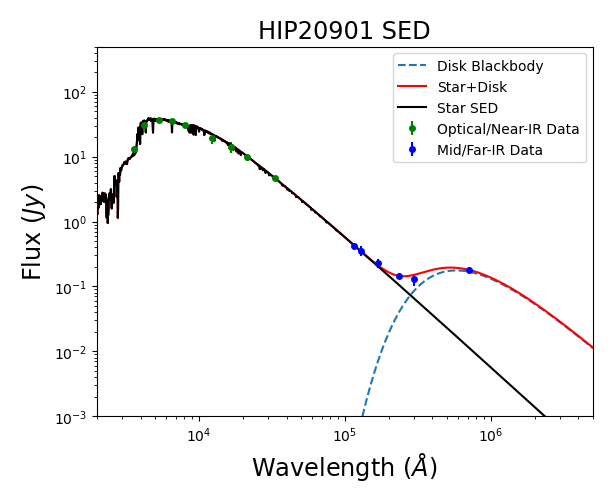}}\\
	
	\subfloat[Hip 21547]{\includegraphics[width=0.3\textwidth]{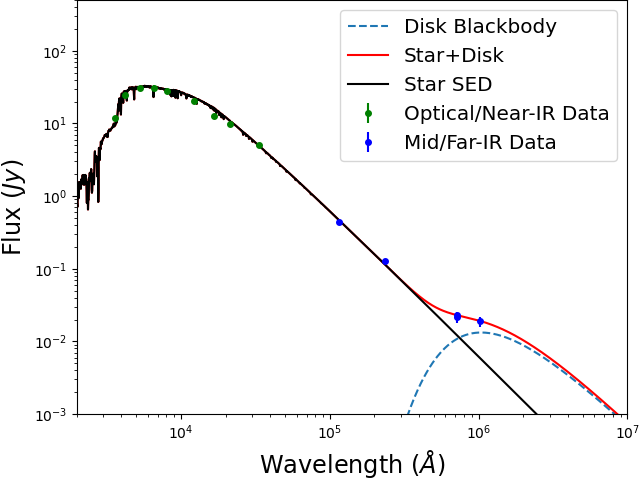}} & \subfloat[Hip 22263]{\includegraphics[width=0.3\textwidth]{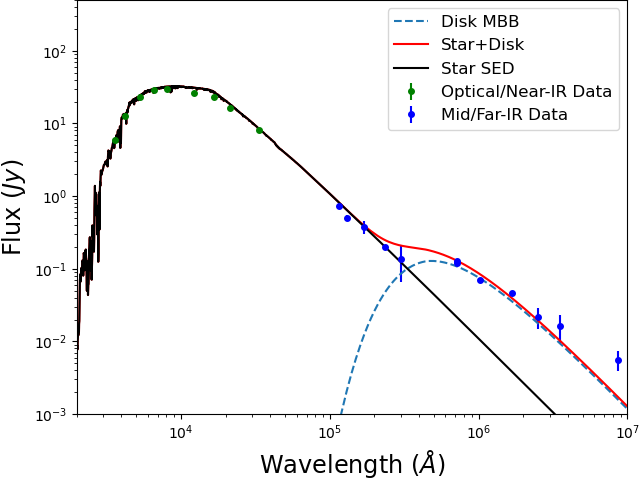}} &
	\subfloat[Hip 22845]{\includegraphics[width=0.3\textwidth]{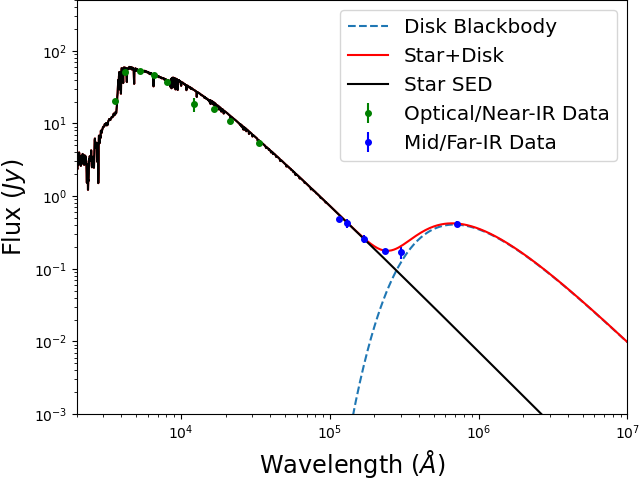}}\\

	\end{tabular}
	\caption{ Best-fit SEDs for star--disk systems with infrared excess detections in two or more wavelengths. Note that model flux in the legends means synthetic flux.}
	\label{fig:allSEDs1}
\end{figure}

\begin{figure}
	\centering
	\captionsetup[subfloat]{labelformat=empty}
	\hspace*{-1.0cm}
	\begin{tabular}{ccc}
	\subfloat[Hip 23693]{\includegraphics[width=0.3\textwidth]{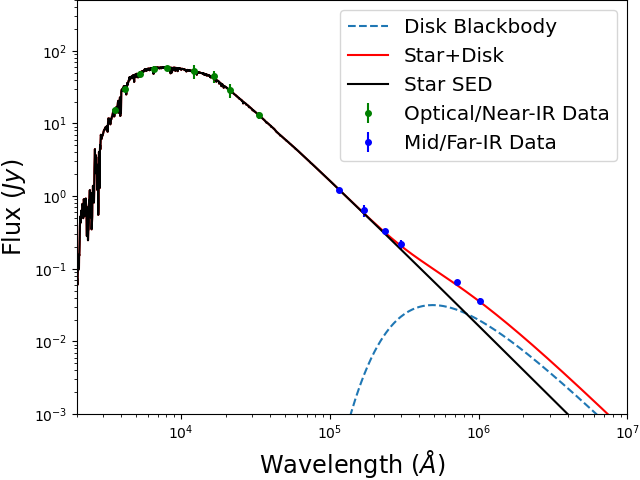}} &
	\subfloat[Hip 41307]{\includegraphics[width=0.3\textwidth]{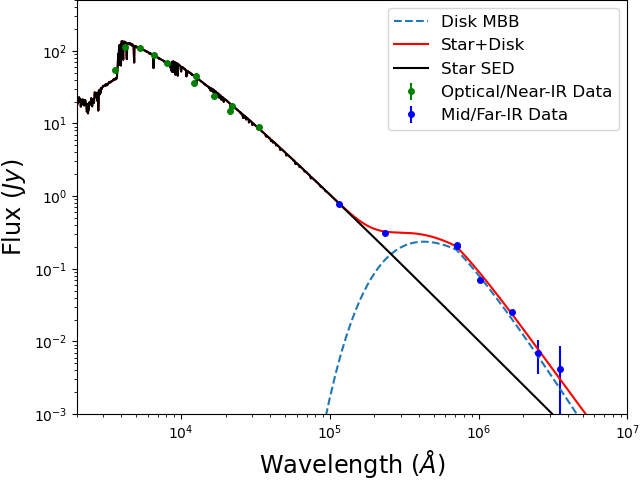}} & 
	\subfloat[Hip 42438]{\includegraphics[width=0.3\textwidth]{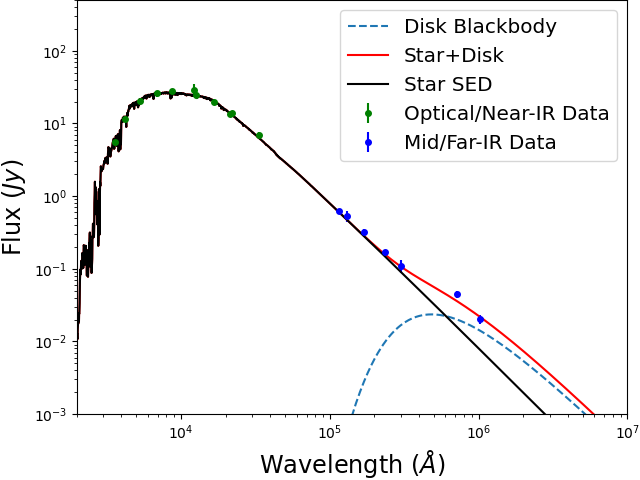}}\\ 
	
	\subfloat[Hip 43726]{\includegraphics[width=0.3\textwidth]{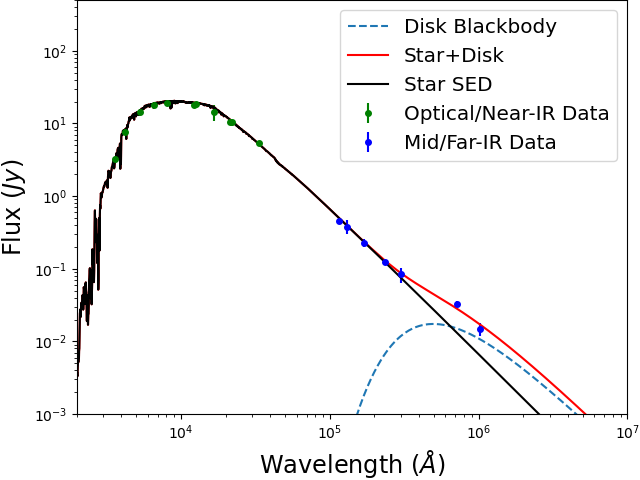}} &
	\subfloat[Hip 61960]{\includegraphics[width=0.3\textwidth]{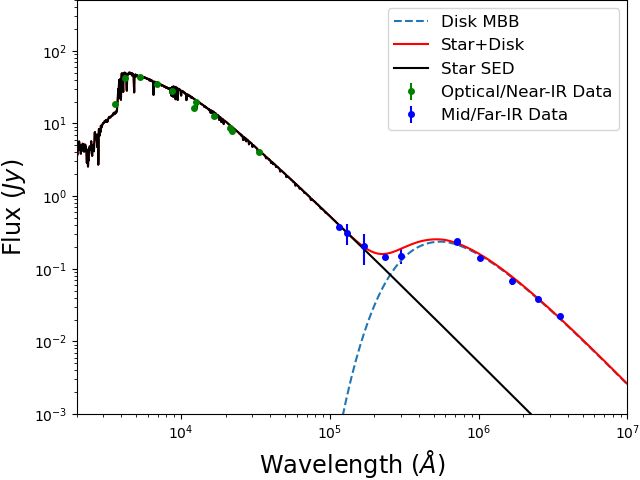}} &
	\subfloat[Hip 62207]{\includegraphics[width=0.3\textwidth]{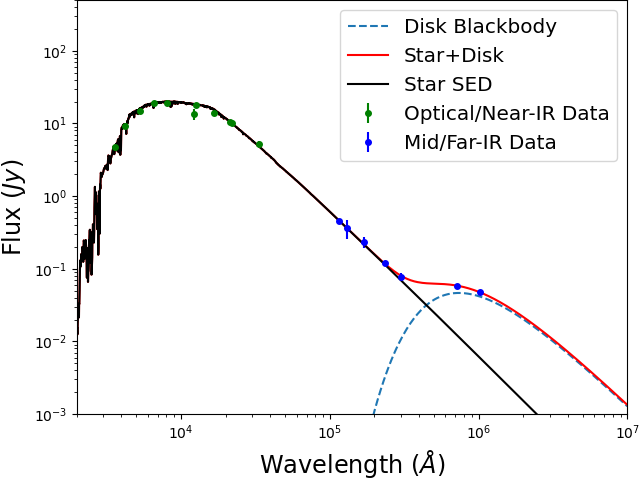}}\\ 
	
	\subfloat[Hip 65721]{\includegraphics[width=0.3\textwidth]{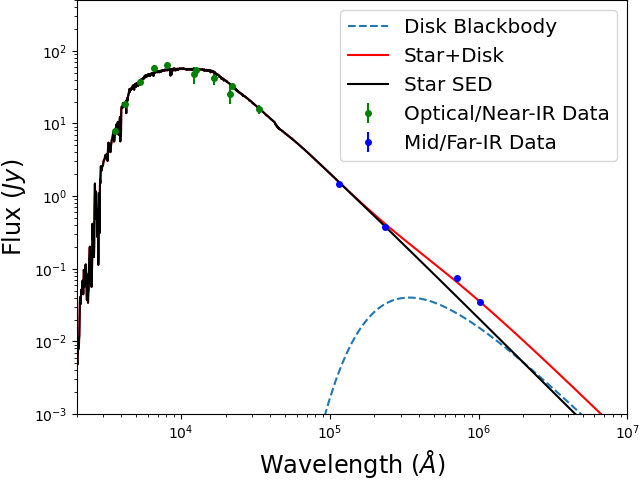}} & 
	\subfloat[Hip 69732]{\includegraphics[width=0.3\textwidth]{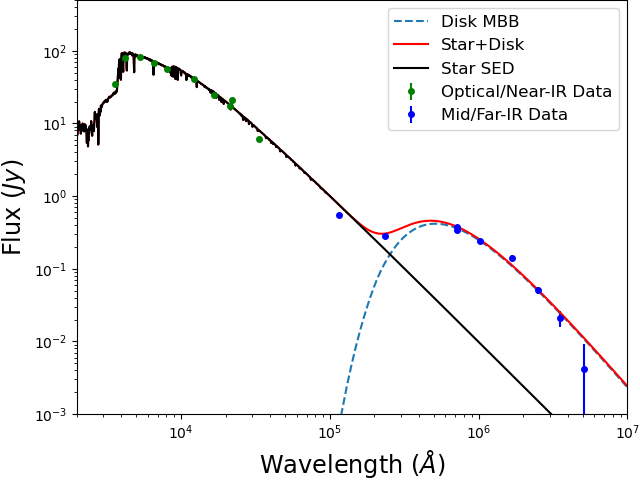}} &
	\subfloat[Hip 86305]{\includegraphics[width=0.3\textwidth]{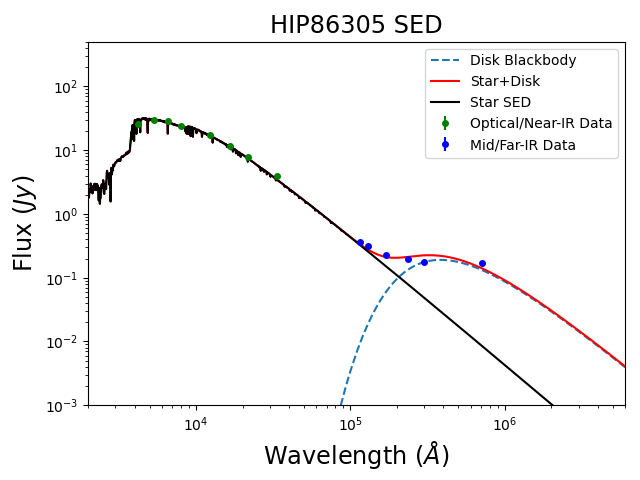}}\\
	
	\subfloat[Hip 87108]{\includegraphics[width=0.3\textwidth]{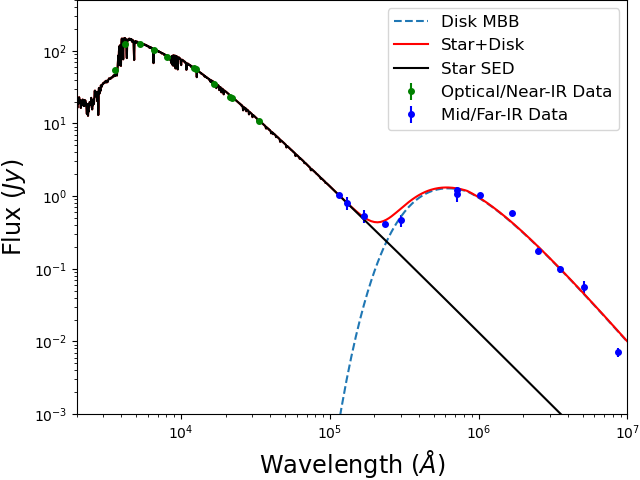}} & \subfloat[Hip 98495]{\includegraphics[width=0.3\textwidth]{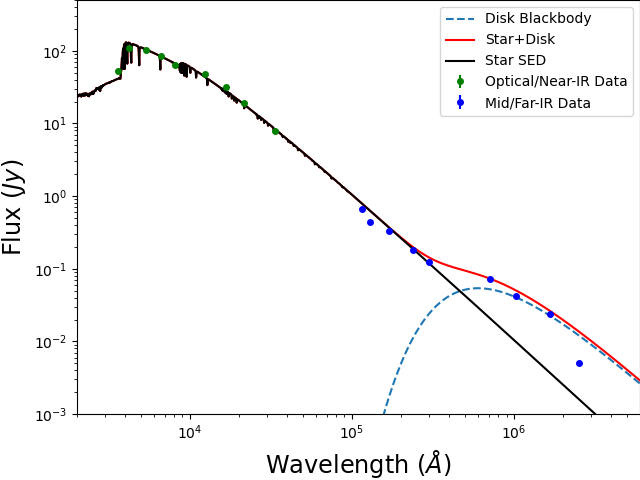}} & \subfloat[Hip 99742]{\includegraphics[width=0.3\textwidth]{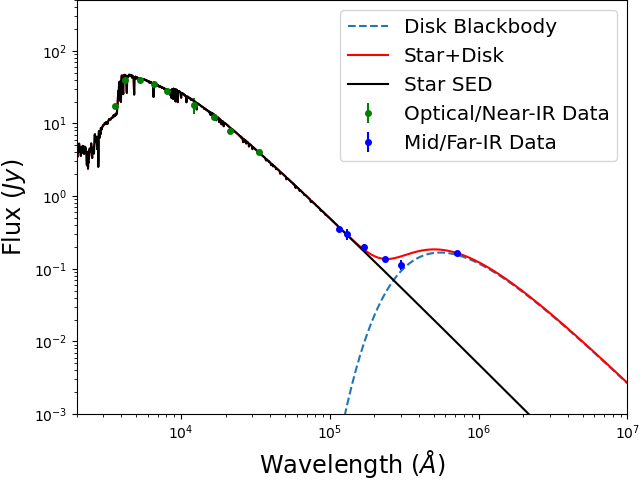}}\\
	
	\subfloat[Hip 101612]{\includegraphics[width=0.3\textwidth]{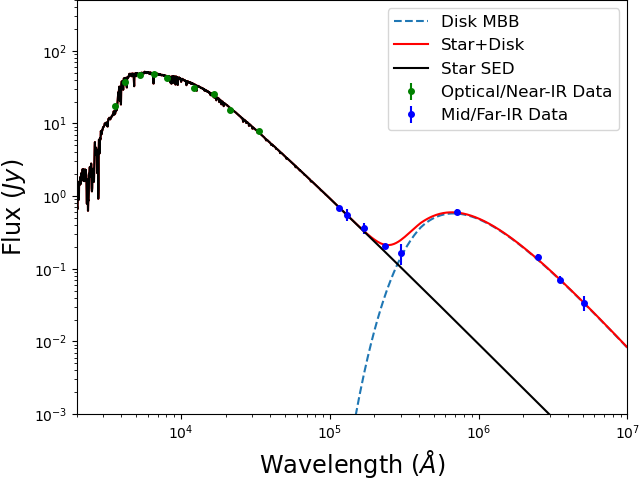}} & \subfloat[Hip 103389]{\includegraphics[width=0.3\textwidth]{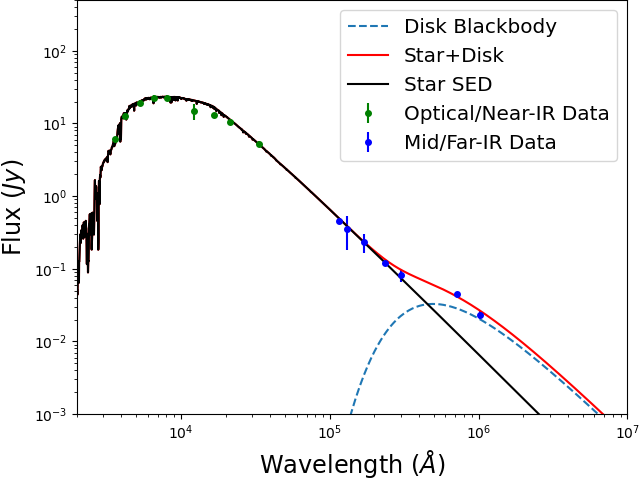}} &
	\subfloat[Hip 111188]{\includegraphics[width=0.3\textwidth]{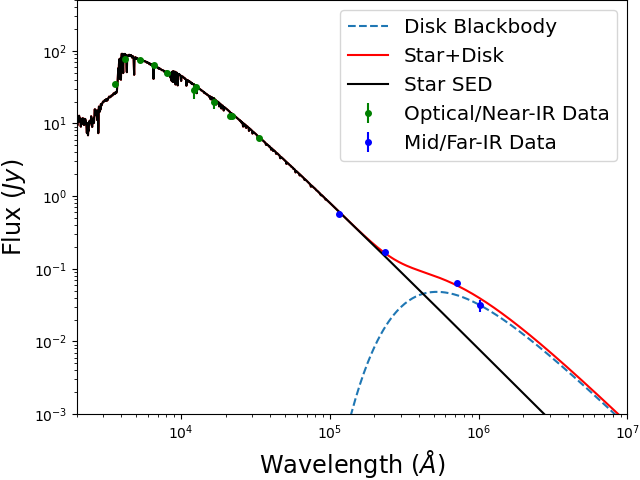}}\\ 

	\end{tabular}
	\caption{ Best-fit SEDs for star--disk systems with infrared excess detections in two or more wavelengths. Note that model flux in the legends means synthetic flux.}
	\label{fig:allSEDs2}
\end{figure}

\begin{figure}
	\centering
	\captionsetup[subfloat]{labelformat=empty}
	\hspace*{-1.0cm}
	\begin{tabular}{ccc}
	\subfloat[Hip 117452]{\includegraphics[width=0.3\textwidth]{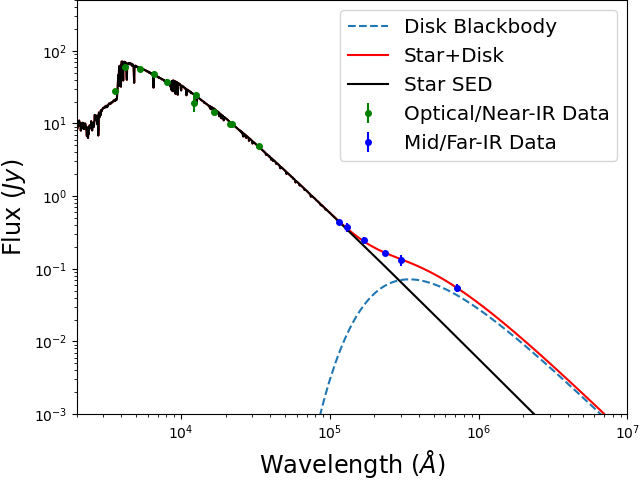}} & & \\
	\end{tabular}
	\caption{ Best-fit SEDs for star--disk systems with infrared excess detections in two or more wavelengths. Note that model flux in the legend means synthetic flux.}
	\label{fig:allSEDs3}
\end{figure}

	From our initial sample of 62 stars with reported debris disks, we found 12 stars with no infrared excess above the $3 \sigma_x$ level in any filter (Table \ref{tab:zeroexcess}), 18 stars with infrared excess in one filter (Table \ref{tab:oneexcess}), 19 stars with infrared excess in two filters (Table \ref{tab:twoexcess}), and 13 stars with infrared excess in three or more filters (Tables \ref{tab:multiexcess} and \ref{tab:mbb}). Figures \ref{fig:oneorlessexcessSED1} and \ref{fig:oneorlessexcessSED2} show SEDs for stars with either no infrared excess or excess at only one wavelength. Figures \ref{fig:allSEDs1}, \ref{fig:allSEDs2}, and \ref{fig:allSEDs3} show all SEDs of stars fitted with BB or modified blackbody disk models (i.e., infrared excess at two or more wavelengths).  

Some targets have ``extra excess'' at $\lambda \geq 160 \micron$ over and above a single BB, 2BB, or MBB model, which suggests supercold outer disks with $T \la 20$~K if interpreted as dust emission. Given the possibility of background contamination in the far-IR \citep[e.g.,][]{krivov13}, we exclude PACS and SPIRE
data that are not consistent with BB-based models from our standard dust SED fits. ``Extra excess'' observations excluded from BB/2BB/MBB fitting are Hip 2578 PACS $160 \micron$, Hip 61960 SPIRE $500 \micron$, and Hip 87108 SPIRE $500 \micron$, which \citet{moor15} identified as being contaminated by a background ridge of submillimeter emission.
	
	
\subsection{Nondetections}
\label{subsec:nondetections}



		
	We first revisit the targets for which we found no infrared excess at or above the $3 \sigma_x$ level in any filter (Table \ref{tab:zeroexcess}). For Hip 20635, Hip 95347, and Hip 109857, the excesses reported by \citet{chen14} are marginal, so it is not surprising they do not quite reach the $3 \sigma_x$ significance threshold in our study. On the other hand, three of the objects for which our study found no significant infrared excess---Hip 33302, Hip 45571, and Hip 75304---have excess ratios $F_{o,x} / F_x \geq 1.3$ in the studies from which the objects were drawn. For these targets, our photosphere models predict higher star fluxes than models in the literature; for example, \citet{chen14} report that Hip~33302 has $T_{\rm eff} = 7000$~K and $F_* = 27.1$~mJy at $70 \micron$, while we find $T_{\rm eff} = 6750$~K and $F_* = 28.7$~mJy at $70  \micron$. With an observational uncertainty of $\sigma_{o,70 \micron} = 5.0$~mJy, the seemingly large excess of $F_{o,70 \micron} / F_{70 \micron} = 1.34$ does not meet our $3 \sigma_{70 \micron}$ threshold for excess detection. It is important to point out, though, that our excess detection criteria would also result in a nondetection when using the Chen et al. photosphere model. The reason these differences exist is that \citet{chen14} used multiple literature sources that already diagnosed the excess by using population statistics of the $F_{o,\lambda} / F_{\star,\lambda}$ values of the sample stars \citep[e.g.][]{beichman06, bryden06, trilling07}. The studies use different excess detection criteria and arrive at different mid- and far-IR synthetic fluxes. For example, the original detection of a debris disk surrounding Hip 33302 came from Trilling et al. (2007), who reported excess in MIPS $70 \micron$ using their photosphere model with $T_{\rm eff} = 7250$~K and $F_* = 23$~mJy and a $2 \sigma_x$ detection threshold.
	We are not claiming that a nondetection in this study invalidates a previous detection; we merely suggest that our nondetections should receive further follow-up before being included in an expensive high-contrast imaging campaign.

\subsection{Infrared Excess Detections at A Single wavelength}
\label{subsec:singlefilter}	
	
	Now we consider targets with infrared excess in one filter (Table \ref{tab:oneexcess}). The fact that all blackbodies approach the same Rayleigh-Jeans power law for $\lambda \gg \lambda_{\rm max}$ means that systems with infrared excess at only one wavelength almost always have that excess at the longest observed wavelength. In the rare cases where there is no observed excess at a longer wavelength than where the excess is found, it is because of low signal-to-noise ratio in the long-wavelength photometry. The disks in Table \ref{tab:oneexcess} have all been detected at either W4 only, in which case we assume $\lambda_{\rm max} = 22 \micron$ and derive $T_{\rm dust} \approx 231$~K; MIPS $24 \micron$ only, which yields $T_{\rm dust} \approx 216$~K; or MIPS/PACS $70 \micron$ only, with $T_{\rm dust} \approx 71.4$~K. 
	
	We note that for three objects in our samples, Hip 5862, Hip 53954, and Hip 57757, they have photometry at longer wavelengths without excess detection. As an example, for Hip 53954, which has an excess at MIPS $24 \micron$, the total star+disk flux was undetected at the $3\sigma$ level in the MIPS $70 \micron$ data ($F_{\nu} = 26.11 \pm 7.76$~mJy) and in the PACS $100 \micron$ photometry ($F_{\nu} = 6.92 \pm 5.90$~mJy).


\subsection{Infrared excess detections in two wavelengths}
\label{subsec:twofilter}

We next turn to systems with infrared excess at two different wavelengths, for which it is possible to estimate the debris disk temperature, luminosity $L_{\rm dust}$, and distance from the star $R_d$,
	\begin{equation}
		\frac{R_d}{1 \; {\rm AU}} = \left(\frac{L_*}{L_{\sun}}\right)^{0.5} \left( \frac{278.3 \: \mathrm{K}}{T_{\rm dust}} \right)^2,
		\label{eq:diskradius}
	\end{equation}
where $L_*$ is the star luminosity. Table \ref{tab:twoexcess} lists the best-fit BB dust temperature and fractional luminosity $\log (L_{\rm dust} / L_*)$, along with their uncertainties;
the estimated dust ring radius given by Equation \ref{eq:diskradius}; and the angular radius $\theta$ of the disk on the sky corresponding to $R_d$.

With an estimate of $R_d$ in hand, we can assess the prospects for direct imaging with JWST and ground-based coronagraphs. Here we focus primarily on two instruments from JWST, MIRI and NIRCam. MIRI has a field of view of $30 \arcsec \times 30 \arcsec$ and a Lyot coronagraph with a $23 \micron$ filter and inner working angle (IWA) of $2.16 \arcsec$ that was designed, in part, for debris disk observations \citet{gardner04}. There is also a standard imaging mode (no coronagraph) with a spatial resolution given by the FWHM of the PSFs, $0.674 \arcsec$ and $0.803 \arcsec$ for the $21 \micron$ (F2100W) and $25.5 \micron$ (F2550W) filters, respectively \citep{bouchet15}. NIRCam MASK430R, with a coronograph IWA of $0.81 \arcsec$ at $4 \micron$, can resolve any disk observable with MIRI so long as the disk is sufficiently bright in scattered light. The NIRCam field of view is $20 \times 20 \arcsec$, and the angular resolution is $0.063 \arcsec$ at $4 \micron$ \citep{riekeb05,NASA_2023}.

\begin{figure}
\centering
    \begin{tabular}{cc}
	\includegraphics[width=0.49\textwidth]{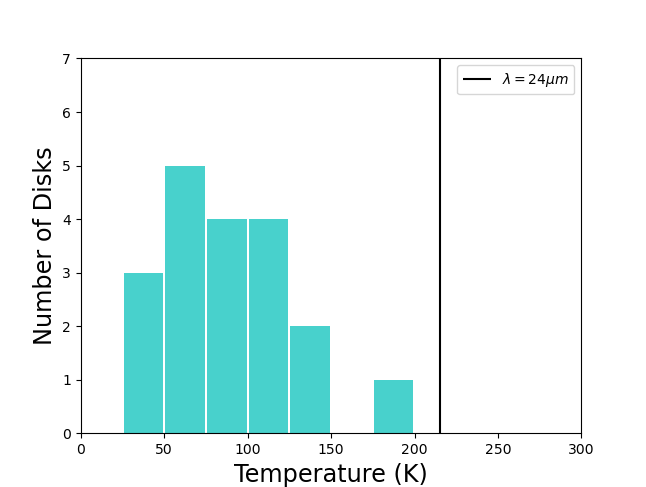} &
	\includegraphics[width=0.49\textwidth]{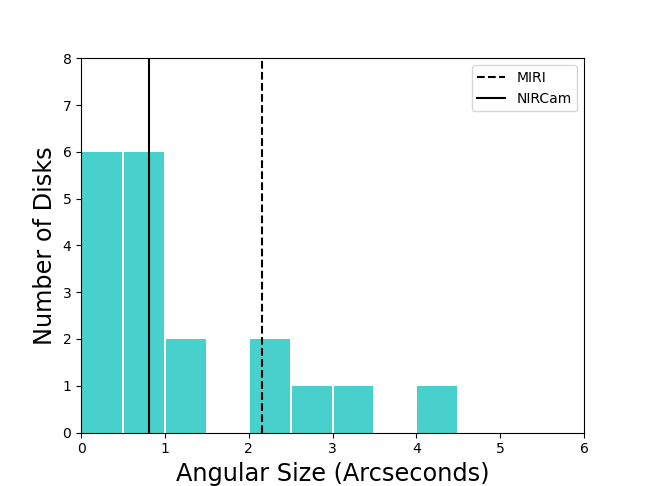} \\
    \end{tabular}
\caption{Characteristics of disks with infrared excess at two wavelengths. Left: disk temperature distribution. The vertical line marks $T = 216$~K, at which $\lambda_{\rm max} = 24 \micron$. Right: disk angular size distribution. Vertical lines mark the IWA of JWST's NIRCam MASK430R (solid) and MIRI $23 \micron$ (dashed).}
\label{fig:disktemphist_2pt}
\end{figure}

Figure \ref{fig:disktemphist_2pt} shows the dust temperature distribution (left) and the angular size distribution (right) for the systems with infrared excess at two wavelengths. The vertical line on the temperature distribution marks $T_{\rm dust} = 216$~K, for which the flux density $F_{\nu}$ peaks at $24 \micron$. On the angular size distribution, the vertical solid line marks the IWA of NIRCam at $4 \micron$ while the vertical dashed line shows the resolution limit of the MIRI coronagraph at $23 \micron$. Twelve of the systems with infrared excess in two wavelengths are predicted to be resolvable with NIRCam at $4 \micron$.

Our results suggest that Hip 8497 ($\chi$~Cet), Hip 15371 ($\zeta^2$ Ret), and Hip 21547 (51 Eri) are potential candidates for high-resolution imaging, with $\theta = 4.2 \arcsec$, $\theta = 3.5  \arcsec$, and $\theta = 2.6 \arcsec$.

Our SED fit to Hip 8497, which yields $T_{\rm dust} = 44 \pm 6$~K and $\log(L_{\rm dust} / L_*) = -5.15 \pm 0.04$, does not agree with that of \citet{absil21}, who found $T_{\rm dust} = 100$~K and $\log(L_{\rm dust} / L_*) = -5.0$ based on a photosphere model built using 2MASS photometry. Our study and \citet{absil21} both conclude that Hip 8497 has infrared excess detected only at $\lambda \geq 70 \micron$, so MIRI observations are not advisable. The dust was also not detected in an interferometric $H$-band survey using the PIONEER instrument on VLTI \citep{ertel14}. Scattered light may be observable with instruments such as NIRCam, SPHERE, or STIS. 

Hip 15371 appears to be a good direct imaging target based on the SED alone, but follow-up of the resolved Herschel PACS images \citep{eiroa10, faramaz14} with ALMA Bands 6 and 7 suggested that the far-IR excess emission might be the result of background confusion \citep{faramaz18}.

Hip 21547 is an F0V star \citep{abt95} with excess in $70$ and $100 \micron$. While GPI (Gemini Planet Imager) direct imaging observation has been done on the star, no disk signal was observed on the near-IR \citep{macintosh15}. Considering that the planet detected in the study has a fractional luminosity of $1.6 - 4.0 \times 10^{-6}$ \citep{macintosh15}, a potential detection of a debris disk would be difficult. Combine that with the lack of excess in the mid-IR at W3 and W4, and we do not recommend the star for JWST observation.




In conclusion, though 12 of the 19 disks detected in two wavelengths are promising targets for scattered-light observations, none should be considered for mid-IR follow-up with MIRI.

\subsection{Blackbody and modified blackbody models for systems with infrared excess in three or more filters}
\label{subsec:blackbody}

 The remainder of our sample consists of systems with infrared excess detections at three or more wavelengths. With the BB model overconstrained, predictions of disk angular sizes sufficiently large enough to be observable by JWST or ground-based instruments provide strong motivation for follow-up imaging. Table \ref{tab:multiexcess} summarizes the properties of disks fit with the standard BB model (Equation \ref{eq:blackbody}). Of the disks with infrared excess in four or more filters, nine
have SEDs that are best fit by the modified blackbody, according to the likelihood ratio test.\footnote{We note that Hip 22263 has a large dust temperature uncertainty of $28$\%. The best-fit MBB SED is close to the simple BB, but tends to underestimate photometry beyond $100 \micron$. We believe this is in part due to background source contamination, which can be observed in the SPIRE images.} Their dust properties are summarized in Table \ref{tab:mbb}.

Interestingly, the likelihood ratio test described in \S \ref{subsec:diskseds} indicates that all except one of the SEDs in our sample are best fit by either the single BB or the MBB. The one remaining system, Hip~10670, has a best-fit two-component model, BB-MBB. We suspect that the MBB is generally preferred over the 2BB because every target with infrared excess in enough wavelengths to fit the MBB and two-component models has at least one excess detection at $\lambda \geq 100 \micron$. The long-wavelength observations are strongly affected by the poor emissivity of small grains, so replacing 2BB dust rings with one MBB dust ring is likely to produce a significant SED fit improvement.


The fact that only one SED supports the BB-MBB model is interesting, given that two-component debris disk models commonly produce good SED fits in the literature \citep[e.g.,][]{morales11, ballering13, pawellek14}. We suspect that our modeling procedure favors the MBB over the BB-MBB because of our photosphere models’ tendency to have larger error bars than in previous literature. Since the star's fractional contribution to the total flux is much larger at $24 \micron$ than at $70 \micron$ and beyond, our augmented error bars reduce the significance of our mid-IR excess detections more than our far-IR detections. With mid-IR excess detections being marginal, improving the part of the dust SED that traces warm dust produces only a modest improvement in $\mathcal{R}$---not enough to overcome the likelihood ratio test's penalty for increasing the number of free parameters. Again, we are not claiming that a nondetection of a warm component in this study invalidates a previous result. 

\begin{figure}[ht]
		\centering
		\includegraphics[width=0.9\textwidth]{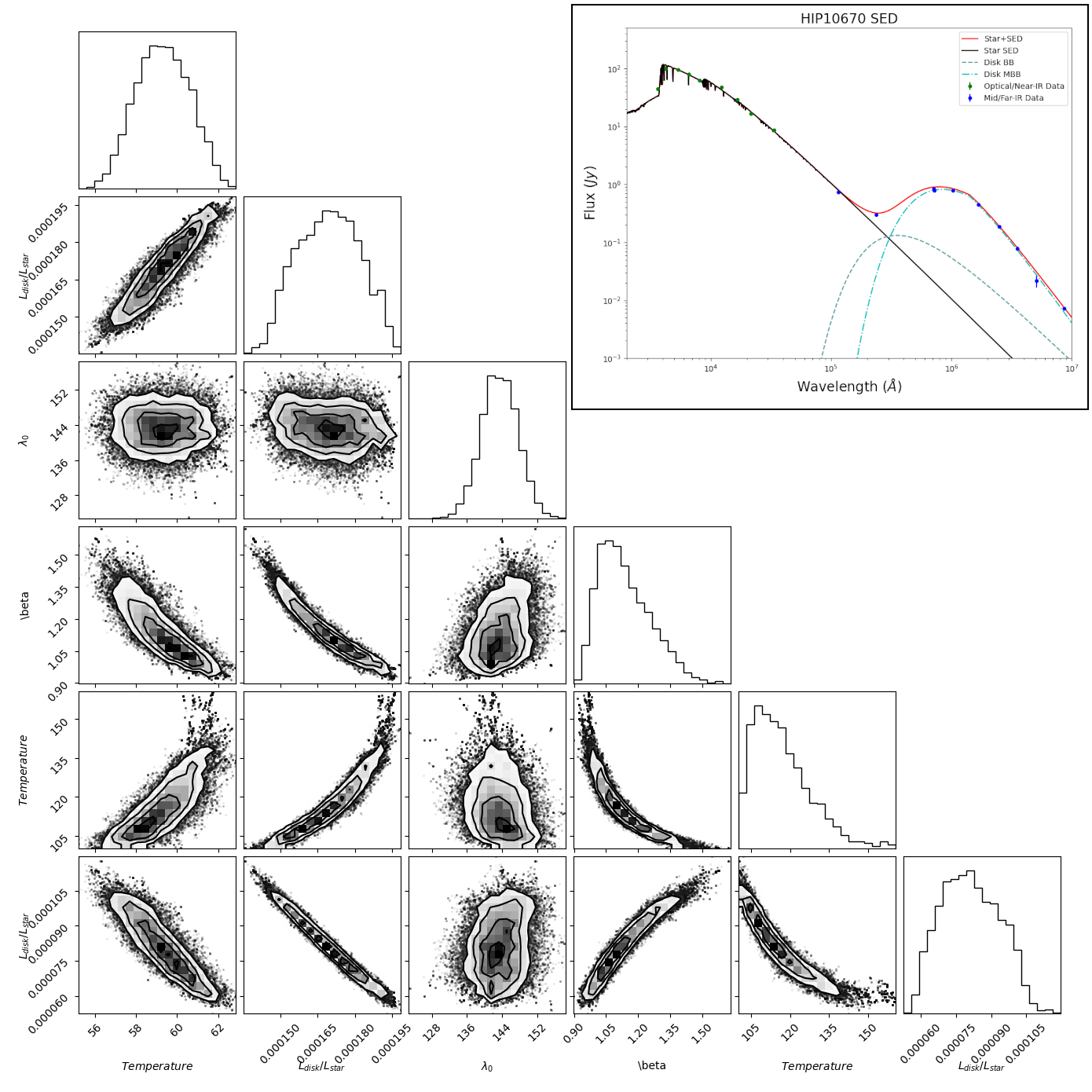}
		\caption{MCMC posteriors and parameter covariances from the Hip 10670 MBB+BB model. Plot generated with \texttt{corner.py}, \citet{foremanmackey16}. The upper right panel is the resulting SED model of Hip 10670.}
		\label{fig:mbbcorner}
	\end{figure}

Plots of posterior distributions and covariances of free parameters for all disks in Tables \ref{tab:multiexcess} and \ref{tab:mbb} are in the accompanying Zenodo archive \citep{zenodo} at doi:10.5281/zenodo.21327122. Hip 10670 is an example of a well-defined disk model: the posterior distributions for each parameter are skew normal, and the degeneracy between the parameters is minimal (Figure \ref{fig:mbbcorner}). 

\begin{figure}[ht]
	\centering
		\includegraphics[width=0.8\textwidth]{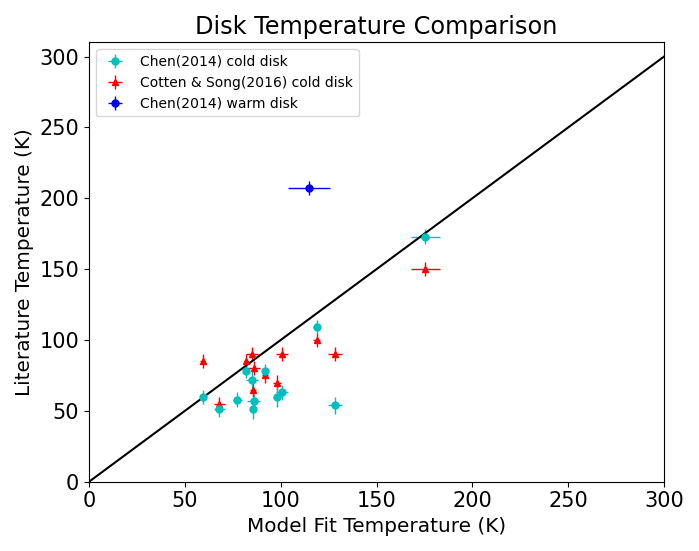} 
	\caption{A comparison of the best-fit disk temperature from this study (x-axis) and previous literature results (y-axis). Our inclusion of photometry with $\lambda > 100 \micron$ leads the likelihood ratio test to favor MBB over 2BB models, which translates into warmer temperatures in this study compared with previous work.}
\label{fig:disktemplitcompare}
\end{figure}

In Figure \ref{fig:disktemplitcompare}, we use the subset of systems with infrared excess at three or more wavelengths to compare our MBB dust temperatures with literature BB/2BB models by \citet{chen14} and \citet{cotten16}. Our MBB temperatures tend to be higher than literature temperatures from either single BB models or the colder components of 2BB models. This is because MBB models consist of dust with a single temperature producing all nonphotospheric emission. The MBB temperature therefore has to be high enough to emit efficiently at the shortest wavelength in which infrared excess is detected. In contrast, a 2BB model can consign nearly all the short-wavelength emission to a warm disk so that the cold component need only cover wavelengths $\lambda \ga 70 \micron$. As explained above, this study generally favors MBB models over BB/2BB because we included photometry at $\lambda \geq 100 \micron$, with increased mid-IR infrared excess detection thresholds from photosphere model error propagation also playing a role.
Figure \ref{fig:disktemplitcompare} shows that differences in the photometry used to build SEDs and the photosphere modeling procedure yield substantially different disk temperatures for many systems.

\subsubsection{Angular size predictions}
\label{subsubsec:angularsize}

\begin{figure}[h]
	\centering
	\begin{tabular}{cc}
		\includegraphics[width=0.52\textwidth]{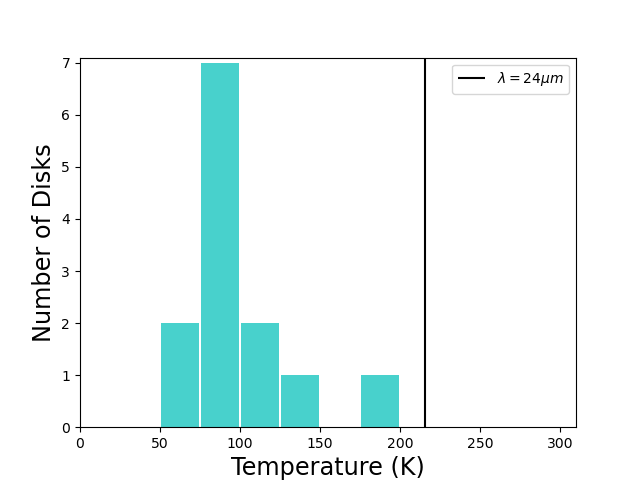} &
		\includegraphics[width=0.5\textwidth]{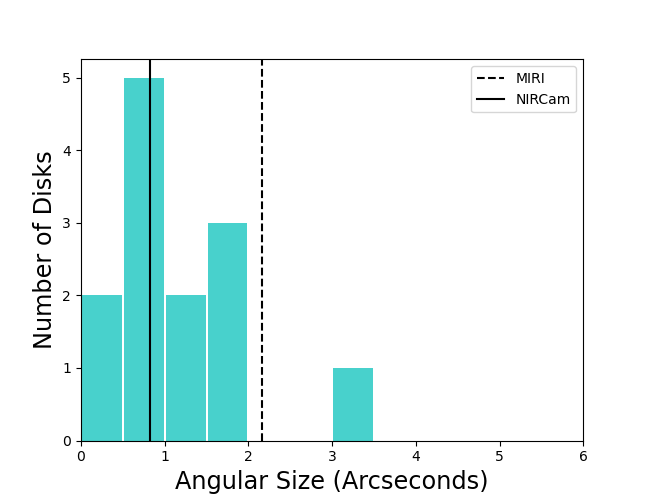} \\
	\end{tabular}
	\caption{Characteristics of disks with infrared excess at three or more wavelengths and best-fit temperatures measured from \texttt{emcee} simulations. Left: disk temperature distributions. The vertical line marks $T = 212.5$~K, at which $\lambda_{\rm max} = 24 \micron$. Right: disk angular size distributions. Vertical lines mark the IWA of JWST's NIRCam MASK430R (solid) and MIRI $23 \micron$ (dashed).}
	\label{fig:disktemphist}
\end{figure}

From our BB/ MBB models, we can identify the disks with large enough angular extent for their thermal emission or scattered light to be resolved with JWST. Figure \ref{fig:disktemphist} shows the dust temperature distribution (left) and the angular size distribution (right) for the systems with infrared excess at three or more wavelengths. As in Figure \ref{fig:disktemphist_2pt}, the vertical lines on the temperature distribution plots mark $T_{\rm dust} = 216$~K, for which the flux density $F_{\nu}$ peaks at $24 \micron$. 
On the angular size distribution plots, vertical lines mark the IWA of NIRCam mask MASK430R covering $4.10$--$4.60 \micron$ \citep[$0.82 \arcsec$;][]{riekeb05} and MIRI with the Lyot coronagraph at $23 \micron$ \citep[$2.16 \arcsec$;][]{gardner04, boccaletti15}.

From Figure \ref{fig:disktemphist}, we see that one disk with infrared excess seen in three or more wavelengths has predicted angular sizes large enough to be resolved with MIRI at $23 \micron$. 
Hip 10670 ($\gamma$ Tri) has $> 14 \sigma$ excess in both W4 and MIPS $24 \micron$. We will return to Hip 10670 after assessing the scattered-light observability of the disks in our sample. 
Another disk, Hip~87108 ($\gamma$ Oph), has a predicted angular size of $1.9 \arcsec$, which is slightly less than the size of the MIRI $23 \micron$ Lyot stop. But the disk flux exceeds the star flux at $24 \micron$, which is why the object was chosen for MIRI observation. Using F1500W and F2500W filters, \citet{han26} imaged a wide, azimuthally asymmetric disk with a radial extent of around $100$~au with a half-width between $33$~au and $48$~au. So even if the MBB model underestimates disk radii, such as in the case of Hip 87108, other considerations such as mid-IR excess and previously resolved images in Herschel or ALMA can be used to determine eligibility for observations.


Eight of the disks detected at three or more wavelengths could be resolved with NIRCam MASK430R provided that they are sufficiently bright in scattered light; these disks are marked with 'Y' in the NIRCam columns of Tables \ref{tab:multiexcess} and \ref{tab:mbb}. Disks that can be resolved with NIRCam are also good candidates for STIS on board the Hubble Space Telescope (HST), which can reach an IWA of $0.2 \arcsec$ in a $52 \arcsec \times 52 \arcsec$ field of view using BAR5 on the 50CORON aperture \citep{debes19A}. The disks that are less extended on the sky can be followed up with SPHERE on the Very Large Telescope (VLT), which has an IWA of $0.1 \arcsec$ in the $H$ band \citep{carbillet11, guerri11} and a field of view of $3.2 \arcsec$ \citep{beuzit19}; SCExAO at Subaru Telescope \citep{jovanovic15}; or GPI 2.0 after it is commissioned \citep{chilcote20}. As for the contrast level needed to observe these disks, scattered-light observations from STIS \citep{schneider14} and SPHERE \citep{engler18,engler19} have detected disks with $L/L_* = 10^{-5}$--$10^{-6}$. As for GPI, all selected targets have $L / L_{\star} > 10^{-5}$ \citep{esposito20}. This qualifies all of our NIRCam targets for STIS and SPHERE and all but Hip 98495 for GPI. Since BB/MBB models give lower limits on the dust ring radii---especially for scattered-light observations that trace the small grains that are most vulnerable to radiation pressure---our models provide strong motivation for including the NIRCam-resolvable disks in direct imaging campaigns that may pave the way for JWST observations.

The NIRCam-resolvable disks---Hip 10670, 19893, 41307, 61960, 69732, 87108, 98495, and 101612---have resolved images from Herschel PACS \citep{booth13,broekhoven13,ertel14,ertel14b,thureau14,moor15,morales16}, though in many cases the source is only marginally extended beyond the PACS PSF and/or the three PACS channels yield different best-fit disk radii, as expected for an extended disk instead of a narrow ring. In agreement with \citet{pawellek14}, we find that the disk sizes measured from Herschel images are larger than the sizes estimated from our SED models for every target except Hip 41307, which may have a warm inner disk extending from $\sim 2$--$3$~au as well as a cold component \citep{booth13}. The potential warm inner disk was revealed by T-ReCS images at $10.4$ and $18.3 \micron$; \citet{moerchen10} found that both images showed extended structure beyond the PSF, but their work and this study both find that the excess flux in T-ReCS wavelengths is marginal ($\sim 2 \sigma$). Lacking strong constraints on the $10$--$20 \micron$ thermal emission, we find that the MBB is a better fit to the SED than the 2BB and recover a disk radius of $39.5$~au, in agreement with the cold disk radius measured from Herschel images by \citet{booth13}. Published Herschel images and their associated ring models can be used as guidelines for determining the appropriate field of view for scattered-light observations.

\subsubsection{Potential MIRI target: Hip~10670}
\label{subsubsec:miritargs}

Finally, we discuss one target that has thermal emission that may be resolvable with MIRI. For such targets, multiwavelength direct imaging provides a detailed map of the dust distribution, tracing both large and small grains and possibly revealing planet--disk interactions. In addition, for warm disks, mid-infrared imaging with an instrument such as MIRI would trace dust that has not moved far from the planetesimal birth ring \citep{strubbe06}, giving a view into the underlying planetesimal population.

	
	
	
	$\gamma$ Tri (Hip 10670) is an A1Vnn star \citep{phillips10} at a distance of $33.96 \pm 0.47$ pc \citep{brown18} for which our best-fit model has two components: a cold disk with $\theta = 3.4 \arcsec$ and a warm disk with $\theta = 0.9 \arcsec$ (Table \ref{tab:mbb}). \citet{pericaud17} used the nondetection of CO molecular line emission plus their DiskFit code to place an upper limit of $0.008 M_{\oplus}$ on the disk's H$_2$ mass, indicating that the infrared excess truly belongs to a debris disk rather than a dispersing protoplanetary disk. The Herschel DEBRIS survey resolved the cold disk at $70$, $100$, and $160 \micron$ and detected it as a point source in all SPIRE wavelengths. Forward modeling of the PACS images by \citet{booth13} suggests that the outer dust ring extends from 120 to 170~au. The same authors fit the SED with separate cold and warm components, reporting temperatures of $60.6 \pm 2.1$ and $183 \pm 38$ K respectively. The \citet{booth13} cold-component temperature closely matches our fit of $59.3 \pm 1.4$ K, while our inner disk temperature of $115 \pm 11$ K differs from their value of $183 \pm 38$~K at the over $3 \sigma$ level. Given the large error bars on both estimates of the inner disk temperature, the discrepancy may be attributable to random noise. But to the extent that the different temperature estimates reveal a true physical difference between the two models, it presumably comes from the difference in input photometry at wavelengths near $10 \micron$, where we use only IRS, while they use both Akari and IRAS. Recall from Section 2.3 that we suspect there is a zero-point offset between Akari and WISE W4; if we are correct, the \citet{booth13} warm disk temperature of $183$~K may be an artifact of a spurious Akari-created local maximum in their SED.
	
	The warm component is too compact to be resolved by either MIRI $23 \micron$ or NIRCam $4 \micron$. However, $15.5 \micron$ observations with the MIRI 4QPM coronagraph revealed flux leakage from the HR~8799 inner disk---which has inferred $T_{\rm dust} \approx 150$~K and $\theta = 0.5 \arcsec$ \citep{su09}, similar to our model of the $\gamma$~Tri inner disk---in the form of a diffraction pattern \citep{boccaletti24}. Since 2BBs may be signposts of (multi)planetary systems \citep[e.g.,][]{matthews18}, we recommend any type of observation that could confirm the existence of the inner disk. Furthermore, if there is substantial mid-IR emission coming from the cold outer disk, it will be resolvable with MIRI $23 \micron$ coronagraphy. This is the case for Fomalhaut's Kuiper Belt analog, which has a best-fit BB model temperature of 50--60~K (similar to our model of $\gamma$~Tri) and was detected with MIRI at both $23$ and $25.5 \micron$ \citep{gaspar23}.
	
	The cold outer disk is also a viable target for scattered-light observations with GPI~2.0 \citep{carbillet11,guerri11,chilcote20}. Although the 2007 and 2008 coronagraphic observations from Gemini North's NIRI did not detect the disk, they used the spectral differential imaging technique designed for substellar companion searches and were processed by algorithms optimized for point-source detection \citep{galicher16}. Most recently, the REASONS survey resolved the debris disk using ALMA $1.27$~mm. Their modeling gave a disk radius of $180$~au \citep{matra25}. We strongly suggest multiwavelength follow-up with MIRI, NIRCam, STIS on board HST, and/or ground-based instruments such as SPHERE and GPI~2.0.

\section{Conclusion}
\label{sec:conclusion}

We used the VOSA SED fitter \citep{bayo08} to create new photosphere models for 62 nearby debris disk host stars with saturated 2MASS photometry. To characterize the Rayleigh-Jeans side of the star SEDs, we replaced the 2MASS data with archival $R_C I_C J H K$ photometry from \citet{ducati02}, which draws primarily from the NASA CIO, and saturation-corrected W1 photometry from the WISE survey \citep{cutri12, patel14}. We found excellent agreement between our model photosphere temperatures and temperatures measured from high-resolution spectroscopy. Our results suggest that the older CIO photometry may be more reliable than 2MASS for nearby stars brighter than $K = 5$. 


With the robust error propagation from VOSA, we found 50 targets that had statistically significant IR excess of $> 3 \sigma_x$ in at least one filter. For the targets with IR excess in two or more filters, we fitted BB models to the excess SEDs. For stars with excess in four or more filters, we also performed MBB and 2BB fits. For every such star except Hip~10670 ($\gamma$~Tri), the likelihood ratio test indicated that a single-component MBB was a better fit to the infrared excess SED than a 2BB or BB+MBB model. Our results differ from similar SED-fitting studies \citep[e.g.,][]{chen14, moor15, cotten16} because we included Herschel PACS and SPIRE plus submillimeter photometry spanning wavelengths $70$--$850 \micron$: We found that the goodness-of-fit penalty for overpredicting the long-wavelength fluxes with the 2BB model, as opposed to the MBB, outweighed the fit improvements the 2BB may have provided at wavelengths $\lambda \leq 70 \micron$.

We used the best-fit dust temperatures to place lower limits on the dust ring radii and the angular extent of the disks on the sky. Eight disks are wider than the IWA of JWST's NIRCam MASK430R and so are compelling targets for high-contrast imaging. To pave the way for eventual JWST observations, the less extended disks could be observed with ground-based instruments such as GPI 2.0, SPHERE, and SCExAO, which have fields of view of $2 \arcsec$--$11 \arcsec$. The more extended disks, such as Hip~10670 ($\theta \geq 3 \arcsec$), could be observed with STIS on board HST. The target disks are moderately resolved in Hershel, but the wide PACS PSF makes radius measurements uncertain. Data with higher spatial resolution would allow us to better characterize the dust properties and search for structure such as brightness asymmetries, warps, gaps, and spirals. Finally, we note that our estimates of the disk radii are probably conservative, and in reality more disks could be resolved by MIRI than just $\gamma$~Tri. This is because the MBB model assumes that grains emit inefficiently at higher wavelengths, which in turn causes dust grains to warm up to higher temperatures than the theoretical BB temperature in order to maintain thermal equilibrium. Considering the overall higher temperatures of our best-fit disk models compared to previous literature, the observed disk radii will likely be larger than any prediction in the literature, including this work.

Our models suggest that disks surrounding $\gamma$~Tri (Hip 10670) will be resolvable with JWST's MIRI coronagraph at $23 \micron$. With high-resolution images of thermal emission that trace the planetesimal birth ring and scattered-light observations of smaller grains, $\gamma$~Tri could become a test bed for collisional cascades. Multiwavelength observing campaigns could also reveal external perturbers such as planets and drag from the ISM.

\clearpage

\startlongtable
\begin{deluxetable*}{ccccccccccc}
	\tablecaption{Optical and Near-IR Photometry \label{tab:optical}}
	\tablewidth{0pt}
	\tabletypesize{\scriptsize}
	\tablehead{
		\colhead{Hipparcos number} & 
		\colhead{Star Name} &
		\colhead{Spectral Type} & 
		\colhead{$U$} & 
		\colhead{$B_T$} & 
		\colhead{$V_T$} &
		\colhead{$R_C$} &
		\colhead{$I_C$} &
		\colhead{$J$} &
		\colhead{$K$} &
		\colhead{W1} \\ 
		\colhead{} & \colhead{} & \colhead{} &  \colhead{(mag)} & \colhead{(mag)} & \colhead{(mag)} &
		\colhead{(mag)} & \colhead{(mag)} & \colhead{(mag)} & \colhead{(mag)} & \colhead{(mag)}
	} 
	\startdata
	1473     & * sig And & A2V&  4.64 & 4.590 & 	4.512 &   4.49 &   4.48 &    4.41 &    4.38 &  4.299\\ 
	2472     & * lam01 Phe & A1Va& 4.820 & 4.785 & 	4.755 &  4.762 &  4.755 & $\cdots$  & $\cdots$ &  4.788\\ 
	2505     & * lam Cas & B8Vn& 4.490 & 4.652 & 	4.756 &   4.75 &   4.82 & $\cdots$  & $\cdots$ &  4.937\\ 
	2578     & * bet03 Tuc & A0V&  5.15 & 5.117 & 	5.069 & $\cdots$ & $\cdots$ &    5.04 &    5.02 &  5.010\\ 
	5862     & * nu. Phe & F9VFe+0.4&  5.65 & 5.638 & 	5.023 &  4.648 &  4.350 &    4.00 &    3.67 &  3.714\\ 
	8497     & * chi Cet & F0V& 5.030 & 5.054 & 	4.694 &  4.452 &  4.273 & $\cdots$ & $\cdots$  &  3.859\\ 
	10670    & * gam Tri & A1Vnn& 4.040 & 4.025 & 	3.999 &  4.02  &  4.01  & $\cdots$  & $\cdots$  &  3.900\\ 
	13209    & * 41 Ari & B8Vn&       & 3.478 & 	3.579 &  3.67  &  3.74  & $\cdots$ & $\cdots$  &  3.842\\
\enddata	
\end{deluxetable*}
 
\newpage

\startlongtable
\begin{longrotatetable}
\begin{deluxetable*}{cccccccccccccccccc}
	\tablecaption{Mid-IR photometry, effective temperature, and metallicity \label{tab:IR1}}
	\setlength{\tabcolsep}{3pt} 
	\tablewidth{0pt}
	\tabletypesize{\tiny}
	\tablehead{
		\colhead{Hipparcos No.} & 
		\colhead{W3} & 
		\colhead{Error} & 
		\colhead{W4} & 
		\colhead{Error} &
		\colhead{IRS $13 \micron$} & 
		\colhead{Error} & 
		\colhead{IRS $17 \micron$} & 
		\colhead{Error} & 
		\colhead{MIPS $24 \micron$} & 
		\colhead{Error} & 
		\colhead{IRS $30 \micron$} &
		\colhead{Error} & 
		\colhead{$T_{\rm eff}^a$} &
		\colhead{$\log g^a$} &
		\colhead{[Fe/H]} &
		\colhead{Literature $T_{\rm eff}$} &
		\colhead{$T_{\rm eff}$ References} \\
		\colhead{} & 
		\colhead{SED W3} & 
		\colhead{} & 
		\colhead{SED W4} & 
		\colhead{} &
		\colhead{SED IRS $13 \micron$} & 
		\colhead{} & 
		\colhead{SED IRS $17 \micron$} & 
		\colhead{} & 
		\colhead{SED MIPS $24 \micron$} & 
		\colhead{} & 
		\colhead{SED IRS $30 \micron$} &
		\colhead{} & 
		\colhead{} &
		\colhead{} &
		\colhead{} &
		\colhead{} &
		\colhead{} \\
		\colhead{ } & \colhead{(Jy)} & \colhead{(Jy)} & \colhead{(Jy)} & \colhead{(Jy)} & \colhead{(Jy)} & \colhead{(Jy)} & \colhead{(Jy)} & \colhead{(Jy)} & \colhead{(Jy)} & \colhead{(Jy)} & \colhead{(Jy)} & \colhead{(Jy)} & \colhead{(K)} & \colhead{} & \colhead{ } & \colhead{(K)} & \colhead{ }  }
\startdata
 1473 & 0.518 & 0.024& 0.182 & 0.011&0.383 & 0.064& 0.246 & 0.037&0.15641 & 0.00313&0.139 & 0.031&8500.0&4.0&0.5&$\cdots$ &$\cdots$ \\ 
1473 & 0.572 & 0.037 & 0.159 & 0.010 &0.537 & 0.034 & 0.305 & 0.020 & 0.150 & 0.010 & 0.100 & 0.00637 & $\cdots$ &$\cdots$ &$\cdots$ &$\cdots$ &$\cdots$ \\ 
\midrule[0.01mm] 
2472 & 0.371 & 0.018& 0.130 & 0.008&0.293 & 0.084& 0.193 & 0.033&0.11481 & 0.00230&0.091 & 0.025&8750.0&3.5&0.5&5774.0&(1)\\ 
2472 & 0.455 & 0.009 & 0.126 & 0.002 &0.427 & 0.009 & 0.243 & 0.005 & 0.119 & 0.002 & 0.079 & 0.00154 & $\cdots$ &$\cdots$ &$\cdots$ &$\cdots$ & $\cdots$\\ 
\midrule[0.01mm] 
2505 & 0.280 & 0.013& 0.100 & 0.006&0.3473 & 0.00568& $\cdots$  & $\cdots$ & $\cdots$ & $\cdots$ &0.0583 & 0.00527&10,500.0&4.0&0.5&$\cdots$ & $\cdots$\\ 
2505 & 0.340 & 0.011 & 0.094 & 0.003 &0.319 & 0.010 & 0.181 & 0.006 & 0.089 & 0.003 & 0.059 & 0.00178 & $\cdots$ &$\cdots$ &$\cdots$ &$\cdots$ & $\cdots$\\ 
\midrule[0.01mm] 
2578 & 0.327 & 0.015& 0.255* & 0.015&0.279 & 0.098& 0.243 & 0.061&0.23749* & 0.00475&0.224 & 0.063&8750.0&4.0&0.5&$\cdots$ & $\cdots$ \\ 
2578 & 0.344 & 0.006 & 0.095 & 0.002 &0.323 & 0.006 & 0.183 & 0.003 & 0.090 & 0.002 & 0.060 & 0.00106 &  $\cdots$ &$\cdots$ &$\cdots$ &$\cdots$ & $\cdots$ \\
\enddata
\tablecomments{Mid-IR photometry up to $30 \micron$ used for identifying infrared excesses, plus metallicity and temperature from this study and previous literature. For each star, there are two rows. The first row contains the observed flux at each wavelength, and an asterisk to the right of the flux value denotes an infrared excess detection. The second row contains the predicted star flux. References. (1) \cite{dasilva15}; (2) \cite{bensby14}; (3) \cite{takeda05}; (4) \cite{delgado17}; (5) \cite{santos04}; (6) \cite{gebran10}; (7) \cite{ramirez14}; (8) \cite{maldonado12}; (9) \cite{sitnova15}; (10) \cite{dasilva12}; (11) \cite{ghezzi10}; (12) \cite{prugniel11}; (13) \cite{brewer16}.} 
	\tablenotetext{a}{Best-fit values of $T_{\rm eff}$ and $\log g$ from model photosphere Set 1 (see \S \ref{sec:photospheres})}
\end{deluxetable*}
\end{longrotatetable}

\newpage

\startlongtable
\begin{longrotatetable}
\begin{deluxetable*}{ccccccccccccccccc}
	\setlength{\tabcolsep}{3pt} 
	\tablecaption{Far-IR photometry \label{tab:IR2}}
	\tablewidth{0pt}
	\tabletypesize{\tiny}
	\tablehead{
		\colhead{Hipparcos No.} & 
		\colhead{MIPS 70} &
		\colhead{Error} &
		\colhead{PACS 70} &
		\colhead{Error} &
		\colhead{PACS 100} &
		\colhead{Error} &
		\colhead{PACS 160} &
		\colhead{Error} &
		\colhead{SPIRE 250} &
		\colhead{Error} &
		\colhead{SPIRE 350} &
		\colhead{Error} &
		\colhead{SPIRE 500} &
		\colhead{Error} &
		\colhead{850 $\micron$} &
		\colhead{Error}\\	
		\colhead{} & 
		\colhead{SED MIPS 70} &
		\colhead{} &
		\colhead{SED PACS 70} &
		\colhead{} &
		\colhead{SED PACS 100} &
		\colhead{} &
		\colhead{SED PACS 160} &
		\colhead{} &
		\colhead{SED SPIRE 250} &
		\colhead{} &
		\colhead{SED SPIRE 350} &
		\colhead{} &
		\colhead{SED SPIRE 500} &
		\colhead{} &
		\colhead{SED 850 $\micron$} &
		\colhead{}\\	 
		\colhead{} & \colhead{(Jy)} & \colhead{(Jy)} & \colhead{(Jy)} & \colhead{(Jy)} &
		\colhead{(Jy)} & \colhead{(Jy)} & \colhead{(Jy)}& \colhead{(Jy)} & \colhead{(Jy)} & \colhead{(Jy)} & \colhead{(Jy)} & \colhead{(Jy)} & \colhead{(Jy)} & \colhead{(Jy)} & \colhead{(Jy)} & \colhead{(Jy)}
	} 
	\startdata
	1473 & 0.04382* & 0.00612&  &   &0.02778* & 0.00268& $\cdots$  & $\cdots$ & $\cdots$  & $\cdots$ & $\cdots$ & $\cdots$ &$\cdots$  & $\cdots$  & $\cdots$  & $\cdots$ \\ 
	1473 & 0.015 & 0.0003 & 0.015 & 0.0003 & 0.0075 & 0.00017 & 0.0030 & 0.00007 & 0.00116 & 0.000029 & 0.000583 & 0.0000152 &0.0002920 & 0.00000802 & 0.0000882 & 0.00000265\\ 
	\midrule[0.01mm] 
	2472 & 0.07170* & 0.00502&  $\cdots$ &  $\cdots$  &  $\cdots$ & $\cdots$  &  $\cdots$  & $\cdots$ &   $\cdots$ &  $\cdots$ & $\cdots$  & $\cdots$ & $\cdots$  &   $\cdots$ &  $\cdots$  &  \\ 
	2472 & 0.011 & 0.0002 & 0.011 & 0.0002 & 0.0056 & 0.00009 & 0.0022 & 0.00004 & 0.00085 & 0.000015 & 0.000429 & 0.0000083 &0.0002147 & 0.00000441 & 0.0000647 & 0.00000148\\ 
	\midrule[0.01mm] 
	2505 &  $\cdots$  & $\cdots$ &  $\cdots$ &   $\cdots$ &  $\cdots$ &  $\cdots$ &  $\cdots$  & $\cdots$ &   $\cdots$ & $\cdots$  & $\cdots$  & $\cdots$ &  $\cdots$ &  $\cdots$  &  $\cdots$  &  $\cdots$ \\ 
	2505 & 0.009 & 0.0002 & 0.009 & 0.0002 & 0.0042 & 0.00010 & 0.0017 & 0.00005 & 0.00065 & 0.000020 & 0.000329 & 0.0000108 &0.0001647 & 0.00000595 & 0.0000498 & 0.00000208\\ 
	\midrule[0.01mm] 
	2578 & 0.06220* & 0.00510&0.0597* & 0.00380 &0.01900* & 0.00240& 0.0182* &0.00182&  $\cdots$  &  $\cdots$ &  $\cdots$ & $\cdots$ &  $\cdots$ &  $\cdots$  &  $\cdots$  &  $\cdots$ \\ 
	2578 & 0.009 & 0.0001 & 0.009 & 0.0001 & 0.0042 & 0.00007 & 0.0017 & 0.00003 & 0.00064 & 0.000013 & 0.000322 & 0.0000069 &0.0001615 & 0.00000375 & 0.0000488 & 0.00000128\\
\enddata
	\tablecomments{Mid- and far-IR photometry used for identifying infrared excesses. This table includes the PACS, IRS, and SPIRE photometry. For each star, there are two rows. The first row contains the observed flux at each wavelength, and an asterisk to the right of the flux value denotes an infrared excess detection. The second row contains the predicted star flux.}
\end{deluxetable*}
\end{longrotatetable}

\clearpage

\begin{deluxetable*}{ccccc}
	\tablecaption{Binary stars with $K$-band flux ratio $F_{K,2} / F_{K,1} > 0.01$ \label{tab:binary}}
        \tablewidth{0pt}
        \tabletypesize{\normalsize}
	\tablehead{
		\colhead{Hipparcos Number} & 
		\colhead{$F_{K,2} / F_{K,1}$} & 
		\colhead{$(F_{K,2} + F_{K,1})/F_{K,{\rm SED}}$} &
		\colhead{Separation (")} &
		\colhead{Separation reference\tablenotemark{a}}
	}
	\startdata
	1473 & 0.033 & 1.002 & 7.00 & WDS \\ 
	2472 & 0.012 & 0.833 & 35.4 & WDS \\
	2505 & 0.853 & 1.417 & 0.446 & ET \\ 
	2578 & 0.010 & 0.920 & 2.4 & ET \\ 
	8497 & 0.276 & 1.131 & 192.90 & WDS\\ 
	14954 & 0.056 & 0.910 & 2.20 & WDS \\ 
	15197 & 0.522 & 1.572 & 999.9 & ET \\ 
	15371 & 0.857 & 1.834 & 309.1 & WDS\\ 
	19990 & 0.096 & 1.019 & 179.90 & WDS \\ 
	20635 & 0.513 & 1.434 & 339.40 & WDS \\ 
	21547 & 0.072 & 0.967 & 66.70 & WDS \\ 
	22845 & 0.108 & 1.080 & 171.60 & WDS \\ 
	23497 & 1.000 & 1.356 & 0.4 & ET \\ 
	23693 & 0.141 & 1.049 & 321.60 & WDS\\ 
	33302 & 0.087 & 1.035 & 11.60 & WDS \\ 
	45571 & 0.437 & 1.262 & 0.124 & BSC \\ 
	55705 & 0.128 & 1.161 & 4.40 & WDS \\ 
	57757 & 0.057 & 1.026 & 326.60 & WDS \\ 
	74824 & 0.010 & 1.007 & 217.60 & WDS \\ 
	75304 & 0.230 & 1.103 & \nodata & \nodata \\ 
	109857 & 0.094 & 0.963 & 0.30 & WDS \\ 
	111188 & 0.218 & 1.086 & 30.40& WDS \\ 
	117452 & 0.032 & 0.972 & 3.40& WDS \\ 
	\enddata
	 \tablenotetext{a}{References for the angular separations are \citet[ET;][]{eggleton08}, the Washington Visual Double Star Catalog \citep[WDS;][]{mason17}, and the Bright Star Catalog \citep[BSC;][]{hoffleit91}. Hip 75304, which has no entry in column 4, has a companion that is inferred but not directly detected.}
\end{deluxetable*}

\clearpage

\begin{deluxetable*}{c}
	\tablecaption{Stars with no statistically significant infrared excess \label{tab:zeroexcess}}
	\tablewidth{6in}
	\tabletypesize{\normalsize}
	\tablehead{
		\colhead{Hipparcos Number} 
	} 
	\startdata
	2505\\
	20635\\ 
	22509\\ 
	23497\\ 
	33302\\ 
	45571\\ 
	66234\\ 
	75304\\ 
	79881\\ 
	95347\\ 
	109857\\ 
	116631\\   
	\enddata	
	\tablecomments{Since no significant IR excess has been detected for these stars, no disk property estimation is possible.}
\end{deluxetable*}

\newpage

\begin{deluxetable*}{ccccc}
	\tablecaption{Constraints on dust models for sources with IR excess in one filter \label{tab:oneexcess}}
	\tablewidth{0pt}
	\tablehead{
		\colhead{Hipparcos Number} & 
		\colhead{Maximum $T_{\rm dust}$} & 
		\colhead{Minimum $L_{\rm dust} / L_*$} \\ 
		\colhead{} & \colhead{(K)} & \colhead{} 
	} 
	\startdata
	2472 & 71.4 & $1.3 \times 10^{-5}$ \\ 
	5862 & 71.4 & $4.8 \times 10^{-6}$ \\ 
	13209 & 231.0 & $9.0 \times 10^{-6}$\\
	18975 & 71.4 & $6.3 \times 10^{-5}$ \\ 
	19990 & 71.4 & $8.0 \times 10^{-6}$ \\ 
	27435 & 71.4 & $8.3 \times 10^{-5}$ \\ 
	53954 & 215.5 & $2.7 \times 10^{-5}$ \\ 
	55705 & 71.4 & $3.3 \times 10^{-6}$ \\ 
	57757 & 71.4 & $2.4 \times 10^{-6}$ \\ 
	66765 & 71.4 & $3.0 \times 10^{-5}$ \\ 
	74702 & 71.4 & $1.3 \times 10^{-4}$ \\ 
	74824 & 215.5 & $3.3 \times 10^{-5}$ \\ 
	76829 & 71.4 & $8.7 \times 10^{-5}$ \\ 
	78554 & 70.9 & $4.1 \times 10^{-5}$ \\ 
	82673 & 231.0 & $1.8 \times 10^{-5}$ \\ 
	93542 & 231.0 & $8.0 \times 10^{-5}$ \\ 
	113044 & 71.4 & $1.8 \times 10^{-5}$ \\ 
	115738 & 215.5 & $2.1 \times 10^{-5}$ \\  
	\enddata	
	\tablecomments{Dust temperature upper limits and luminosity lower limits for sources with infrared excess in one filter.}
\end{deluxetable*}

\clearpage

\begin{deluxetable*}{ccccccc}
	\tablecaption{Best-fit dust models for sources with infrared excess in two filters \label{tab:twoexcess}}
	\tablewidth{0pt}
	\tablehead{
		\colhead{Hipparcos Number} & 
		\colhead{$T_{disk}$} & 
		\colhead{$T_{disk}$ err} & 
		\colhead{$\log(L_{disk} / L_{\star})$} & 
		\colhead{$\log(L_{disk} / L_{\star})$ err} & 
		\colhead{R}  &
		 \colhead{$\theta$} \\ 
		\colhead{} & \colhead{(K)} & \colhead{(K)} & \colhead{} & 
		\colhead{} & \colhead{(AU)} & \colhead{($\arcsec$)}
	} 
	\startdata
	1473 & 95.3 & \nodata &-5.16 & \nodata & 40.4 & 0.96 \\ 
	8497 & 43.5 & 6.2 &-5.15 & 0.05 & 98.0 & 4.20 \\ 
	13402 & 72.3 & 15.6 &-4.55 & 0.17 & 9.2 & 0.89 \\ 
	14954 & 57.5 & 16.4 &-5.31 & 0.29 & 47.2 & 2.10 \\ 
	15371 & 42.8 & 7.1 &-5.28 & 0.05 & 41.7 & 3.47 \\ 
	16852 & 65.4 & 9.1 &-5.15 & 0.09 & 31.6 & 2.26 \\ 
	20901 & 82.7 & 2.5 &-4.24 & 0.04 & 47.0 & 0.97 \\ 
	21547 & 49.1 & 11.2 &-5.44 & 0.57 & 77.0 & 2.58 \\ 
	22845 & 74.4 & 1.9 &-4.07 & 0.04 & 49.4 & 1.44 \\ 
	23693 & 103.6 & \nodata &-5.05 & \nodata & 8.7 & 0.75 \\ 
	42438 & 105.3 & \nodata &-4.82 & \nodata & 7.0 & 0.48 \\ 
	43726 & 102.7 & \nodata &-4.81 & \nodata & 7.0 & 0.42 \\ 
	49593 & 185.3 & 34.4 &-4.66 & 0.60 & 6.7 & 0.25 \\ 
	62207 & 68.5 & 8.8 &-4.57 & 0.07 & 17.3 & 0.98 \\ 
	65721 & 148.7 & \nodata &-4.75 & \nodata & 6.1 & 0.34 \\ 
	99742 & 92.3 & 2.0 &-4.26 & 0.02 & 40.3 & 0.84 \\ 
	103389 & 102.9 & \nodata &-4.64 & \nodata & 10.2 & 0.48 \\ 
	111188 & 97.2 & \nodata &-5.07 & \nodata & 45.3 & 1.05 \\ 
	117452 & 148.2 & 10.5 &-4.61 & 0.02 & 17.6 & 0.41 \\ 
	\enddata	
\end{deluxetable*}

\clearpage

\begin{deluxetable*}{ccccccccc}
	\tablecaption{Best-fit blackbody dust models for systems with infrared excess at three or more wavelengths \label{tab:multiexcess}}
	\tablewidth{0pt}
	\tabletypesize{\normalsize}
	\tablehead{
		\colhead{Hipparcos Number} & 
		\colhead{$T_{disk}$} & 
		\colhead{$T_{disk}$ err} & 
		\colhead{$L_{disk} / L_{\star}$} & 
		\colhead{$L_{disk} / L_{\star}$ err} &
		\colhead{R} &
		\colhead{$\theta$} & 
		\colhead{MIRI\tablenotemark{a}} & 
		\colhead{NIRCam\tablenotemark{b}} \\ 
		\colhead{} & \colhead{(K)} & \colhead{(K)} & \colhead{} & 
		\colhead{} & \colhead{(AU)} & \colhead{(")}
	} 
	\startdata
	86305 & 128.3 & 3.8 & $3.7 \times 10^{-4}$ & $1.0 \times 10^{-5}$ & 15.7 & 0.4 & N & N \\ 
	98495 & 84.9 & 3.1 & $1.5 \times 10^{-5}$ & $1.2 \times 10^{-6}$ & 52.8 & 1.7 & N & Y \\ 
	\enddata	
	\tablenotetext{a}{``Y'' indicates that the disk is resolvable with the MIRI coronagraph at $23 \micron$. ``N'' indicates that MIRI cannot resolve the disk at $23 \micron$.}
	\tablenotetext{b}{``Y'' indicates that the disk is resolvable with the NIRCam coronagraph MASK201R, which covers 1.82--$2.12 \micron$. ``N'' indicates that NIRCam MASK201R cannot resolve the disk unless there is an as-yet-undetected halo of small grains.}
\end{deluxetable*}

\begin{deluxetable*}{cccccccccccc}
	\tablecaption{Best-fit modified blackbody dust models for systems with infrared excess at four or more wavelengths \label{tab:mbb}}
	\tablewidth{0pt}
	\tabletypesize{\scriptsize}
	\tablehead{
		\colhead{Hip Number} & 
		\colhead{Model} &
		\colhead{$T$} & 
		\colhead{$T$ err} & 
		\colhead{$L / L_{\star}$} & 
		\colhead{$L / L_{\star}$ err} &
		\colhead{$\lambda_0$} & 
		\colhead{$\beta$} &
		\colhead{$R_d$} &
		\colhead{$\theta$} & 
		\colhead{MIRI} & 
		\colhead{NIRCam} \\ 
		\colhead{} & \colhead{} & \colhead{(K)} & \colhead{(K)} & \colhead{} & 
		\colhead{} & \colhead{} & \colhead{} & \colhead{(AU)} & \colhead{(")}
	} 
	\startdata
	HIP2578 & MBB & 175.5  & 7.4 & $4.0 \times 10^{-4}$ & $3.2\times 10^{-5}$ & 35.0 & 0.7 & 10.1 & 0.2  & N & N\\ 
	HIP10670 & BB+MBB & 59.3 & 1.4 & $1.7 \times 10^{-4}$ & $1.3\times 10^{-5}$ & 143.4 & 1.1 & 114.6 & 3.4  & Y & Y\\ 
	&   & 114.7 & 10.9 &$8.1 \times 10^{-5}$ & $1.3\times 10^{-5}$ &  &  & 30.6 & 0.9  & N & Y\\ 
	HIP15197 & MBB & 98.3  & 2.1 & $1.2 \times 10^{-4}$ & $5.5\times 10^{-6}$ & 42.2 & 0.0 & 27.0 & 0.8  & N & N\\ 
	HIP17439 & MBB & 77.4  & 2.2 & $4.4 \times 10^{-4}$ & $3.9\times 10^{-5}$ & 41.5 & 0.1 & 8.1 & 0.5  & N & N\\ 
	HIP19893 & MBB & 85.5  & 1.4 & $1.1 \times 10^{-4}$ & $3.2\times 10^{-6}$ & 44.0 & 0.0 & 26.9 & 1.3  & N & Y\\ 
	HIP22263 & MBB & 100.7  & 3.1 & $2.2 \times 10^{-4}$ & $1.7\times 10^{-5}$ & 44.8 & 0.1 & 7.3 & 0.6  & N & N\\ 
	HIP41307 & MBB & 118.8  & 1.9 & $1.0 \times 10^{-4}$ & $2.2\times 10^{-6}$ & 71.1 & 1.2 & 32.8 & 0.9  & N & Y\\
\enddata	
	\tablecomments{The ``model'' column indicates whether the best-fit model is a modified blackbody (MBB) or a two-component model with a blackbody inner disk and a modified blackbody outer disk (BB+MBB). The entries under MIRI and and NIRCam denote whether the objects are resolvable by the respective instruments, with ``Y" meaning yes and ``N" meaning no. The parameters for the two-component model SED of Hip~10670 ($\gamma$~Tri) are given two rows in the table, with the cold outer component on top and the warm inner component underneath.}
	
\end{deluxetable*}
\clearpage

\section{Acknowledgements}

This publication makes use of VOSA, developed under the Spanish Virtual Observatory project supported by the Spanish MICINN through grant AyA2011-24052. This research has made use of the SVO Filter Profile Service "Carlos Rodrigo", funded by MCIN/AEI/10.13039/501100011033/ through grant PID2023-146210NB-I00. This research has made use of the SIMBAD database, operated at CDS, Strasbourg, France \citep{wenger00}. This research has made use of the VizieR catalogue access tool, CDS, Strasbourg, France (DOI: 10.26093/cds/vizier). The original description of the VizieR service was published in \citep{ochsenbein00}. This publication makes use of data products from the Two Micron All Sky Survey, which is a joint project of the University of Massachusetts and the Infrared Processing and Analysis Center/California Institute of Technology, funded by the National Aeronautics and Space Administration and the National Science Foundation. This research has made use of the NASA/IPAC Infrared Science Archive, which is funded by the National Aeronautics and Space Administration and operated by the California Institute of Technology. This work was supported by NSF grant 1519216.

\bibliography{2mass_sat_paper_source2}

@inproceedings{backman93,
	adsurl = {https://ui.adsabs.harvard.edu/abs/1993prpl.conf.1253B},
	author = {{Backman}, Dana E. and {Paresce}, Francesco},
	booktitle = {Protostars and Planets III},
	editor = {{Levy}, Eugene H. and {Lunine}, Jonathan I.},
	month = jan,
	pages = {1253},
	title = {{Main-Sequence Stars with Circumstellar Solid Material - the VEGA Phenomenon}},
	publisher = {Univ. of Arizona Press},
	year = {1993}}

@MISC{cutri12,
       author = {{Cutri}, R.~M. and {Wright}, E.~L. and {Conrow}, T. and {Bauer}, J. and {Benford}, D. and {Brandenburg}, H. and {Dailey}, J. and {Eisenhardt}, P.~R.~M. and {Evans}, T. and {Fajardo-Acosta}, S. and {Fowler}, J. and {Gelino}, C. and {Grillmair}, C. and {Harbut}, M. and {Hoffman}, D. and {Jarrett}, T. and {Kirkpatrick}, J.~D. and {Leisawitz}, D. and {Liu}, W. and {Mainzer}, A. and {Marsh}, K. and {Masci}, F. and {McCallon}, H. and {Padgett}, D. and {Ressler}, M.~E. and {Royer}, D. and {Skrutskie}, M.~F. and {Stanford}, S.~A. and {Wyatt}, P.~L. and {Tholen}, D. and {Tsai}, C.~W. and {Wachter}, S. and {Wheelock}, S.~L. and {Yan}, L. and {Alles}, R. and {Beck}, R. and {Grav}, T. and {Masiero}, J. and {McCollum}, B. and {McGehee}, P. and {Papin}, M. and {Wittman}, M.},
        title = "{Explanatory Supplement to the WISE All-Sky Data Release Products}",
 howpublished = {Explanatory Supplement to the WISE All-Sky Data Release Products},
         year = 2012,
        month = mar,
        pages = {1},
       adsurl = {https://ui.adsabs.harvard.edu/abs/2012wise.rept....1C},
}

@MISC{cutri12data,
       author = {{Cutri}, R.~M. and {et al.}},
        title = "{VizieR Online Data Catalog: WISE All-Sky Data Release (Cutri+ 2012)}",
 howpublished = {VizieR On-line Data Catalog: II/311.  Originally published in: 2012wise.rept....1C},
         year = 2012,
        month = apr,
          eid = {II/311},
       adsurl = {https://ui.adsabs.harvard.edu/abs/2012yCat.2311....0C}
}

@ARTICLE{yang24,
       author = {{Yang}, Qiancheng and {Liu}, Qiong and {Kennedy}, Grant M. and {Wyatt}, Mark C. and {Dodson-Robinson}, Sarah and {Akeson}, Rachel and {Liao}, Nenghui},
        title = "{First ALMA observations of the HD 105211 debris disc: A warm dust component close to a gigayear-old star}",
      journal = {\aap},
         year = 2024,
        month = jun,
       volume = {686},
          eid = {A206},
        pages = {A206},
          doi = {10.1051/0004-6361/202449280},
archivePrefix = {arXiv},
       eprint = {2404.04508},
 primaryClass = {astro-ph.EP},
       adsurl = {https://ui.adsabs.harvard.edu/abs/2024A&A...686A.206Y}
}

@ARTICLE{adams18,
       author = {{Adams}, J.~D. and {Herter}, T.~L. and {Lau}, R.~M. and {Trinh}, C. and {Hankins}, M.},
        title = "{Dust Production Rates in the Fomalhaut Debris Disk from SOFIA/FORCAST Mid-infrared Imaging}",
      journal = {\apj},
         year = 2018,
        month = aug,
       volume = {862},
       number = {2},
          eid = {161},
        pages = {161},
          doi = {10.3847/1538-4357/aacf96},
archivePrefix = {arXiv},
       eprint = {1806.09746},
 primaryClass = {astro-ph.EP},
       adsurl = {https://ui.adsabs.harvard.edu/abs/2018ApJ...862..161A}
}

@ARTICLE{bouchet15,
       author = {{Bouchet}, Patrice and {Garc{\'\i}a-Mar{\'\i}n}, Macarena and {Lagage}, P. -O. and {Amiaux}, J{\'e}rome and {Augu{\'e}res}, J. -L. and {Bauwens}, Eva and {Blommaert}, J.~A.~D.~L. and {Chen}, C.~H. and {Detre}, {\"O}. H. and {Dicken}, Dan and {Dubreuil}, D. and {Galdemard}, Ph. and {Gastaud}, R. and {Glasse}, A. and {Gordon}, K.~D. and {Gougnaud}, F. and {Guillard}, Phillippe and {Justtanont}, K. and {Krause}, Oliver and {Leboeuf}, Didier and {Longval}, Yuying and {Martin}, Laurant and {Mazy}, Emmanuel and {Moreau}, Vincent and {Olofsson}, G{\"o}ran and {Ray}, T.~P. and {Rees}, J. -M. and {Renotte}, Etienne and {Ressler}, M.~E. and {Ronayette}, Samuel and {Salasca}, Sophie and {Scheithauer}, Silvia and {Sykes}, Jon and {Thelen}, M.~P. and {Wells}, Martyn and {Wright}, David and {Wright}, G.~S.},
        title = "{The Mid-Infrared Instrument for the James Webb Space Telescope, III: MIRIM, The MIRI Imager}",
      journal = {\pasp},
         year = 2015,
        month = jul,
       volume = {127},
       number = {953},
        pages = {612},
          doi = {10.1086/682254},
archivePrefix = {arXiv},
       eprint = {1508.02488},
 primaryClass = {astro-ph.IM},
       adsurl = {https://ui.adsabs.harvard.edu/abs/2015PASP..127..612B}
}

@ARTICLE{sawicki12,
       author = {{Sawicki}, Marcin},
        title = "{SEDfit: Software for Spectral Energy Distribution Fitting of Photometric Data}",
      journal = {\pasp},
         year = 2012,
        month = nov,
       volume = {124},
       number = {921},
        pages = {1208},
          doi = {10.1086/668636},
archivePrefix = {arXiv},
       eprint = {1210.0285},
 primaryClass = {astro-ph.IM},
       adsurl = {https://ui.adsabs.harvard.edu/abs/2012PASP..124.1208S}
}

@ARTICLE{gaspar23,
       author = {{G{\'a}sp{\'a}r}, Andr{\'a}s and {Wolff}, Schuyler Grace and {Rieke}, George H. and {Leisenring}, Jarron M. and {Morrison}, Jane and {Su}, Kate Y.~L. and {Ward-Duong}, Kimberly and {Aguilar}, Jonathan and {Ygouf}, Marie and {Beichman}, Charles and {Llop-Sayson}, Jorge and {Bryden}, Geoffrey},
        title = "{Spatially resolved imaging of the inner Fomalhaut disk using JWST/MIRI}",
      journal = {Nature Astronomy},
         year = 2023,
        month = jul,
       volume = {7},
        pages = {790-798},
          doi = {10.1038/s41550-023-01962-6},
archivePrefix = {arXiv},
       eprint = {2305.03789},
 primaryClass = {astro-ph.EP},
       adsurl = {https://ui.adsabs.harvard.edu/abs/2023NatAs...7..790G}
}

@ARTICLE{matthews18,
       author = {{Matthews}, Elisabeth and {Hinkley}, Sasha and {Vigan}, Arthur and {Kennedy}, Grant and {Sutlieff}, Ben and {Wickenden}, Dawn and {Treves}, Sam and {David}, Trevor and {Meshkat}, Tiffany and {Mawet}, Dimitri and {Morales}, Farisa and {Shannon}, Andrew and {Stapelfeldt}, Karl},
        title = "{Constraining the presence of giant planets in two-belt debris disc systems with VLT/SPHERE direct imaging and dynamical arguments}",
      journal = {\mnras},
         year = 2018,
        month = oct,
       volume = {480},
       number = {2},
        pages = {2757-2783},
          doi = {10.1093/mnras/sty1778},
archivePrefix = {arXiv},
       eprint = {1807.01326},
 primaryClass = {astro-ph.EP},
       adsurl = {https://ui.adsabs.harvard.edu/abs/2018MNRAS.480.2757M}
}

@ARTICLE{su09,
       author = {{Su}, K.~Y.~L. and {Rieke}, G.~H. and {Stapelfeldt}, K.~R. and {Malhotra}, R. and {Bryden}, G. and {Smith}, P.~S. and {Misselt}, K.~A. and {Moro-Martin}, A. and {Williams}, J.~P.},
        title = "{The Debris Disk Around HR 8799}",
      journal = {\apj},
         year = 2009,
        month = nov,
       volume = {705},
       number = {1},
        pages = {314-327},
          doi = {10.1088/0004-637X/705/1/314},
archivePrefix = {arXiv},
       eprint = {0909.2687},
 primaryClass = {astro-ph.EP},
       adsurl = {https://ui.adsabs.harvard.edu/abs/2009ApJ...705..314S},
}

@INPROCEEDINGS{bailey23,
       author = {{Bailey}, Vanessa P. and {Bendek}, Eduardo and {Monacelli}, Brian and {Baker}, Caleb and {Bedrosian}, Gasia and {Cady}, Eric and {Douglas}, Ewan S. and {Groff}, Tyler and {Hildebrandt}, Sergi R. and {Kasdin}, N. Jeremy and {Krist}, John and {Macintosh}, Bruce and {Mennesson}, Bertrand and {Morrissey}, Patrick and {Poberezhskiy}, Ilya and {Subedi}, Hari B. and {Rhodes}, Jason and {Roberge}, Aki and {Ygouf}, Marie and {Zellem}, Robert T. and {Zhao}, Feng and {Zimmerman}, Neil T.},
        title = "{Nancy Grace Roman Space Telescope coronagraph instrument overview and status}",
    booktitle = {Society of Photo-Optical Instrumentation Engineers (SPIE) Conference Series},
         year = 2023,
       series = {Society of Photo-Optical Instrumentation Engineers (SPIE) Conference Series},
       volume = {12680},
        month = oct,
          eid = {126800T},
        pages = {126800T},
          doi = {10.1117/12.2679036},
archivePrefix = {arXiv},
       eprint = {2309.08672},
 primaryClass = {astro-ph.IM},
       adsurl = {https://ui.adsabs.harvard.edu/abs/2023SPIE12680E..0TB}
}

@INPROCEEDINGS{anche24,
       author = {{Anche}, Ramya M. and {Milani}, Kian and {Doelman}, David and {Hom}, Justin and {Millar-Blanchaer}, Maxwell A. and {Wolff}, Schuyler and {Douglas}, Ewan and {Snik}, Frans and {Ashcraft}, Jaren N.},
        title = "{High-contrast polarimetric observations of debris disks through the Roman Coronagraph Instrument}",
    booktitle = {Space Telescopes and Instrumentation 2024: Optical, Infrared, and Millimeter Wave},
         year = 2024,
       editor = {{Coyle}, Laura E. and {Matsuura}, Shuji and {Perrin}, Marshall D.},
       series = {Society of Photo-Optical Instrumentation Engineers (SPIE) Conference Series},
       volume = {13092},
        month = aug,
          eid = {1309257},
        pages = {1309257},
          doi = {10.1117/12.3020643},
       adsurl = {https://ui.adsabs.harvard.edu/abs/2024SPIE13092E..57A}
}

@ARTICLE{chen20,
       author = {{Chen}, Christine and {Mazoyer}, Johan and {Poteet}, Charles A. and {Ren}, Bin and {Duch{\^e}ne}, Gaspard and {Hom}, Justin and {Arriaga}, Pauline and {Millar-Blanchaer}, Maxwell A. and {Arnold}, Jessica and {Bailey}, Vanessa P. and {Bruzzone}, Juan Sebasti{\'a}n and {Chilcote}, Jeffrey and {Choquet}, {\'E}lodie and {De Rosa}, Robert J. and {Draper}, Zachary H. and {Esposito}, Thomas M. and {Fitzgerald}, Michael P. and {Follette}, Katherine B. and {Hibon}, Pascale and {Hines}, Dean C. and {Kalas}, Paul and {Marchis}, Franck and {Matthews}, Brenda and {Milli}, Julien and {Patience}, Jennifer and {Perrin}, Marshall D. and {Pueyo}, Laurent and {Rajan}, Abhijith and {Rantakyr{\"o}}, Fredrik T. and {Rodigas}, Timothy J. and {Roudier}, Gael M. and {Schneider}, Glenn and {Soummer}, R{\'e}mi and {Stark}, Christopher and {Wang}, Jason J. and {Ward-Duong}, Kimberly and {Weinberger}, Alycia J. and {Wilner}, David J. and {Wolff}, Schuyler},
        title = "{Multiband GPI Imaging of the HR 4796A Debris Disk}",
      journal = {\apj},
         year = 2020,
        month = jul,
       volume = {898},
       number = {1},
          eid = {55},
        pages = {55},
          doi = {10.3847/1538-4357/ab9aba},
archivePrefix = {arXiv},
       eprint = {2006.16131},
 primaryClass = {astro-ph.SR},
       adsurl = {https://ui.adsabs.harvard.edu/abs/2020ApJ...898...55C}
}

@ARTICLE{singh21,
       author = {{Singh}, G. and {Bhowmik}, T. and {Boccaletti}, A. and {Th{\'e}bault}, P. and {Kral}, Q. and {Milli}, J. and {Mazoyer}, J. and {Pantin}, E. and {van Holstein}, R.~G. and {Olofsson}, J. and {Boukrouche}, R. and {Di Folco}, E. and {Janson}, M. and {Langlois}, M. and {Maire}, A. -L. and {Vigan}, A. and {Benisty}, M. and {Augereau}, J. -C. and {Perrot}, C. and {Gratton}, R. and {Henning}, T. and {M{\'e}nard}, F. and {Rickman}, E. and {Wahhaj}, Z. and {Zurlo}, A. and {Biller}, B. and {Bonnefoy}, M. and {Chauvin}, G. and {Delorme}, P. and {Desidera}, S. and {D'Orazi}, V. and {Feldt}, M. and {Hagelberg}, J. and {Keppler}, M. and {Kopytova}, T. and {Lagadec}, E. and {Lagrange}, A. -M. and {Mesa}, D. and {Meyer}, M. and {Rouan}, D. and {Sissa}, E. and {Schmidt}, T.~O.~B. and {Jaquet}, M. and {Fusco}, T. and {Pavlov}, A. and {Rabou}, P.},
        title = "{Revealing asymmetrical dust distribution in the inner regions of HD 141569}",
      journal = {\aap},
         year = 2021,
        month = sep,
       volume = {653},
          eid = {A79},
        pages = {A79},
          doi = {10.1051/0004-6361/202140319},
archivePrefix = {arXiv},
       eprint = {2107.07570},
 primaryClass = {astro-ph.EP},
       adsurl = {https://ui.adsabs.harvard.edu/abs/2021A&A...653A..79S}
}

@ARTICLE{arnold22,
       author = {{Arnold}, Jessica A. and {Weinberger}, Alycia J. and {Videen}, Gorden and {Zubko}, Evgenij S.},
        title = "{Stumbling over Planetary Building Blocks: AU Microscopii as an Example of the Challenge of Retrieving Debris-disk Dust Properties}",
      journal = {\apj},
         year = 2022,
        month = may,
       volume = {930},
       number = {2},
          eid = {123},
        pages = {123},
          doi = {10.3847/1538-4357/ac63a9},
archivePrefix = {arXiv},
       eprint = {2105.12264},
 primaryClass = {astro-ph.EP},
       adsurl = {https://ui.adsabs.harvard.edu/abs/2022ApJ...930..123A}
}

@ARTICLE{rebollido24,
       author = {{Rebollido}, Isabel and {Stark}, Christopher C. and {Kammerer}, Jens and {Perrin}, Marshall D. and {Lawson}, Kellen and {Pueyo}, Laurent and {Chen}, Christine and {Hines}, Dean and {Girard}, Julien H. and {Worthen}, Kadin and {Ingerbretsen}, Carl and {Betti}, Sarah and {Clampin}, Mark and {Golimowski}, David and {Hoch}, Kielan and {Lewis}, Nikole K. and {Lu}, Cicero X. and {van der Marel}, Roeland P. and {Rickman}, Emily and {Seager}, Sara and {Soummer}, R{\'e}mi and {Valenti}, Jeff A. and {Ward-Duong}, Kimberly and {Mountain}, C. Matt},
        title = "{JWST-TST High Contrast: Asymmetries, Dust Populations, and Hints of a Collision in the {\ensuremath{\beta}} Pictoris Disk with NIRCam and MIRI}",
      journal = {\aj},
         year = 2024,
        month = feb,
       volume = {167},
       number = {2},
          eid = {69},
        pages = {69},
          doi = {10.3847/1538-3881/ad1759},
archivePrefix = {arXiv},
       eprint = {2401.05271},
 primaryClass = {astro-ph.EP},
       adsurl = {https://ui.adsabs.harvard.edu/abs/2024AJ....167...69R}
}

@ARTICLE{thebault23,
       author = {{Thebault}, P. and {Olofsson}, J. and {Kral}, Q.},
        title = "{A reinvestigation of debris disc halos}",
      journal = {\aap},
         year = 2023,
        month = jun,
       volume = {674},
          eid = {A51},
        pages = {A51},
          doi = {10.1051/0004-6361/202345995},
archivePrefix = {arXiv},
       eprint = {2303.17434},
 primaryClass = {astro-ph.EP},
       adsurl = {https://ui.adsabs.harvard.edu/abs/2023A&A...674A..51T}
}

@ARTICLE{matthews23,
       author = {{Matthews}, E.~C. and {Bonnefoy}, M. and {Xie}, C. and {Desgrange}, C. and {Desidera}, S. and {Delorme}, P. and {Milli}, J. and {Olofsson}, J. and {Barbato}, D. and {Ceva}, W. and {Augereau}, J. -C. and {Biller}, B.~A. and {Chen}, C.~H. and {Faramaz-Gorka}, V. and {Galicher}, R. and {Hinkley}, S. and {Lagrange}, A. -M. and {M{\'e}nard}, F. and {Pinte}, C. and {Stapelfeldt}, K.~R.},
        title = "{The first scattered light images of HD 112810, a faint debris disk in the Sco-Cen association}",
      journal = {\aap},
         year = 2023,
        month = nov,
       volume = {679},
          eid = {A58},
        pages = {A58},
          doi = {10.1051/0004-6361/202347335},
archivePrefix = {arXiv},
       eprint = {2309.15158},
 primaryClass = {astro-ph.EP},
       adsurl = {https://ui.adsabs.harvard.edu/abs/2023A&A...679A..58M}
}

@ARTICLE{boccaletti24,
       author = {{Boccaletti}, Anthony and {M{\^a}lin}, Mathilde and {Baudoz}, Pierre and {Tremblin}, Pascal and {Perrot}, Cl{\'e}ment and {Rouan}, Daniel and {Lagage}, Pierre-Olivier and {Whiteford}, Niall and {Molli{\`e}re}, Paul and {Waters}, Rens and {Henning}, Thomas and {Decin}, Leen and {G{\"u}del}, Manuel and {Vandenbussche}, Bart and {Absil}, Olivier and {Argyriou}, Ioannis and {Bouwman}, Jeroen and {Cossou}, Christophe and {Coulais}, Alain and {Gastaud}, Ren{\'e} and {Glasse}, Alistair and {Glauser}, Adrian M. and {Kamp}, Inga and {Kendrew}, Sarah and {Krause}, Oliver and {Lahuis}, Fred and {Mueller}, Michael and {Olofsson}, Goran and {Patapis}, Polychronis and {Pye}, John and {Royer}, Pierre and {Serabyn}, Eugene and {Scheithauer}, Silvia and {Colina}, Luis and {van Dishoeck}, Ewine F. and {Ostlin}, G{\"o}ran and {Ray}, Tom P. and {Wright}, Gillian},
        title = "{Imaging detection of the inner dust belt and the four exoplanets in the HR 8799 system with JWST's MIRI coronagraph}",
      journal = {\aap},
         year = 2024,
        month = jun,
       volume = {686},
          eid = {A33},
        pages = {A33},
          doi = {10.1051/0004-6361/202347912},
archivePrefix = {arXiv},
       eprint = {2310.13414},
 primaryClass = {astro-ph.EP},
       adsurl = {https://ui.adsabs.harvard.edu/abs/2024A&A...686A..33B}
}

@article{ballering13,
	adsurl = {https://ui.adsabs.harvard.edu/abs/2013ApJ...775...55B},
	archiveprefix = {arXiv},
	author = {{Ballering}, Nicholas P. and {Rieke}, George H. and {Su}, Kate Y.~L. and {Montiel}, Edward},
	doi = {10.1088/0004-637X/775/1/55},
	eid = {55},
	eprint = {1308.2223},
	journal = {\apj},
	month = sep,
	number = {1},
	pages = {55},
	primaryclass = {astro-ph.EP},
	title = {{A Trend between Cold Debris Disk Temperature and Stellar Type: Implications for the Formation and Evolution of Wide-orbit Planets}},
	volume = {775},
	year = 2013}

@article{morales11,
	adsurl = {https://ui.adsabs.harvard.edu/abs/2011ApJ...730L..29M},
	author = {{Morales}, Farisa Y. and {Rieke}, G.~H. and {Werner}, M.~W. and {Bryden}, G. and {Stapelfeldt}, K.~R. and {Su}, K.~Y.~L.},
	doi = {10.1088/2041-8205/730/2/L29},
	eid = {L29},
	journal = {\apjl},
	month = apr,
	number = {2},
	pages = {L29},
	title = {{Common Warm Dust Temperatures Around Main-sequence Stars}},
	volume = {730},
	year = 2011}

@article{su13,
	adsurl = {https://ui.adsabs.harvard.edu/abs/2013ApJ...763..118S},
	archiveprefix = {arXiv},
	author = {{Su}, Kate Y.~L. and {Rieke}, George H. and {Malhotra}, Renu and {Stapelfeldt}, Karl R. and {Hughes}, A. Meredith and {Bonsor}, Amy and {Wilner}, David J. and {Balog}, Zoltan and {Watson}, Dan M. and {Werner}, Michael W. and {Misselt}, Karl A.},
	doi = {10.1088/0004-637X/763/2/118},
	eid = {118},
	eprint = {1301.1331},
	journal = {\apj},
	month = feb,
	number = {2},
	pages = {118},
	primaryclass = {astro-ph.EP},
	title = {{Asteroid Belts in Debris Disk Twins: Vega and Fomalhaut}},
	volume = {763},
	year = 2013}

@misc{NASA_2023,
	author = {NASA, Space Telescope Science Institute},
	journal = {Space Telescope Science Institute},
	month = {Jun},
	publisher = {NASA},
	title = {JWST pocket guide, Space Telescope Science Institute},
	url = {https://www.stsci.edu/files/live/sites/www/files/home/jwst/instrumentation/_documents/jwst-pocket-guide.pdf},
	year = {2026}}

@misc{zenodo,
	author = {Yea, Inbok and Dodson-Robinson, Sarah},
	doi = {10.5281/zenodo.21327122},
	month = jul,
	publisher = {Zenodo},
	title = {{Plots for The Nearest, Brightest Debris-Disk Host Stars: Updated Photosphere Models and Dust Properties for Stars with Saturated 2MASS Photometry, v11}},
	year = 2026}

@article{beichman05,
	adsurl = {https://ui.adsabs.harvard.edu/abs/2005ApJ...622.1160B},
	archiveprefix = {arXiv},
	author = {{Beichman}, C.~A. and {Bryden}, G. and {Rieke}, G.~H. and {Stansberry}, J.~A. and {Trilling}, D.~E. and {Stapelfeldt}, K.~R. and {Werner}, M.~W. and {Engelbracht}, C.~W. and {Blaylock}, M. and {Gordon}, K.~D. and {Chen}, C.~H. and {Su}, K.~Y.~L. and {Hines}, D.~C.},
	doi = {10.1086/428115},
	eprint = {astro-ph/0412265},
	journal = {\apj},
	month = apr,
	number = {2},
	pages = {1160-1170},
	primaryclass = {astro-ph},
	title = {{Planets and Infrared Excesses: Preliminary Results from a Spitzer MIPS Survey of Solar-Type Stars}},
	volume = {622},
	year = 2005}

@inproceedings{chilcote20,
	adsurl = {https://ui.adsabs.harvard.edu/abs/2020SPIE11447E..1SC},
	author = {{Chilcote}, Jeffrey and {Konopacky}, Quinn and {De Rosa}, Robert J. and {Hamper}, Randall and {Macintosh}, Bruce and {Marois}, Christian and {Perrin}, Marshall D. and {Savransky}, Dmitry and {Soummer}, R{\'e}mi and {V{\'e}ran}, Jean-Pierre and {Agapito}, Guido and {Aleman}, Arlene and {Ammons}, S. Mark and {Bonaglia}, Marco and {Boucher}, Marc-Andre and {Curliss}, Maeve and {Dunn}, Jennifer and {Esposito}, Simone and {Filion}, Guillaume and {Fitzsimmons}, Joeleff and {Kain}, Isabel and {Kerley}, Dan and {Landry}, Jean-Thomas and {Lardiere}, Olivier and {Lemoine-Busserolle}, Marie and {Li}, Duan and {Limbach}, Mary Anne and {Madurowicz}, Alex and {Maire}, Jerome and {N'Diaye}, Mamadou and {Nielsen}, Eric L. and {Poyneer}, Lisa and {Pueyo}, Laurent and {Summey}, Kaitlyn and {Thomas}, Coleman},
	booktitle = {Society of Photo-Optical Instrumentation Engineers (SPIE) Conference Series},
	doi = {10.1117/12.2562578},
	eid = {114471S},
	month = dec,
	pages = {114471S},
	series = {Society of Photo-Optical Instrumentation Engineers (SPIE) Conference Series},
	title = {{GPI 2.0: upgrading the Gemini Planet Imager}},
	volume = {11447},
	year = 2020}

@article{aumann84,
	author = {{Aumann}, H.~H. and {Gillett}, F.~C. and {Beichman}, C.~A. and {de Jong}, T. and {Houck}, J.~R. and {Low}, F.~J. and {Neugebauer}, G. and {Walker}, R.~G. and {Wesselius}, P.~R.},
	doi = {10.1086/184214},
	journal = {\apjl},
	month = {Mar},
	pages = {L23-L27},
	title = {{Discovery of a shell around alpha Lyrae.}},
	volume = {278},
	year = {1984}}

@article{patel14,
	author = {{Patel}, Rahul I. and {Metchev}, Stanimir A. and {Heinze}, Aren},
	doi = {10.1088/0067-0049/212/1/10},
	eid = {10},
	journal = {\apjs},
	month = {May},
	number = {1},
	pages = {10},
	title = {{A Sensitive Identification of Warm Debris Disks in the Solar Neighborhood through Precise Calibration of Saturated WISE Photometry}},
	volume = {212},
	year = {2014}}

@article{galicher16,
	adsurl = {https://ui.adsabs.harvard.edu/abs/2016A&A...594A..63G},
	archiveprefix = {arXiv},
	author = {{Galicher}, R. and {Marois}, C. and {Macintosh}, B. and {Zuckerman}, B. and {Barman}, T. and {Konopacky}, Q. and {Song}, I. and {Patience}, J. and {Lafreni{\`e}re}, D. and {Doyon}, R. and {Nielsen}, E.~L.},
	doi = {10.1051/0004-6361/201527828},
	eid = {A63},
	eprint = {1607.08239},
	journal = {\aap},
	month = oct,
	pages = {A63},
	primaryclass = {astro-ph.EP},
	title = {{The International Deep Planet Survey. II. The frequency of directly imaged giant exoplanets with stellar mass}},
	volume = {594},
	year = 2016}

@article{su06,
	author = {{Su}, K.~Y.~L. and {Rieke}, G.~H. and {Stansberry}, J.~A. and {Bryden}, G. and {Stapelfeldt}, K.~R. and {Trilling}, D.~E. and {Muzerolle}, J. and {Beichman}, C.~A. and {Moro-Martin}, A. and {Hines}, D.~C. and {Werner}, M.~W.},
	doi = {10.1086/508649},
	journal = {\apj},
	month = {Dec},
	number = {1},
	pages = {675-689},
	title = {{Debris Disk Evolution around A Stars}},
	volume = {653},
	year = {2006}}

@article{schneider18,
	adsurl = {https://ui.adsabs.harvard.edu/abs/2018AJ....155...77S},
	archiveprefix = {arXiv},
	author = {{Schneider}, Glenn and {Debes}, John H. and {Grady}, Carol A. and {G{\'a}sp{\'a}r}, Andras and {Henning}, Thomas and {Hines}, Dean C. and {Kuchner}, Marc J. and {Perrin}, Marshall and {Wisniewski}, John P.},
	doi = {10.3847/1538-3881/aaa3f3},
	eid = {77},
	eprint = {1712.08599},
	journal = {\aj},
	month = feb,
	number = {2},
	pages = {77},
	primaryclass = {astro-ph.SR},
	title = {{The HR 4796A Debris System: Discovery of Extensive Exo-ring Dust Material}},
	volume = {155},
	year = 2018}

@article{hengst17,
	author = {{Hengst}, S. and {Marshall}, J.~P. and {Horner}, J. and {Marsden}, S.~C.},
	doi = {10.1093/mnras/stx753},
	journal = {\mnras},
	month = {Jul},
	number = {4},
	pages = {4725-4734},
	title = {{A Herschel resolved debris disc around HD 105211}},
	volume = {468},
	year = {2017}}

@article{moerchen10,
	adsurl = {https://ui.adsabs.harvard.edu/abs/2010ApJ...723.1418M},
	archiveprefix = {arXiv},
	author = {{Moerchen}, Margaret M. and {Telesco}, Charles M. and {Packham}, Christopher},
	doi = {10.1088/0004-637X/723/2/1418},
	eprint = {1011.1410},
	journal = {\apj},
	month = nov,
	number = {2},
	pages = {1418-1435},
	primaryclass = {astro-ph.SR},
	title = {{High Spatial Resolution Imaging of Thermal Emission from Debris Disks}},
	volume = {723},
	year = 2010}

@article{redfield04,
	adsurl = {https://ui.adsabs.harvard.edu/abs/2004ApJ...613.1004R},
	archiveprefix = {arXiv},
	author = {{Redfield}, Seth and {Linsky}, Jeffrey L.},
	doi = {10.1086/423311},
	eprint = {astro-ph/0406464},
	journal = {\apj},
	month = oct,
	number = {2},
	pages = {1004-1022},
	primaryclass = {astro-ph},
	title = {{The Structure of the Local Interstellar Medium. III. Temperature and Turbulence}},
	volume = {613},
	year = 2004}

@article{ishihara17,
	author = {{Ishihara}, Daisuke and {Takeuchi}, Nami and {Kobayashi}, Hiroshi and {Nagayama}, Takahiro and {Kaneda}, Hidehiro and {Inutsuka}, Shu-ichiro and {Fujiwara}, Hideaki and {Onaka}, Takashi},
	doi = {10.1051/0004-6361/201526215},
	eid = {A72},
	journal = {\aap},
	month = {May},
	pages = {A72},
	title = {{Faint warm debris disks around nearby bright stars explored by AKARI and IRSF}},
	volume = {601},
	year = {2017}}

@article{mann15,
	author = {{Mann}, Andrew W. and {von Braun}, Kaspar},
	doi = {10.1086/680012},
	journal = {\pasp},
	month = {Feb},
	number = {948},
	pages = {102},
	title = {{Revised Filter Profiles and Zero Points for Broadband Photometry}},
	volume = {127},
	year = {2015}}

@article{jovanovic15,
	adsurl = {https://ui.adsabs.harvard.edu/abs/2015PASP..127..890J},
	archiveprefix = {arXiv},
	author = {{Jovanovic}, N. and {Martinache}, F. and {Guyon}, O. and {Clergeon}, C. and {Singh}, G. and {Kudo}, T. and {Garrel}, V. and {Newman}, K. and {Doughty}, D. and {Lozi}, J. and {Males}, J. and {Minowa}, Y. and {Hayano}, Y. and {Takato}, N. and {Morino}, J. and {Kuhn}, J. and {Serabyn}, E. and {Norris}, B. and {Tuthill}, P. and {Schworer}, G. and {Stewart}, P. and {Close}, L. and {Huby}, E. and {Perrin}, G. and {Lacour}, S. and {Gauchet}, L. and {Vievard}, S. and {Murakami}, N. and {Oshiyama}, F. and {Baba}, N. and {Matsuo}, T. and {Nishikawa}, J. and {Tamura}, M. and {Lai}, O. and {Marchis}, F. and {Duchene}, G. and {Kotani}, T. and {Woillez}, J.},
	doi = {10.1086/682989},
	eprint = {1507.00017},
	journal = {\pasp},
	month = sep,
	number = {955},
	pages = {890},
	primaryclass = {astro-ph.IM},
	title = {{The Subaru Coronagraphic Extreme Adaptive Optics System: Enabling High-Contrast Imaging on Solar-System Scales}},
	volume = {127},
	year = 2015}

@article{bayo08,
	author = {{Bayo}, A. and {Rodrigo}, C. and {Barrado y Navascues}, D. and {Solano}, E. and {Gutierrez}, R. and {Morales-Calderon}, M. and {Allard}, F.},
	eid = {J/A+A/492/277},
	journal = {VizieR Online Data Catalog},
	month = {Nov},
	pages = {J/A+A/492/277},
	title = {{VizieR Online Data Catalog: VOSA: virtual observatory SED analyzer (Bayo+, 2008)}},
	year = {2008}}

@article{hog00,
	author = {{H{\o}g}, E. and {Fabricius}, C. and {Makarov}, V.~V. and {Urban}, S. and {Corbin}, T. and {Wycoff}, G. and {Bastian}, U. and {Schwekendiek}, P. and {Wicenec}, A.},
	journal = {\aap},
	month = {Mar},
	pages = {L27-L30},
	title = {{The Tycho-2 catalogue of the 2.5 million brightest stars}},
	volume = {355},
	year = {2000}}

@article{mermiliod97,
	author = {{Mermilliod}, J. -C. and {Mermilliod}, M. and {Hauck}, B.},
	doi = {10.1051/aas:1997197},
	journal = {\aaps},
	month = {Aug},
	pages = {349-352},
	title = {{The General Catalogue of Photometric Data (GCPD). II.}},
	volume = {124},
	year = {1997}}

@article{mallama14,
	author = {{Mallama}, A.},
	journal = {Journal of the American Association of Variable Star Observers (JAAVSO)},
	month = {Dec},
	number = {2},
	pages = {443},
	title = {{Sloan Magnitudes for the Brightest Stars}},
	volume = {42},
	year = {2014}}

@article{moor15,
	adsurl = {https://ui.adsabs.harvard.edu/abs/2015MNRAS.447..577M},
	archiveprefix = {arXiv},
	author = {{Mo{\'o}r}, A. and {K{\'o}sp{\'a}l}, {\'A}. and {{\'A}brah{\'a}m}, P. and {Apai}, D. and {Balog}, Z. and {Grady}, C. and {Henning}, Th. and {Juh{\'a}sz}, A. and {Kiss}, Cs. and {Krivov}, A.~V. and {Pawellek}, N. and {Szab{\'o}}, Gy. M.},
	doi = {10.1093/mnras/stu2442},
	eprint = {1411.5829},
	journal = {\mnras},
	month = feb,
	number = {1},
	pages = {577-597},
	primaryclass = {astro-ph.SR},
	title = {{Stirring in massive, young debris discs from spatially resolved Herschel images}},
	volume = {447},
	year = 2015}

@article{cousins76,
	author = {{Cousins}, A.~W.~J.},
	journal = {\memras},
	month = {Jan},
	pages = {25},
	title = {{VRI standards in the E regions}},
	volume = {81},
	year = {1976}}

@article{ducati02,
	author = {{Ducati}, J.~R.},
	journal = {VizieR Online Data Catalog},
	month = {Jan},
	title = {{VizieR Online Data Catalog: Catalogue of Stellar Photometry in Johnson's 11-color system.}},
	year = {2002}}

@article{morales16,
	adsurl = {https://ui.adsabs.harvard.edu/abs/2016ApJ...831...97M},
	author = {{Morales}, F.~Y. and {Bryden}, G. and {Werner}, M.~W. and {Stapelfeldt}, K.~R.},
	doi = {10.3847/0004-637X/831/1/97},
	eid = {97},
	journal = {\apj},
	month = nov,
	number = {1},
	pages = {97},
	title = {{Herschel-resolved Outer Belts of Two-belt Debris Disks{\textemdash}Evidence of Icy Grains}},
	volume = {831},
	year = 2016}

@article{wright10,
	author = {{Wright}, Edward L. and {Eisenhardt}, Peter R.~M. and {Mainzer}, Amy K. and {Ressler}, Michael E. and {Cutri}, Roc M. and {Jarrett}, Thomas and {Kirkpatrick}, J. Davy and {Padgett}, Deborah and {McMillan}, Robert S. and {Skrutskie}, Michael and {Stanford}, S.~A. and {Cohen}, Martin and {Walker}, Russell G. and {Mather}, John C. and {Leisawitz}, David and {Gautier}, Thomas N., III and {McLean}, Ian and {Benford}, Dominic and {Lonsdale}, Carol J. and {Blain}, Andrew and {Mendez}, Bryan and {Irace}, William R. and {Duval}, Valerie and {Liu}, Fengchuan and {Royer}, Don and {Heinrichsen}, Ingolf and {Howard}, Joan and {Shannon}, Mark and {Kendall}, Martha and {Walsh}, Amy L. and {Larsen}, Mark and {Cardon}, Joel G. and {Schick}, Scott and {Schwalm}, Mark and {Abid}, Mohamed and {Fabinsky}, Beth and {Naes}, Larry and {Tsai}, Chao-Wei},
	doi = {10.1088/0004-6256/140/6/1868},
	journal = {\aj},
	month = {Dec},
	number = {6},
	pages = {1868-1881},
	title = {{The Wide-field Infrared Survey Explorer (WISE): Mission Description and Initial On-orbit Performance}},
	volume = {140},
	year = {2010}}

@INPROCEEDINGS{castelli04,
	author = {{Castelli}, F. and {Kurucz}, R.~L.},
	booktitle = {Modelling of Stellar Atmospheres},
	editor = {{Piskunov}, N. and {Weiss}, W.~W. and {Gray}, D.~F.},
	month = {Jan},
	pages = {A20},
	series = {IAU Symposium},
	title = {{New Grids of ATLAS9 Model Atmospheres}},
	volume = {210},
	year = {2003}}

@inproceedings{su14,
	adsurl = {https://ui.adsabs.harvard.edu/abs/2014IAUS..299..318S},
	archiveprefix = {arXiv},
	author = {{Su}, Kate Y.~L. and {Rieke}, G.~H.},
	booktitle = {Exploring the Formation and Evolution of Planetary Systems},
	doi = {10.1017/S1743921313008764},
	editor = {{Booth}, Mark and {Matthews}, Brenda C. and {Graham}, James R.},
	eprint = {1307.1735},
	month = jan,
	pages = {318-321},
	primaryclass = {astro-ph.EP},
	title = {{Signposts of Multiple Planets in Debris Disks}},
	volume = {299},
	year = 2014}

@article{ballering17,
	adsurl = {https://ui.adsabs.harvard.edu/abs/2017ApJ...845..120B},
	archiveprefix = {arXiv},
	author = {{Ballering}, Nicholas P. and {Rieke}, George H. and {Su}, Kate Y.~L. and {G{\'a}sp{\'a}r}, Andr{\'a}s},
	doi = {10.3847/1538-4357/aa8037},
	eid = {120},
	eprint = {1708.07151},
	journal = {\apj},
	month = aug,
	number = {2},
	pages = {120},
	primaryclass = {astro-ph.EP},
	title = {{What Sets the Radial Locations of Warm Debris Disks?}},
	volume = {845},
	year = 2017}

@article{chen14,
	author = {{Chen}, Christine H. and {Mittal}, Tushar and {Kuchner}, Marc and {Forrest}, William J. and {Lisse}, Carey M. and {Manoj}, P. and {Sargent}, Benjamin A. and {Watson}, Dan M.},
	doi = {10.1088/0067-0049/211/2/25},
	eid = {25},
	journal = {\apjs},
	month = {Apr},
	number = {2},
	pages = {25},
	title = {{The Spitzer Infrared Spectrograph Debris Disk Catalog. I. Continuum Analysis of Unresolved Targets}},
	volume = {211},
	year = {2014}}

@article{cotten16,
	author = {{Cotten}, Tara H. and {Song}, Inseok},
	doi = {10.3847/0067-0049/225/1/15},
	eid = {15},
	journal = {\apjs},
	month = {Jul},
	number = {1},
	pages = {15},
	title = {{A Comprehensive Census of Nearby Infrared Excess Stars}},
	volume = {225},
	year = {2016}}

@article{foreman13,
	author = {{Foreman-Mackey}, Daniel and {Hogg}, David W. and {Lang}, Dustin and {Goodman}, Jonathan},
	doi = {10.1086/670067},
	journal = {\pasp},
	month = {Mar},
	number = {925},
	pages = {306},
	title = {{emcee: The MCMC Hammer}},
	volume = {125},
	year = {2013}}

@article{brown18,
	author = {{Gaia Collaboration} and {Brown}, A.~G.~A. and {Vallenari}, A. and {Prusti}, T. and {de Bruijne}, J.~H.~J. and {Babusiaux}, C. and {Bailer-Jones}, C.~A.~L. and {Biermann}, M. and {Evans}, D.~W. and {Eyer}, L. and {Jansen}, F. and {Jordi}, C. and {Klioner}, S.~A. and {Lammers}, U. and {Lindegren}, L. and {Luri}, X. and {Mignard}, F. and {Panem}, C. and {Pourbaix}, D. and {Randich}, S. and {Sartoretti}, P. and {Siddiqui}, H.~I. and {Soubiran}, C. and {van Leeuwen}, F. and {Walton}, N.~A. and {Arenou}, F. and {Bastian}, U. and {Cropper}, M. and {Drimmel}, R. and {Katz}, D. and {Lattanzi}, M.~G. and {Bakker}, J. and {Cacciari}, C. and {Casta{\~n}eda}, J. and {Chaoul}, L. and {Cheek}, N. and {De Angeli}, F. and {Fabricius}, C. and {Guerra}, R. and {Holl}, B. and {Masana}, E. and {Messineo}, R. and {Mowlavi}, N. and {Nienartowicz}, K. and {Panuzzo}, P. and {Portell}, J. and {Riello}, M. and {Seabroke}, G.~M. and {Tanga}, P. and {Th{\'e}venin}, F. and {Gracia-Abril}, G. and {Comoretto}, G. and {Garcia-Reinaldos}, M. and {Teyssier}, D. and {Altmann}, M. and {Andrae}, R. and {Audard}, M. and {Bellas-Velidis}, I. and {Benson}, K. and {Berthier}, J. and {Blomme}, R. and {Burgess}, P. and {Busso}, G. and {Carry}, B. and {Cellino}, A. and {Clementini}, G. and {Clotet}, M. and {Creevey}, O. and {Davidson}, M. and {De Ridder}, J. and {Delchambre}, L. and {Dell'Oro}, A. and {Ducourant}, C. and {Fern{\'a}ndez-Hern{\'a}ndez}, J. and {Fouesneau}, M. and {Fr{\'e}mat}, Y. and {Galluccio}, L. and {Garc{\'\i}a-Torres}, M. and {Gonz{\'a}lez-N{\'u}{\~n}ez}, J. and {Gonz{\'a}lez-Vidal}, J.~J. and {Gosset}, E. and {Guy}, L.~P. and {Halbwachs}, J.-L. and {Hambly}, N.~C. and {Harrison}, D.~L. and {Hern{\'a}ndez}, J. and {Hestroffer}, D. and {Hodgkin}, S.~T. and {Hutton}, A. and {Jasniewicz}, G. and {Jean-Antoine-Piccolo}, A. and {Jordan}, S. and {Korn}, A.~J. and {Krone-Martins}, A. and {Lanzafame}, A.~C. and {Lebzelter}, T. and {L{\"o}ffler}, W. and {Manteiga}, M. and {Marrese}, P.~M. and {Mart{\'\i}n-Fleitas}, J.~M. and {Moitinho}, A. and {Mora}, A. and {Muinonen}, K. and {Osinde}, J. and {Pancino}, E. and {Pauwels}, T. and {Petit}, J.-M. and {Recio-Blanco}, A. and {Richards}, P.~J. and {Rimoldini}, L. and {Robin}, A.~C. and {Sarro}, L.~M. and {Siopis}, C. and {Smith}, M. and {Sozzetti}, A. and {S{\"u}veges}, M. and {Torra}, J. and {van Reeven}, W. and {Abbas}, U. and {Abreu Aramburu}, A. and {Accart}, S. and {Aerts}, C. and {Altavilla}, G. and {{\'A}lvarez}, M.~A. and {Alvarez}, R. and {Alves}, J. and {Anderson}, R.~I. and {Andrei}, A.~H. and {Anglada Varela}, E. and {Antiche}, E. and {Antoja}, T. and {Arcay}, B. and {Astraatmadja}, T.~L. and {Bach}, N. and {Baker}, S.~G. and {Balaguer-N{\'u}{\~n}ez}, L. and {Balm}, P. and {Barache}, C. and {Barata}, C. and {Barbato}, D. and {Barblan}, F. and {Barklem}, P.~S. and {Barrado}, D. and {Barros}, M. and {Barstow}, M.~A. and {Bartholom{\'e} Mu{\~n}oz}, S. and {Bassilana}, J.-L. and {Becciani}, U. and {Bellazzini}, M. and {Berihuete}, A. and {Bertone}, S. and {Bianchi}, L. and {Bienaym{\'e}}, O. and {Blanco-Cuaresma}, S. and {Boch}, T. and {Boeche}, C. and {Bombrun}, A. and {Borrachero}, R. and {Bossini}, D. and {Bouquillon}, S. and {Bourda}, G. and {Bragaglia}, A. and {Bramante}, L. and {Breddels}, M.~A. and {Bressan}, A. and {Brouillet}, N. and {Br{\"u}semeister}, T. and {Brugaletta}, E. and {Bucciarelli}, B. and {Burlacu}, A. and {Busonero}, D. and {Butkevich}, A.~G. and {Buzzi}, R. and {Caffau}, E. and {Cancelliere}, R. and {Cannizzaro}, G. and {Cantat-Gaudin}, T. and {Carballo}, R. and {Carlucci}, T. and {Carrasco}, J.~M. and {Casamiquela}, L. and {Castellani}, M. and {Castro-Ginard}, A. and {Charlot}, P. and {Chemin}, L. and {Chiavassa}, A. and {Cocozza}, G. and {Costigan}, G. and {Cowell}, S. and {Crifo}, F. and {Crosta}, M. and {Crowley}, C. and {Cuypers}, J. and {Dafonte}, C. and {Damerdji}, Y. and {Dapergolas}, A. and {David}, P. and {David}, M. and {de Laverny}, P. and {De Luise}, F.},
	doi = {10.1051/0004-6361/201833051},
	eid = {A1},
	journal = {\aap},
	month = {Aug},
	pages = {A1},
	title = {Gaia Data Release 2. Summary of the contents and survey properties},
	volume = {616},
	year = {2018},
}

@misc{rodrigo1,
	author = {{Rodrigo}, Carlos and {Solano}, Enrique and {Bayo}, Amelia},
	title = "{SVO Filter Profile Service Version 1.0}",
	howpublished = {IVOA Working Draft 15 October 2012},
	year = 2012,
	month = oct,
	pages = {1015},
	doi = {10.5479/ADS/bib/2012ivoa.rept.1015R},
}

@misc{rodrigo2,
	author = {{Rodrigo}, C. and {Solano}, E.},
	title = "{The SVO Filter Profile Service}",
	booktitle = {XIV.0 Scientific Meeting (virtual) of the Spanish Astronomical Society},
	year = 2020,
	month = jul,
	eid = {182},
	pages = {182},
}

@ARTICLE{rodrigo24,
	author = {{Rodrigo}, Carlos and {Cruz}, Patricia and {Aguilar}, John F. and {Aller}, Alba and {Solano}, Enrique and {G{\'a}lvez-Ortiz}, Maria Cruz and {Jim{\'e}nez-Esteban}, Francisco and {Mas-Buitrago}, Pedro and {Bayo}, Amelia and {Cort{\'e}s-Contreras}, Miriam and {Murillo-Ojeda}, Raquel and {Bonoli}, Silvia and {Cenarro}, Javier and {Dupke}, Renato and {L{\'o}pez-Sanjuan}, Carlos and {Mar{\'\i}n-Franch}, Antonio and {de Oliveira}, Claudia Mendes and {Moles}, Mariano and {Taylor}, Keith and {Varela}, Jes{\'u}s and {Rami{\'o}}, H{\'e}ctor V{\'a}zquez},
	title = "{Photometric segregation of dwarf and giant FGK stars using the SVO Filter Profile Service and photometric tools}",
	journal = {\aap},
	year = 2024,
	month = sep,
	volume = {689},
	eid = {A93},
	pages = {A93},
	doi = {10.1051/0004-6361/202449998},
}

@article{dasilva15,
	author = {{da Silva}, Ronaldo and {Milone}, Andr{\'e} de C. and {Rocha-Pinto}, Helio J.},
	doi = {10.1051/0004-6361/201525770},
	eid = {A24},
	journal = {\aap},
	month = {Aug},
	pages = {A24},
	title = {{Homogeneous abundance analysis of FGK dwarf, subgiant, and giant stars with and without giant planets}},
	volume = {580},
	year = {2015}}

@article{dasilva12,
	author = {{da Silva}, R. and {Porto de Mello}, G.~F. and {Milone}, A.~C. and {da Silva}, L. and {Ribeiro}, L.~S. and {Rocha-Pinto}, H.~J.},
	doi = {10.1051/0004-6361/201118751},
	eid = {A84},
	journal = {\aap},
	month = jun,
	pages = {A84},
	title = {{Accurate and homogeneous abundance patterns in solar-type stars of the solar neighbourhood: a chemo-chronological analysis}},
	volume = {542},
	year = 2012}

@article{santos04,
	author = {{Santos}, N.~C. and {Israelian}, G. and {Mayor}, M.},
	doi = {10.1051/0004-6361:20034469},
	journal = {\aap},
	month = mar,
	pages = {1153-1166},
	title = {{Spectroscopic [Fe/H] for 98 extra-solar planet-host stars. Exploring the probability of planet formation}},
	volume = {415},
	year = 2004}

@article{carbillet11,
	adsurl = {https://ui.adsabs.harvard.edu/abs/2011ExA....30...39C},
	author = {{Carbillet}, Marcel and {Bendjoya}, Philippe and {Abe}, Lyu and {Guerri}, G{\'e}raldine and {Boccaletti}, Anthony and {Daban}, Jean-Baptiste and {Dohlen}, Kjetil and {Ferrari}, Andr{\'e} and {Robbe-Dubois}, Sylvie and {Douet}, Richard and {Vakili}, Farrokh},
	doi = {10.1007/s10686-011-9219-4},
	journal = {Experimental Astronomy},
	month = may,
	number = {1},
	pages = {39-58},
	title = {{Apodized Lyot coronagraph for SPHERE/VLT. I. Detailed numerical study}},
	volume = {30},
	year = 2011}

@article{guerri11,
	adsurl = {https://ui.adsabs.harvard.edu/abs/2011ExA....30...59G},
	author = {{Guerri}, G{\'e}raldine and {Daban}, Jean-Baptiste and {Robbe-Dubois}, Sylvie and {Douet}, Richard and {Abe}, Lyu and {Baudrand}, Jacques and {Carbillet}, Marcel and {Boccaletti}, Anthony and {Bendjoya}, Philippe and {Gouvret}, Carole and {Vakili}, Farrokh},
	doi = {10.1007/s10686-011-9220-y},
	journal = {Experimental Astronomy},
	month = may,
	number = {1},
	pages = {59-81},
	title = {{Apodized Lyot coronagraph for SPHERE/VLT: II. Laboratory tests and performance}},
	volume = {30},
	year = 2011}

@article{strubbe06,
	adsurl = {https://ui.adsabs.harvard.edu/abs/2006ApJ...648..652S},
	archiveprefix = {arXiv},
	author = {{Strubbe}, Linda E. and {Chiang}, Eugene I.},
	doi = {10.1086/505736},
	eprint = {astro-ph/0510527},
	journal = {\apj},
	month = sep,
	number = {1},
	pages = {652-665},
	primaryclass = {astro-ph},
	title = {{Dust Dynamics, Surface Brightness Profiles, and Thermal Spectra of Debris Disks: The Case of AU Microscopii}},
	volume = {648},
	year = 2006}

@article{maldonado12,
	author = {{Maldonado}, J. and {Eiroa}, C. and {Villaver}, E. and {Montesinos}, B. and {Mora}, A.},
	doi = {10.1051/0004-6361/201218800},
	eid = {A40},
	journal = {\aap},
	month = may,
	pages = {A40},
	title = {{Metallicity of solar-type stars with debris discs and planets{\ensuremath{\star}}}},
	volume = {541},
	year = 2012}

@article{gebran10,
	author = {{Gebran}, M. and {Vick}, M. and {Monier}, R. and {Fossati}, L.},
	doi = {10.1051/0004-6361/200913273},
	eid = {A71},
	journal = {\aap},
	month = nov,
	pages = {A71},
	title = {{Chemical composition of A and F dwarfs members of the Hyades open cluster}},
	volume = {523},
	year = 2010}

@article{beuzit19,
	adsurl = {https://ui.adsabs.harvard.edu/abs/2019A&A...631A.155B},
	archiveprefix = {arXiv},
	author = {{Beuzit}, J. -L. and {Vigan}, A. and {Mouillet}, D. and {Dohlen}, K. and {Gratton}, R. and {Boccaletti}, A. and {Sauvage}, J. -F. and {Schmid}, H.~M. and {Langlois}, M. and {Petit}, C. and {Baruffolo}, A. and {Feldt}, M. and {Milli}, J. and {Wahhaj}, Z. and {Abe}, L. and {Anselmi}, U. and {Antichi}, J. and {Barette}, R. and {Baudrand}, J. and {Baudoz}, P. and {Bazzon}, A. and {Bernardi}, P. and {Blanchard}, P. and {Brast}, R. and {Bruno}, P. and {Buey}, T. and {Carbillet}, M. and {Carle}, M. and {Cascone}, E. and {Chapron}, F. and {Charton}, J. and {Chauvin}, G. and {Claudi}, R. and {Costille}, A. and {De Caprio}, V. and {de Boer}, J. and {Delboulb{\'e}}, A. and {Desidera}, S. and {Dominik}, C. and {Downing}, M. and {Dupuis}, O. and {Fabron}, C. and {Fantinel}, D. and {Farisato}, G. and {Feautrier}, P. and {Fedrigo}, E. and {Fusco}, T. and {Gigan}, P. and {Ginski}, C. and {Girard}, J. and {Giro}, E. and {Gisler}, D. and {Gluck}, L. and {Gry}, C. and {Henning}, T. and {Hubin}, N. and {Hugot}, E. and {Incorvaia}, S. and {Jaquet}, M. and {Kasper}, M. and {Lagadec}, E. and {Lagrange}, A. -M. and {Le Coroller}, H. and {Le Mignant}, D. and {Le Ruyet}, B. and {Lessio}, G. and {Lizon}, J. -L. and {Llored}, M. and {Lundin}, L. and {Madec}, F. and {Magnard}, Y. and {Marteaud}, M. and {Martinez}, P. and {Maurel}, D. and {M{\'e}nard}, F. and {Mesa}, D. and {M{\"o}ller-Nilsson}, O. and {Moulin}, T. and {Moutou}, C. and {Orign{\'e}}, A. and {Parisot}, J. and {Pavlov}, A. and {Perret}, D. and {Pragt}, J. and {Puget}, P. and {Rabou}, P. and {Ramos}, J. and {Reess}, J. -M. and {Rigal}, F. and {Rochat}, S. and {Roelfsema}, R. and {Rousset}, G. and {Roux}, A. and {Saisse}, M. and {Salasnich}, B. and {Santambrogio}, E. and {Scuderi}, S. and {Segransan}, D. and {Sevin}, A. and {Siebenmorgen}, R. and {Soenke}, C. and {Stadler}, E. and {Suarez}, M. and {Tiph{\`e}ne}, D. and {Turatto}, M. and {Udry}, S. and {Vakili}, F. and {Waters}, L.~B.~F.~M. and {Weber}, L. and {Wildi}, F. and {Zins}, G. and {Zurlo}, A.},
	doi = {10.1051/0004-6361/201935251},
	eid = {A155},
	eprint = {1902.04080},
	journal = {\aap},
	month = nov,
	pages = {A155},
	primaryclass = {astro-ph.IM},
	title = {{SPHERE: the exoplanet imager for the Very Large Telescope}},
	volume = {631},
	year = 2019}

@article{prugniel11,
	author = {{Prugniel}, Ph. and {Vauglin}, I. and {Koleva}, M.},
	doi = {10.1051/0004-6361/201116769},
	eid = {A165},
	journal = {\aap},
	month = jul,
	pages = {A165},
	title = {{The atmospheric parameters and spectral interpolator for the MILES stars}},
	volume = {531},
	year = 2011}

@article{debes19A,
	adsurl = {https://ui.adsabs.harvard.edu/abs/2019JATIS...5c5003D},
	archiveprefix = {arXiv},
	author = {{Debes}, John H. and {Ren}, Bin and {Schneider}, Glenn},
	doi = {10.1117/1.JATIS.5.3.035003},
	eid = {035003},
	eprint = {1905.06838},
	journal = {Journal of Astronomical Telescopes, Instruments, and Systems},
	month = jul,
	pages = {035003},
	primaryclass = {astro-ph.IM},
	title = {{Pushing the limits of the coronagraphic occulters on Hubble Space Telescope/Space Telescope Imaging Spectrograph}},
	volume = {5},
	year = 2019}

@article{brewer16,
	author = {{Brewer}, John M. and {Fischer}, Debra A. and {Valenti}, Jeff A. and {Piskunov}, Nikolai},
	doi = {10.3847/0067-0049/225/2/32},
	eid = {32},
	journal = {\apjs},
	month = aug,
	number = {2},
	pages = {32},
	title = {{Spectral Properties of Cool Stars: Extended Abundance Analysis of 1,617 Planet-search Stars}},
	volume = {225},
	year = 2016}

@article{bensby14,
	author = {{Bensby}, T. and {Feltzing}, S. and {Oey}, M.~S.},
	doi = {10.1051/0004-6361/201322631},
	eid = {A71},
	journal = {\aap},
	month = feb,
	pages = {A71},
	title = {{Exploring the Milky Way stellar disk. A detailed elemental abundance study of 714 F and G dwarf stars in the solar neighbourhood}},
	volume = {562},
	year = 2014}

@article{takeda05,
	author = {{Takeda}, Yoichi and {Ohkubo}, Michiko and {Sato}, Bun'ei and {Kambe}, Eiji and {Sadakane}, Kozo},
	doi = {10.1093/pasj/57.1.27},
	journal = {\pasj},
	month = feb,
	pages = {27-43},
	title = {{Spectroscopic Study on the Atmospheric Parameters of Nearby F--K Dwarfs and Subgiants}},
	volume = {57},
	year = 2005}

@article{delgado17,
	author = {{Delgado Mena}, E. and {Tsantaki}, M. and {Adibekyan}, V. Zh. and {Sousa}, S.~G. and {Santos}, N.~C. and {Gonz{\'a}lez Hern{\'a}ndez}, J.~I. and {Israelian}, G.},
	doi = {10.1051/0004-6361/201730535},
	eid = {A94},
	journal = {\aap},
	month = oct,
	pages = {A94},
	title = {{Chemical abundances of 1111 FGK stars from the HARPS GTO planet search program. II. Cu, Zn, Sr, Y, Zr, Ba, Ce, Nd, and Eu}},
	volume = {606},
	year = 2017}

@article{ramirez14,
	author = {{Ram{\'\i}rez}, I. and {Mel{\'e}ndez}, J. and {Bean}, J. and {Asplund}, M. and {Bedell}, M. and {Monroe}, T. and {Casagrande}, L. and {Schirbel}, L. and {Dreizler}, S. and {Teske}, J. and {Tucci Maia}, M. and {Alves-Brito}, A. and {Baumann}, P.},
	doi = {10.1051/0004-6361/201424244},
	eid = {A48},
	journal = {\aap},
	month = dec,
	pages = {A48},
	title = {{The Solar Twin Planet Search. I. Fundamental parameters of the stellar sample}},
	volume = {572},
	year = 2014}

@article{sitnova15,
	author = {{Sitnova}, T. and {Zhao}, G. and {Mashonkina}, L. and {Chen}, Y. and {Liu}, F. and {Pakhomov}, Yu. and {Tan}, K. and {Bolte}, M. and {Alexeeva}, S. and {Grupp}, F. and {Shi}, J. -R. and {Zhang}, H. -W.},
	doi = {10.1088/0004-637X/808/2/148},
	eid = {148},
	journal = {\apj},
	month = aug,
	number = {2},
	pages = {148},
	title = {{Systematic Non-LTE Study of the -2.6 \&lt; [Fe/H] \&lt; 0.2 F and G dwarfs in the Solar Neighborhood. I. Stellar Atmosphere Parameters}},
	volume = {808},
	year = 2015}

@article{ghezzi10,
	author = {{Ghezzi}, L. and {Cunha}, K. and {Smith}, V.~V. and {de Ara{\'u}jo}, F.~X. and {Schuler}, S.~C. and {de la Reza}, R.},
	doi = {10.1088/0004-637X/720/2/1290},
	journal = {\apj},
	month = sep,
	number = {2},
	pages = {1290-1302},
	title = {{Stellar Parameters and Metallicities of Stars Hosting Jovian and Neptunian Mass Planets: A Possible Dependence of Planetary Mass on Metallicity}},
	volume = {720},
	year = 2010}

@article{cutri03,
	author = {{Cutri}, R.~M. and {Skrutskie}, M.~F. and {van Dyk}, S. and {Beichman}, C.~A. and {Carpenter}, J.~M. and {Chester}, T. and {Cambresy}, L. and {Evans}, T. and {Fowler}, J. and {Gizis}, J. and {Howard}, E. and {Huchra}, J. and {Jarrett}, T. and {Kopan}, E.~L. and {Kirkpatrick}, J.~D. and {Light}, R.~M. and {Marsh}, K.~A. and {McCallon}, H. and {Schneider}, S. and {Stiening}, R. and {Sykes}, M. and {Weinberg}, M. and {Wheaton}, W.~A. and {Wheelock}, S. and {Zacarias}, N.},
	journal = {VizieR Online Data Catalog},
	month = jun,
	title = {{VizieR Online Data Catalog: 2MASS All-Sky Catalog of Point Sources (Cutri+ 2003)}},
	year = 2003}

@article{lawler09,
	adsurl = {https://ui.adsabs.harvard.edu/abs/2009ApJ...705...89L},
	archiveprefix = {arXiv},
	author = {{Lawler}, S.~M. and {Beichman}, C.~A. and {Bryden}, G. and {Ciardi}, D.~R. and {Tanner}, A.~M. and {Su}, K.~Y.~L. and {Stapelfeldt}, K.~R. and {Lisse}, C.~M. and {Harker}, D.~E.},
	doi = {10.1088/0004-637X/705/1/89},
	eprint = {0909.0058},
	journal = {\apj},
	month = nov,
	number = {1},
	pages = {89-111},
	primaryclass = {astro-ph.SR},
	title = {{Explorations Beyond the Snow Line: Spitzer/IRS Spectra of Debris Disks Around Solar-type Stars}},
	volume = {705},
	year = 2009}

@article{bryden06,
	author = {{Bryden}, G. and {Beichman}, C.~A. and {Trilling}, D.~E. and {Rieke}, G.~H. and {Holmes}, E.~K. and {Lawler}, S.~M. and {Stapelfeldt}, K.~R. and {Werner}, M.~W. and {Gautier}, T.~N. and {Blaylock}, M. and {Gordon}, K.~D. and {Stansberry}, J.~A. and {Su}, K.~Y.~L.},
	doi = {10.1086/498093},
	journal = {\apj},
	month = jan,
	number = {2},
	pages = {1098-1113},
	title = {{Frequency of Debris Disks around Solar-Type Stars: First Results from a Spitzer MIPS Survey}},
	volume = {636},
	year = 2006}

@article{beichman06,
	author = {{Beichman}, C.~A. and {Bryden}, G. and {Stapelfeldt}, K.~R. and {Gautier}, T.~N. and {Grogan}, K. and {Shao}, M. and {Velusamy}, T. and {Lawler}, S.~M. and {Blaylock}, M. and {Rieke}, G.~H. and {Lunine}, J.~I. and {Fischer}, D.~A. and {Marcy}, G.~W. and {Greaves}, J.~S. and {Wyatt}, M.~C. and {Holland}, W.~S. and {Dent}, W.~R.~F.},
	doi = {10.1086/508449},
	journal = {\apj},
	month = dec,
	number = {2},
	pages = {1674-1693},
	title = {{New Debris Disks around Nearby Main-Sequence Stars: Impact on the Direct Detection of Planets}},
	volume = {652},
	year = 2006}

@article{trilling08,
	author = {{Trilling}, D.~E. and {Bryden}, G. and {Beichman}, C.~A. and {Rieke}, G.~H. and {Su}, K.~Y.~L. and {Stansberry}, J.~A. and {Blaylock}, M. and {Stapelfeldt}, K.~R. and {Beeman}, J.~W. and {Haller}, E.~E.},
	doi = {10.1086/525514},
	journal = {\apj},
	month = feb,
	number = {2},
	pages = {1086-1105},
	title = {{Debris Disks around Sun-like Stars}},
	volume = {674},
	year = 2008}

@article{bryden09,
	author = {{Bryden}, G. and {Beichman}, C.~A. and {Carpenter}, J.~M. and {Rieke}, G.~H. and {Stapelfeldt}, K.~R. and {Werner}, M.~W. and {Tanner}, A.~M. and {Lawler}, S.~M. and {Wyatt}, M.~C. and {Trilling}, D.~E. and {Su}, K.~Y.~L. and {Blaylock}, M. and {Stansberry}, J.~A.},
	doi = {10.1088/0004-637X/705/2/1226},
	journal = {\apj},
	month = nov,
	number = {2},
	pages = {1226-1236},
	title = {{Planets and Debris Disks: Results from a Spitzer/MIPS Search for Infrared Excess}},
	volume = {705},
	year = 2009}

@article{morales09,
	author = {{Morales}, Farisa Y. and {Werner}, M.~W. and {Bryden}, G. and {Plavchan}, P. and {Stapelfeldt}, K.~R. and {Rieke}, G.~H. and {Su}, K.~Y.~L. and {Beichman}, C.~A. and {Chen}, C.~H. and {Grogan}, K. and {Kenyon}, S.~J. and {Moro-Martin}, A. and {Wolf}, S.},
	doi = {10.1088/0004-637X/699/2/1067},
	journal = {\apj},
	month = jul,
	number = {2},
	pages = {1067-1086},
	title = {{Spitzer Mid-IR Spectra of Dust Debris Around A and Late B Type Stars: Asteroid Belt Analogs and Power-Law Dust Distributions}},
	volume = {699},
	year = 2009}

@article{dodsonrobinson11,
	author = {{Dodson-Robinson}, Sarah E. and {Beichman}, C.~A. and {Carpenter}, John M. and {Bryden}, Geoffrey},
	doi = {10.1088/0004-6256/141/1/11},
	eid = {11},
	journal = {\aj},
	month = jan,
	number = {1},
	pages = {11},
	title = {{A Spitzer Infrared Spectrograph Study of Debris Disks Around Planet-host Stars}},
	volume = {141},
	year = 2011}

@article{chen12,
	author = {{Chen}, Christine H. and {Pecaut}, Mark and {Mamajek}, Eric E. and {Su}, Kate Y.~L. and {Bitner}, Martin},
	doi = {10.1088/0004-637X/756/2/133},
	eid = {133},
	journal = {\apj},
	month = sep,
	number = {2},
	pages = {133},
	title = {{A Spitzer MIPS Study of 2.5-2.0 M $_{☉}$ Stars in Scorpius-Centaurus}},
	volume = {756},
	year = 2012}

@article{apai15,
	author = {{Apai}, D{\'a}niel and {Schneider}, Glenn and {Grady}, Carol A. and {Wyatt}, Mark C. and {Lagrange}, Anne-Marie and {Kuchner}, Marc J. and {Stark}, Christopher J. and {Lubow}, Stephen H.},
	doi = {10.1088/0004-637X/800/2/136},
	eid = {136},
	journal = {\apj},
	month = feb,
	number = {2},
	pages = {136},
	title = {{The Inner Disk Structure, Disk-Planet Interactions, and Temporal Evolution in the {\ensuremath{\beta}} Pictoris System: A Two-epoch HST/STIS Coronagraphic Study}},
	volume = {800},
	year = 2015}

@article{kalas95,
	adsurl = {https://ui.adsabs.harvard.edu/abs/1995AJ....110..794K},
	author = {{Kalas}, Paul and {Jewitt}, David},
	doi = {10.1086/117565},
	journal = {\aj},
	month = aug,
	pages = {794},
	title = {{Asymmetries in the Beta Pictoris Dust Disk}},
	volume = {110},
	year = 1995}

@article{kalas05,
	author = {{Kalas}, Paul and {Graham}, James R. and {Clampin}, Mark},
	doi = {10.1038/nature03601},
	journal = {\nat},
	month = jun,
	number = {7045},
	pages = {1067-1070},
	title = {{A planetary system as the origin of structure in Fomalhaut's dust belt}},
	volume = {435},
	year = 2005}

@article{debes19,
	archiveprefix = {arXiv},
	author = {{Debes}, John and {Choquet}, Elodie and {Faramaz}, Virginie C. and {Duchene}, Gaspard and {Hines}, Dean and {Stark}, Chris and {Ygouf}, Marie and {Girard}, Julien and {Moro-Martin}, Amaya and {Arriaga}, Pauline and {Chen}, Christine and {Currie}, Thayne and {Dodson-Robinson}, Sally and {Douglas}, Ewan S. and {Kalas}, Paul and {Lisse}, Carey M. and {Mawet}, Dimitri and {Mazoyer}, Johan and {Mennesson}, Bertrand and {Millar-Blanchaer}, Max A. and {Sivramakrishnan}, Anand and {Wang}, Jason},
	eid = {arXiv:1906.02129},
	journal = {arXiv e-prints},
	month = jun,
	pages = {arXiv:1906.02129},
	title = {{Cold Debris Disks as Strategic Targets for the 2020s}},
	year = 2019}

@INPROCEEDINGS{gardner04,
	author = {{Gardner}, Jonathan P. and {Mather}, John C. and {Clampin}, Mark and {Greenhouse}, Matthew A. and {Hammel}, Heidi B. and {Hutchings}, John B. and {Jakobsen}, Peter and {Lilly}, Simon J. and {Lunine}, Jonathan I. and {McCaughrean}, Mark J. and {Mountain}, Matt and {Rieke}, George H. and {Rieke}, Marcia J. and {Smith}, Eric P. and {Stiavelli}, Massimo and {Stockman}, Hervey S. and {Windhorst}, Rogier A. and {Wright}, Gillian S.},
	booktitle = {\procspie},
	doi = {10.1117/12.550657},
	editor = {{Mather}, John C.},
	pages = {564-575},
	series = {Society of Photo-Optical Instrumentation Engineers (SPIE) Conference Series},
	title = {{The Science Requirements of the James Webb Space Telescope}},
	volume = {5487},
	year = 2004}

@article{smith84,
	author = {{Smith}, B.~A. and {Terrile}, R.~J.},
	doi = {10.1126/science.226.4681.1421},
	journal = {Science},
	month = dec,
	number = {4681},
	pages = {1421-1424},
	title = {{A Circumstellar Disk around {\ensuremath{\beta}} Pictoris}},
	volume = {226},
	year = 1984}

@article{aumann85,
	author = {{Aumann}, H.~H.},
	doi = {10.1086/131620},
	journal = {\pasp},
	month = oct,
	pages = {885-891},
	title = {{IRAS observations of matter around nearby stars.}},
	volume = {97},
	year = 1985}

@article{morel78,
	author = {{Morel}, M. and {Magnenat}, P.},
	journal = {\aaps},
	month = dec,
	pages = {477},
	title = {{UBVRIJKLMNH photoelectric photometric catalogue. (Magnetic tape).}},
	volume = {34},
	year = 1978}

@article{phillips10,
	author = {{Phillips}, N.~M. and {Greaves}, J.~S. and {Dent}, W.~R.~F. and {Matthews}, B.~C. and {Holland}, W.~S. and {Wyatt}, M.~C. and {Sibthorpe}, B.},
	doi = {10.1111/j.1365-2966.2009.15641.x},
	journal = {\mnras},
	month = apr,
	number = {3},
	pages = {1089-1101},
	title = {{Target selection for the SUNS and DEBRIS surveys for debris discs in the solar neighbourhood}},
	volume = {403},
	year = 2010}

@article{absil21,
	adsurl = {https://ui.adsabs.harvard.edu/abs/2021A&A...651A..45A},
	archiveprefix = {arXiv},
	author = {{Absil}, O. and {Marion}, L. and {Ertel}, S. and {Defr{\`e}re}, D. and {Kennedy}, G.~M. and {Romagnolo}, A. and {Le Bouquin}, J. -B. and {Christiaens}, V. and {Milli}, J. and {Bonsor}, A. and {Olofsson}, J. and {Su}, K.~Y.~L. and {Augereau}, J. -C.},
	doi = {10.1051/0004-6361/202140561},
	eid = {A45},
	eprint = {2104.14216},
	journal = {\aap},
	month = jul,
	pages = {A45},
	primaryclass = {astro-ph.EP},
	title = {{A near-infrared interferometric survey of debris-disk stars. VII. The hot-to-warm dust connection}},
	volume = {651},
	year = 2021}

@article{ertel14,
	adsurl = {https://ui.adsabs.harvard.edu/abs/2014A&A...570A.128E},
	archiveprefix = {arXiv},
	author = {{Ertel}, S. and {Absil}, O. and {Defr{\`e}re}, D. and {Le Bouquin}, J. -B. and {Augereau}, J. -C. and {Marion}, L. and {Blind}, N. and {Bonsor}, A. and {Bryden}, G. and {Lebreton}, J. and {Milli}, J.},
	doi = {10.1051/0004-6361/201424438},
	eid = {A128},
	eprint = {1409.6143},
	journal = {\aap},
	month = oct,
	pages = {A128},
	primaryclass = {astro-ph.EP},
	title = {{A near-infrared interferometric survey of debris-disk stars. IV. An unbiased sample of 92 southern stars observed in H band with VLTI/PIONIER}},
	volume = {570},
	year = 2014}

@article{booth13,
	author = {{Booth}, Mark and {Kennedy}, Grant and {Sibthorpe}, Bruce and {Matthews}, Brenda C. and {Wyatt}, Mark C. and {Duch{\^e}ne}, Gaspard and {Kavelaars}, J.~J. and {Rodriguez}, David and {Greaves}, Jane S. and {Koning}, Alice and {Vican}, Laura and {Rieke}, George H. and {Su}, Kate Y.~L. and {Moro-Mart{\'\i}n}, Amaya and {Kalas}, Paul},
	doi = {10.1093/mnras/sts117},
	journal = {\mnras},
	month = jan,
	number = {2},
	pages = {1263-1280},
	title = {{Resolved debris discs around A stars in the Herschel DEBRIS survey}},
	volume = {428},
	year = 2013}

@article{broekhoven13,
	author = {{Broekhoven-Fiene}, Hannah and {Matthews}, Brenda C. and {Kennedy}, Grant M. and {Booth}, Mark and {Sibthorpe}, Bruce and {Lawler}, Samantha M. and {Kavelaars}, J.~J. and {Wyatt}, Mark C. and {Qi}, Chenruo and {Koning}, Alice and {Su}, Kate Y.~L. and {Rieke}, George H. and {Wilner}, David J. and {Greaves}, Jane S.},
	doi = {10.1088/0004-637X/762/1/52},
	eid = {52},
	journal = {\apj},
	month = jan,
	number = {1},
	pages = {52},
	title = {{The Debris Disk around {\ensuremath{\gamma}} Doradus Resolved with Herschel}},
	volume = {762},
	year = 2013}

@article{thureau14,
	author = {{Thureau}, N.~D. and {Greaves}, J.~S. and {Matthews}, B.~C. and {Kennedy}, G. and {Phillips}, N. and {Booth}, M. and {Duch{\^e}ne}, G. and {Horner}, J. and {Rodriguez}, D.~R. and {Sibthorpe}, B. and {Wyatt}, M.~C.},
	doi = {10.1093/mnras/stu1864},
	journal = {\mnras},
	month = dec,
	number = {3},
	pages = {2558-2573},
	title = {{An unbiased study of debris discs around A-type stars with Herschel}},
	volume = {445},
	year = 2014}

@book{gezari93,
	author = {{Gezari}, Daniel Y. and {Schmitz}, Marion and {Pitts}, Patricia S. and {Mead}, Jaylee M.},
	title = {Catalog of Infrared Observations},
	year = 1993,
	publisher = {3rd ed.; NASA},}

@article{gaspar13,
	author = {{G{\'a}sp{\'a}r}, Andr{\'a}s and {Rieke}, George H. and {Balog}, Zolt{\'a}n},
	doi = {10.1088/0004-637X/768/1/25},
	eid = {25},
	journal = {\apj},
	month = may,
	number = {1},
	pages = {25},
	title = {{The Collisional Evolution of Debris Disks}},
	volume = {768},
	year = 2013}

@article{pawellek14,
	author = {{Pawellek}, Nicole and {Krivov}, Alexander V. and {Marshall}, Jonathan P. and {Montesinos}, Benjamin and {{\'A}brah{\'a}m}, P{\'e}ter and {Mo{\'o}r}, Attila and {Bryden}, Geoffrey and {Eiroa}, Carlos},
	eid = {65},
	journal = {\apj},
	month = sep,
	number = {1},
	pages = {65},
	title = {{Disk Radii and Grain Sizes in Herschel-resolved Debris Disks}},
	volume = {792},
	year = 2014}

@article{eiroa10,
	adsurl = {https://ui.adsabs.harvard.edu/abs/2010A&A...518L.131E},
	archiveprefix = {arXiv},
	author = {{Eiroa}, C. and {Fedele}, D. and {Maldonado}, J. and {Gonz{\'a}lez-Garc{\'\i}a}, B.~M. and {Rodmann}, J. and {Heras}, A.~M. and {Pilbratt}, G.~L. and {Augereau}, J. -Ch. and {Mora}, A. and {Montesinos}, B. and {Ardila}, D. and {Bryden}, G. and {Liseau}, R. and {Stapelfeldt}, K. and {Launhardt}, R. and {Solano}, E. and {Bayo}, A. and {Absil}, O. and {Ar{\'e}valo}, M. and {Barrado}, D. and {Beichmann}, C. and {Danchi}, W. and {Del Burgo}, C. and {Ertel}, S. and {Fridlund}, M. and {Fukagawa}, M. and {Guti{\'e}rrez}, R. and {Gr{\"u}n}, E. and {Kamp}, I. and {Krivov}, A. and {Lebreton}, J. and {L{\"o}hne}, T. and {Lorente}, R. and {Marshall}, J. and {Mart{\'\i}nez-Arn{\'a}iz}, R. and {Meeus}, G. and {Montes}, D. and {Morbidelli}, A. and {M{\"u}ller}, S. and {Mutschke}, H. and {Nakagawa}, T. and {Olofsson}, G. and {Ribas}, I. and {Roberge}, A. and {Sanz-Forcada}, J. and {Th{\'e}bault}, P. and {Walker}, H. and {White}, G.~J. and {Wolf}, S.},
	doi = {10.1051/0004-6361/201014594},
	eid = {L131},
	eprint = {1005.3151},
	journal = {\aap},
	month = jul,
	pages = {L131},
	primaryclass = {astro-ph.SR},
	title = {{Cold DUst around NEarby Stars (DUNES). First results. A resolved exo-Kuiper belt around the solar-like star {\ensuremath{\zeta}}$^{2}$ Ret}},
	volume = {518},
	year = 2010}

@article{faramaz14,
	adsurl = {https://ui.adsabs.harvard.edu/abs/2014A&A...563A..72F},
	archiveprefix = {arXiv},
	author = {{Faramaz}, V. and {Beust}, H. and {Th{\'e}bault}, P. and {Augereau}, J. -C. and {Bonsor}, A. and {del Burgo}, C. and {Ertel}, S. and {Marshall}, J.~P. and {Milli}, J. and {Montesinos}, B. and {Mora}, A. and {Bryden}, G. and {Danchi}, W. and {Eiroa}, C. and {White}, G.~J. and {Wolf}, S.},
	doi = {10.1051/0004-6361/201322469},
	eid = {A72},
	eprint = {1312.5146},
	journal = {\aap},
	month = mar,
	pages = {A72},
	primaryclass = {astro-ph.EP},
	title = {{Can eccentric debris disks be long-lived?. A first numerical investigation and application to {\ensuremath{\zeta}}$^{2}$ Reticuli}},
	volume = {563},
	year = 2014}

@article{faramaz18,
	adsurl = {https://ui.adsabs.harvard.edu/abs/2018MNRAS.481...44F},
	archiveprefix = {arXiv},
	author = {{Faramaz}, V. and {Bryden}, G. and {Stapelfeldt}, K.~R. and {Booth}, M. and {Bayo}, A. and {Beust}, H. and {Casassus}, S. and {Cuadra}, J. and {Hales}, A. and {Hughes}, A.~M. and {Olofsson}, J. and {Su}, K.~Y.~L. and {Wilner}, D.~J.},
	doi = {10.1093/mnras/sty2304},
	eprint = {1809.00645},
	journal = {\mnras},
	month = nov,
	number = {1},
	pages = {44-48},
	primaryclass = {astro-ph.EP},
	title = {{Is there really a debris disc around {\ensuremath{\zeta}}$^{2}$ Reticuli?}},
	volume = {481},
	year = 2018}

@article{rieke05,
	author = {{Rieke}, G.~H. and {Su}, K.~Y.~L. and {Stansberry}, J.~A. and {Trilling}, D. and {Bryden}, G. and {Muzerolle}, J. and {White}, B. and {Gorlova}, N. and {Young}, E.~T. and {Beichman}, C.~A. and {Stapelfeldt}, K.~R. and {Hines}, D.~C.},
	doi = {10.1086/426937},
	journal = {\apj},
	month = feb,
	number = {2},
	pages = {1010-1026},
	title = {{Decay of Planetary Debris Disks}},
	volume = {620},
	year = 2005}

@article{bessell90,
	author = {{Bessell}, M.~S.},
	doi = {10.1086/132749},
	journal = {\pasp},
	month = oct,
	pages = {1181-1199},
	title = {{UBVRI passbands.}},
	volume = {102},
	year = 1990}

@inproceedings{perryman97,
	author = {{Perryman}, M.~A.~C.},
	booktitle = {AAS Meeting},
	eid = {68.01},
	month = dec,
	pages = {68.01},
	title = {{Recent Results from HIPPARCOS}},
	volume = {191},
	year = 1997}

@article{pawellek15,
	author = {{Pawellek}, Nicole and {Krivov}, Alexander V.},
	journal = {\mnras},
	month = dec,
	number = {3},
	pages = {3207-3221},
	title = {{The dust grain size-stellar luminosity trend in debris discs}},
	volume = {454},
	year = 2015}

@article{mason17,
	author = {{Mason}, B.~D. and {Wycoff}, G.~L. and {Hartkopf}, W.~I. and {Douglass}, G.~G. and {Worley}, C.~E.},
	journal = {VizieR Online Data Catalog},
	title = {{VizieR Online Data Catalog: The Washington Visual Double Star Catalog (Mason+ 2001-2014)}},
	year = 2017}

@book{kruegel03,
	author = {{Kruegel}, Endrik},
	title = {{The Physics of Interstellar Dust}},
	year = 2003,
	publisher = {IOP Publishing},
	}

@article{sibthorpe18,
	adsurl = {https://ui.adsabs.harvard.edu/abs/2018MNRAS.475.3046S},
	archiveprefix = {arXiv},
	author = {{Sibthorpe}, B. and {Kennedy}, G.~M. and {Wyatt}, M.~C. and {Lestrade}, J. -F. and {Greaves}, J.~S. and {Matthews}, B.~C. and {Duch{\^e}ne}, G.},
	doi = {10.1093/mnras/stx3188},
	eprint = {1803.00072},
	journal = {\mnras},
	month = apr,
	number = {3},
	pages = {3046-3064},
	primaryclass = {astro-ph.EP},
	title = {{Analysis of the Herschel DEBRIS Sun-like star sample}},
	volume = {475},
	year = 2018}

@article{foremanmackey16,
	author = {Daniel Foreman-Mackey},
	journal = {The Journal of Open Source Software},
	month = {jun},
	number = {2},
	pages = {24},
	publisher = {The Open Journal},
	title = {corner.py: Scatterplot matrices in Python},
	url = {https://doi.org/10.21105/joss.00024},
	volume = {1},
	year = 2016,
	doi = {10.21105/joss.00024}}

@inproceedings{riekeb05,
	adsurl = {https://ui.adsabs.harvard.edu/abs/2005SPIE.5904....1R},
	author = {{Rieke}, Marcia J. and {Kelly}, Douglas and {Horner}, Scott},
	booktitle = {Cryogenic Optical Systems and Instruments XI},
	doi = {10.1117/12.615554},
	editor = {{Heaney}, James B. and {Burriesci}, Lawrence G.},
	month = aug,
	pages = {1-8},
	series = {Society of Photo-Optical Instrumentation Engineers (SPIE) Conference Series},
	title = {{Overview of James Webb Space Telescope and NIRCam's Role}},
	volume = {5904},
	year = 2005}

@article{yelverton19,
	author = {{Yelverton}, Ben and {Kennedy}, Grant M. and {Su}, Kate Y.~L. and {Wyatt}, Mark C.},
	doi = {10.1093/mnras/stz1927},
	eprint = {1907.04800},
	journal = {\mnras},
	month = sep,
	number = {3},
	pages = {3588-3606},
	primaryclass = {astro-ph.EP},
	title = {{A statistically significant lack of debris discs in medium separation binary systems}},
	volume = {488},
	year = 2019}

@article{lebouteiller11,
	author = {{Lebouteiller}, V. and {Barry}, D.~J. and {Spoon}, H.~W.~W. and {Bernard-Salas}, J. and {Sloan}, G.~C. and {Houck}, J.~R. and {Weedman}, D.~W.},
	doi = {10.1088/0067-0049/196/1/8},
	eid = {8},
	journal = {\apjs},
	month = sep,
	number = {1},
	pages = {8},
	title = {{CASSIS: The Cornell Atlas of Spitzer/Infrared Spectrograph Sources}},
	volume = {196},
	year = 2011}

@article{lebouteiller10,
	author = {{Lebouteiller}, V. and {Bernard-Salas}, J. and {Sloan}, G.~C. and {Barry}, D.~J.},
	doi = {10.1086/650426},
	journal = {\pasp},
	month = feb,
	number = {888},
	pages = {231},
	title = {{Advanced Optimal Extraction for the Spitzer/IRS}},
	volume = {122},
	year = 2010}

@article{drimmel19,
	author = {Ronald Drimmel and Beatrice Bucciarelli and Laura Inno},
	doi = {10.3847/2515-5172/ab2632},
	journal = {RNAAS},
	month = {jun},
	number = {6},
	pages = {79},
	publisher = {American Astronomical Society},
	title = {Possible Large Systematic Errors of Gaia {DR}2 Parallaxes for Very Bright Stars},
	url = {https://doi.org/10.3847/2515-5172/ab2632},
	volume = {3},
	year = 2019}

@article{wenger00,
	adsurl = {https://ui.adsabs.harvard.edu/abs/2000A&AS..143....9W},
	archiveprefix = {arXiv},
	author = {{Wenger}, M. and {Ochsenbein}, F. and {Egret}, D. and {Dubois}, P. and {Bonnarel}, F. and {Borde}, S. and {Genova}, F. and {Jasniewicz}, G. and {Lalo{\"e}}, S. and {Lesteven}, S. and {Monier}, R.},
	doi = {10.1051/aas:2000332},
	eprint = {astro-ph/0002110},
	journal = {\aaps},
	month = apr,
	pages = {9-22},
	primaryclass = {astro-ph},
	title = {{The SIMBAD astronomical database. The CDS reference database for astronomical objects}},
	volume = {143},
	year = 2000}

@article{krivov13,
	author = {{Krivov}, A.~V. and {Eiroa}, C. and {L{\"o}hne}, T. and {Marshall}, J.~P. and {Montesinos}, B. and {del Burgo}, C. and {Absil}, O. and {Ardila}, D. and {Augereau}, J. -C. and {Bayo}, A. and {Bryden}, G. and {Danchi}, W. and {Ertel}, S. and {Lebreton}, J. and {Liseau}, R. and {Mora}, A. and {Mustill}, A.~J. and {Mutschke}, H. and {Neuh{\"a}user}, R. and {Pilbratt}, G.~L. and {Roberge}, A. and {Schmidt}, T.~O.~B. and {Stapelfeldt}, K.~R. and {Th{\'e}bault}, Ph. and {Vitense}, Ch. and {White}, G.~J. and {Wolf}, S.},
	doi = {10.1088/0004-637X/772/1/32},
	eid = {32},
	journal = {\apj},
	month = jul,
	number = {1},
	pages = {32},
	title = {{Herschel's ``Cold Debris Disks'': Background Galaxies or Quiescent Rims of Planetary Systems?}},
	volume = {772},
	year = 2013}

@article{pericaud17,
	adsurl = {https://ui.adsabs.harvard.edu/abs/2017A&A...600A..62P},
	archiveprefix = {arXiv},
	author = {{P{\'e}ricaud}, J. and {Di Folco}, E. and {Dutrey}, A. and {Guilloteau}, S. and {Pi{\'e}tu}, V.},
	doi = {10.1051/0004-6361/201629371},
	eid = {A62},
	eprint = {1612.06582},
	journal = {\aap},
	month = apr,
	pages = {A62},
	primaryclass = {astro-ph.SR},
	title = {{The hybrid disks: a search and study to better understand evolution of disks}},
	volume = {600},
	year = 2017}

@misc{skrutskie19,
	author = {{Skrutskie}, M. F. and Cutri, R. M. and Stiening, R. and Weinberg, M. D. and Schneider, S. and Carpenter, J. M. and Beichman, C. and Capps, R. and Chester, T. and Elias, J. and Huchra, J. and Liebert, J. and Lonsdale, C. and Monet, D. G. and Price, S. and Seitzer, P. and Jarrett, T. and Kirkpatrick, J. D. and Gizis, J. E. and Howard, E. and Evans, T. and Fowler, J. and Fullmer, L. and Hurt, R. and Light, R. and Kopan, E. L. and Marsh, K. A. and McCallon, H. L. and Tam, R. and Van Dyk, S. and Wheelock, S.},
	doi = {10.26131/IRSA2},
	publisher = {IPAC},
	title = {{2MASS All-Sky Point Source Catalog}},
	url = {https://catcopy.ipac.caltech.edu/dois/doi.php?id=10.26131/IRSA2},
	year = 2019}

@misc{gaia20,
	author = {{Gaia Collaboration}},
	doi = {10.26131/IRSA12},
	publisher = {IPAC},
	title = {Gaia Source Catalogue DR2},
	url = {https://catcopy.ipac.caltech.edu/dois/doi.php?id=10.26131/IRSA12},
	year = {2020}}

@misc{wise20,
	author = {{WISE Team}},
	doi = {10.26131/IRSA142},
	publisher = {IPAC},
	title = {WISE All-Sky Source Catalog},
	url = {https://catcopy.ipac.caltech.edu/dois/doi.php?id=10.26131/IRSA142},
	year = {2020}}

@article{rieke04,
	adsurl = {https://ui.adsabs.harvard.edu/abs/2004ApJS..154...25R},
	author = {{Rieke}, G.~H. and {Young}, E.~T. and {Engelbracht}, C.~W. and {Kelly}, D.~M. and {Low}, F.~J. and {Haller}, E.~E. and {Beeman}, J.~W. and {Gordon}, K.~D. and {Stansberry}, J.~A. and {Misselt}, K.~A. and {Cadien}, J. and {Morrison}, J.~E. and {Rivlis}, G. and {Latter}, W.~B. and {Noriega-Crespo}, A. and {Padgett}, D.~L. and {Stapelfeldt}, K.~R. and {Hines}, D.~C. and {Egami}, E. and {Muzerolle}, J. and {Alonso-Herrero}, A. and {Blaylock}, M. and {Dole}, H. and {Hinz}, J.~L. and {Le Floc'h}, E. and {Papovich}, C. and {P{\'e}rez-Gonz{\'a}lez}, P.~G. and {Smith}, P.~S. and {Su}, K.~Y.~L. and {Bennett}, L. and {Frayer}, D.~T. and {Henderson}, D. and {Lu}, N. and {Masci}, F. and {Pesenson}, M. and {Rebull}, L. and {Rho}, J. and {Keene}, J. and {Stolovy}, S. and {Wachter}, S. and {Wheaton}, W. and {Werner}, M.~W. and {Richards}, P.~L.},
	doi = {10.1086/422717},
	journal = {\apjs},
	month = sep,
	number = {1},
	pages = {25-29},
	title = {{The Multiband Imaging Photometer for Spitzer (MIPS)}},
	volume = {154},
	year = 2004}

@article{poglitsch10,
	author = {{Poglitsch}, A. and {Waelkens}, C. and {Geis}, N. and {Feuchtgruber}, H. and {Vandenbussche}, B. and {Rodriguez}, L. and {Krause}, O. and {Renotte}, E. and {van Hoof}, C. and {Saraceno}, P. and {Cepa}, J. and {Kerschbaum}, F. and {Agn{\`e}se}, P. and {Ali}, B. and {Altieri}, B. and {Andreani}, P. and {Augueres}, J. -L. and {Balog}, Z. and {Barl}, L. and {Bauer}, O.~H. and {Belbachir}, N. and {Benedettini}, M. and {Billot}, N. and {Boulade}, O. and {Bischof}, H. and {Blommaert}, J. and {Callut}, E. and {Cara}, C. and {Cerulli}, R. and {Cesarsky}, D. and {Contursi}, A. and {Creten}, Y. and {De Meester}, W. and {Doublier}, V. and {Doumayrou}, E. and {Duband}, L. and {Exter}, K. and {Genzel}, R. and {Gillis}, J. -M. and {Gr{\"o}zinger}, U. and {Henning}, T. and {Herreros}, J. and {Huygen}, R. and {Inguscio}, M. and {Jakob}, G. and {Jamar}, C. and {Jean}, C. and {de Jong}, J. and {Katterloher}, R. and {Kiss}, C. and {Klaas}, U. and {Lemke}, D. and {Lutz}, D. and {Madden}, S. and {Marquet}, B. and {Martignac}, J. and {Mazy}, A. and {Merken}, P. and {Montfort}, F. and {Morbidelli}, L. and {M{\"u}ller}, T. and {Nielbock}, M. and {Okumura}, K. and {Orfei}, R. and {Ottensamer}, R. and {Pezzuto}, S. and {Popesso}, P. and {Putzeys}, J. and {Regibo}, S. and {Reveret}, V. and {Royer}, P. and {Sauvage}, M. and {Schreiber}, J. and {Stegmaier}, J. and {Schmitt}, D. and {Schubert}, J. and {Sturm}, E. and {Thiel}, M. and {Tofani}, G. and {Vavrek}, R. and {Wetzstein}, M. and {Wieprecht}, E. and {Wiezorrek}, E.},
	doi = {10.1051/0004-6361/201014535},
	eid = {L2},
	journal = {\aap},
	month = jul,
	pages = {L2},
	title = {{The Photodetector Array Camera and Spectrometer (PACS) on the Herschel Space Observatory}},
	volume = {518},
	year = 2010}

@article{griffin10,
	author = {{Griffin}, M.~J. and {Abergel}, A. and {Abreu}, A. and {Ade}, P.~A.~R. and {Andr{\'e}}, P. and {Augueres}, J. -L. and {Babbedge}, T. and {Bae}, Y. and {Baillie}, T. and {Baluteau}, J. -P. and {Barlow}, M.~J. and {Bendo}, G. and {Benielli}, D. and {Bock}, J.~J. and {Bonhomme}, P. and {Brisbin}, D. and {Brockley-Blatt}, C. and {Caldwell}, M. and {Cara}, C. and {Castro-Rodriguez}, N. and {Cerulli}, R. and {Chanial}, P. and {Chen}, S. and {Clark}, E. and {Clements}, D.~L. and {Clerc}, L. and {Coker}, J. and {Communal}, D. and {Conversi}, L. and {Cox}, P. and {Crumb}, D. and {Cunningham}, C. and {Daly}, F. and {Davis}, G.~R. and {de Antoni}, P. and {Delderfield}, J. and {Devin}, N. and {di Giorgio}, A. and {Didschuns}, I. and {Dohlen}, K. and {Donati}, M. and {Dowell}, A. and {Dowell}, C.~D. and {Duband}, L. and {Dumaye}, L. and {Emery}, R.~J. and {Ferlet}, M. and {Ferrand}, D. and {Fontignie}, J. and {Fox}, M. and {Franceschini}, A. and {Frerking}, M. and {Fulton}, T. and {Garcia}, J. and {Gastaud}, R. and {Gear}, W.~K. and {Glenn}, J. and {Goizel}, A. and {Griffin}, D.~K. and {Grundy}, T. and {Guest}, S. and {Guillemet}, L. and {Hargrave}, P.~C. and {Harwit}, M. and {Hastings}, P. and {Hatziminaoglou}, E. and {Herman}, M. and {Hinde}, B. and {Hristov}, V. and {Huang}, M. and {Imhof}, P. and {Isaak}, K.~J. and {Israelsson}, U. and {Ivison}, R.~J. and {Jennings}, D. and {Kiernan}, B. and {King}, K.~J. and {Lange}, A.~E. and {Latter}, W. and {Laurent}, G. and {Laurent}, P. and {Leeks}, S.~J. and {Lellouch}, E. and {Levenson}, L. and {Li}, B. and {Li}, J. and {Lilienthal}, J. and {Lim}, T. and {Liu}, S.~J. and {Lu}, N. and {Madden}, S. and {Mainetti}, G. and {Marliani}, P. and {McKay}, D. and {Mercier}, K. and {Molinari}, S. and {Morris}, H. and {Moseley}, H. and {Mulder}, J. and {Mur}, M. and {Naylor}, D.~A. and {Nguyen}, H. and {O'Halloran}, B. and {Oliver}, S. and {Olofsson}, G. and {Olofsson}, H. -G. and {Orfei}, R. and {Page}, M.~J. and {Pain}, I. and {Panuzzo}, P. and {Papageorgiou}, A. and {Parks}, G. and {Parr-Burman}, P. and {Pearce}, A. and {Pearson}, C. and {P{\'e}rez-Fournon}, I. and {Pinsard}, F. and {Pisano}, G. and {Podosek}, J. and {Pohlen}, M. and {Polehampton}, E.~T. and {Pouliquen}, D. and {Rigopoulou}, D. and {Rizzo}, D. and {Roseboom}, I.~G. and {Roussel}, H. and {Rowan-Robinson}, M. and {Rownd}, B. and {Saraceno}, P. and {Sauvage}, M. and {Savage}, R. and {Savini}, G. and {Sawyer}, E. and {Scharmberg}, C. and {Schmitt}, D. and {Schneider}, N. and {Schulz}, B. and {Schwartz}, A. and {Shafer}, R. and {Shupe}, D.~L. and {Sibthorpe}, B. and {Sidher}, S. and {Smith}, A. and {Smith}, A.~J. and {Smith}, D. and {Spencer}, L. and {Stobie}, B. and {Sudiwala}, R. and {Sukhatme}, K. and {Surace}, C. and {Stevens}, J.~A. and {Swinyard}, B.~M. and {Trichas}, M. and {Tourette}, T. and {Triou}, H. and {Tseng}, S. and {Tucker}, C. and {Turner}, A. and {Vaccari}, M. and {Valtchanov}, I. and {Vigroux}, L. and {Virique}, E. and {Voellmer}, G. and {Walker}, H. and {Ward}, R. and {Waskett}, T. and {Weilert}, M. and {Wesson}, R. and {White}, G.~J. and {Whitehouse}, N. and {Wilson}, C.~D. and {Winter}, B. and {Woodcraft}, A.~L. and {Wright}, G.~S. and {Xu}, C.~K. and {Zavagno}, A. and {Zemcov}, M. and {Zhang}, L. and {Zonca}, E.},
	doi = {10.1051/0004-6361/201014519},
	eid = {L3},
	journal = {\aap},
	month = jul,
	pages = {L3},
	title = {{The Herschel-SPIRE instrument and its in-flight performance}},
	volume = {518},
	year = 2010}

@book{hoffleit91,
	author = {{Hoffleit}, Dorrit and {Jaschek}, Carlos},
	title = {The Bright Star Catalogue},
	year = 1991,
	publisher = {5th ed.; Yale Univ. Observatory},
	}

@article{eggleton08,
	adsurl = {https://ui.adsabs.harvard.edu/abs/2008MNRAS.389..869E},
	archiveprefix = {arXiv},
	author = {{Eggleton}, P.~P. and {Tokovinin}, A.~A.},
	doi = {10.1111/j.1365-2966.2008.13596.x},
	eprint = {0806.2878},
	journal = {\mnras},
	month = sep,
	number = {2},
	pages = {869-879},
	primaryclass = {astro-ph},
	title = {{A catalogue of multiplicity among bright stellar systems}},
	volume = {389},
	year = 2008}

@article{boccaletti15,
	author = {{Boccaletti}, A. and {Lagage}, P. -O. and {Baudoz}, P. and {Beichman}, C. and {Bouchet}, P. and {Cavarroc}, C. and {Dubreuil}, D. and {Glasse}, Alistair and {Glauser}, A.~M. and {Hines}, D.~C. and {Lajoie}, C. -P. and {Lebreton}, J. and {Perrin}, M.~D. and {Pueyo}, L. and {Reess}, J.~M. and {Rieke}, G.~H. and {Ronayette}, S. and {Rouan}, D. and {Soummer}, R. and {Wright}, G.~S.},
	doi = {10.1086/682256},
	journal = {\pasp},
	month = jul,
	number = {953},
	pages = {633},
	title = {{The Mid-Infrared Instrument for the James Webb Space Telescope, V: Predicted Performance of the MIRI Coronagraphs}},
	volume = {127},
	year = 2015}

@article{holland17,
	author = {{Holland}, Wayne S. and {Matthews}, Brenda C. and {Kennedy}, Grant M. and {Greaves}, Jane S. and {Wyatt}, Mark C. and {Booth}, Mark and {Bastien}, Pierre and {Bryden}, Geoff and {Butner}, Harold and {Chen}, Christine H. and {Chrysostomou}, Antonio and {Davies}, Claire L. and {Dent}, William R.~F. and {Di Francesco}, James and {Duch{\^e}ne}, Gaspard and {Gibb}, Andy G. and {Friberg}, Per and {Ivison}, Rob J. and {Jenness}, Tim and {Kavelaars}, JJ and {Lawler}, Samantha and {Lestrade}, Jean-Fran{\c{c}}ois and {Marshall}, Jonathan P. and {Moro-Martin}, Amaya and {Pani{\'c}}, Olja and {Phillips}, Neil and {Serjeant}, Stephen and {Schieven}, Gerald H. and {Sibthorpe}, Bruce and {Vican}, Laura and {Ward-Thompson}, Derek and {van der Werf}, Paul and {White}, Glenn J. and {Wilner}, David and {Zuckerman}, Ben},
	doi = {10.1093/mnras/stx1378},
	journal = {\mnras},
	month = sep,
	number = {3},
	pages = {3606-3663},
	title = {{SONS: The JCMT legacy survey of debris discs in the submillimetre}},
	volume = {470},
	year = 2017}

@article{pawellek21,
	adsurl = {https://ui.adsabs.harvard.edu/abs/2021MNRAS.502.5390P},
	archiveprefix = {arXiv},
	author = {{Pawellek}, Nicole and {Wyatt}, Mark and {Matr{\`a}}, Luca and {Kennedy}, Grant and {Yelverton6}, Ben},
	doi = {10.1093/mnras/stab269},
	eprint = {2101.12049},
	journal = {\mnras},
	month = apr,
	number = {4},
	pages = {5390-5416},
	primaryclass = {astro-ph.EP},
	title = {{A {\ensuremath{\sim}}75 per cent occurrence rate of debris discs around F stars in the {\ensuremath{\beta}} Pic moving group}},
	volume = {502},
	year = 2021}

@article{eiroa13,
	adsurl = {https://ui.adsabs.harvard.edu/abs/2013A&A...555A..11E},
	archiveprefix = {arXiv},
	author = {{Eiroa}, C. and {Marshall}, J.~P. and {Mora}, A. and {Montesinos}, B. and {Absil}, O. and {Augereau}, J. Ch. and {Bayo}, A. and {Bryden}, G. and {Danchi}, W. and {del Burgo}, C. and {Ertel}, S. and {Fridlund}, M. and {Heras}, A.~M. and {Krivov}, A.~V. and {Launhardt}, R. and {Liseau}, R. and {L{\"o}hne}, T. and {Maldonado}, J. and {Pilbratt}, G.~L. and {Roberge}, A. and {Rodmann}, J. and {Sanz-Forcada}, J. and {Solano}, E. and {Stapelfeldt}, K. and {Th{\'e}bault}, P. and {Wolf}, S. and {Ardila}, D. and {Ar{\'e}valo}, M. and {Beichmann}, C. and {Faramaz}, V. and {Gonz{\'a}lez-Garc{\'\i}a}, B.~M. and {Guti{\'e}rrez}, R. and {Lebreton}, J. and {Mart{\'\i}nez-Arn{\'a}iz}, R. and {Meeus}, G. and {Montes}, D. and {Olofsson}, G. and {Su}, K.~Y.~L. and {White}, G.~J. and {Barrado}, D. and {Fukagawa}, M. and {Gr{\"u}n}, E. and {Kamp}, I. and {Lorente}, R. and {Morbidelli}, A. and {M{\"u}ller}, S. and {Mutschke}, H. and {Nakagawa}, T. and {Ribas}, I. and {Walker}, H.},
	doi = {10.1051/0004-6361/201321050},
	eid = {A11},
	eprint = {1305.0155},
	journal = {\aap},
	month = jul,
	pages = {A11},
	primaryclass = {astro-ph.SR},
	title = {{DUst around NEarby Stars. The survey observational results}},
	volume = {555},
	year = 2013}

@article{greaves09,
	adsurl = {https://ui.adsabs.harvard.edu/abs/2009MNRAS.397..757G},
	archiveprefix = {arXiv},
	author = {{Greaves}, J.~S. and {Wyatt}, M.~C. and {Bryden}, G.},
	doi = {10.1111/j.1365-2966.2009.15048.x},
	eprint = {0907.3677},
	journal = {\mnras},
	month = aug,
	number = {2},
	pages = {757-762},
	primaryclass = {astro-ph.GA},
	title = {{Debris discs around nearby solar analogues}},
	volume = {397},
	year = 2009}

@article{jarrett11,
	adsurl = {https://ui.adsabs.harvard.edu/abs/2011ApJ...735..112J},
	author = {{Jarrett}, T.~H. and {Cohen}, M. and {Masci}, F. and {Wright}, E. and {Stern}, D. and {Benford}, D. and {Blain}, A. and {Carey}, S. and {Cutri}, R.~M. and {Eisenhardt}, P. and {Lonsdale}, C. and {Mainzer}, A. and {Marsh}, K. and {Padgett}, D. and {Petty}, S. and {Ressler}, M. and {Skrutskie}, M. and {Stanford}, S. and {Surace}, J. and {Tsai}, C.~W. and {Wheelock}, S. and {Yan}, D.~L.},
	doi = {10.1088/0004-637X/735/2/112},
	eid = {112},
	journal = {\apj},
	month = jul,
	number = {2},
	pages = {112},
	title = {{The Spitzer-WISE Survey of the Ecliptic Poles}},
	volume = {735},
	year = 2011}

@article{mamajek13,
	author = {{Pecaut}, Mark J. and {Mamajek}, Eric E.},
	doi = {10.1088/0067-0049/208/1/9},
	eid = {9},
	journal = {\apjs},
	month = sep,
	number = {1},
	pages = {9},
	title = {{Intrinsic Colors, Temperatures, and Bolometric Corrections of Pre-main-sequence Stars}},
	volume = {208},
	year = 2013}

@ARTICLE{trilling07,
	author = {{Trilling}, D.~E. and {Stansberry}, J.~A. and {Stapelfeldt}, K.~R. and {Rieke}, G.~H. and {Su}, K.~Y.~L. and {Gray}, R.~O. and {Corbally}, C.~J. and {Bryden}, G. and {Chen}, C.~H. and {Boden}, A. and {Beichman}, C.~A.},
	title = "{Debris disks in main-sequence binary systems.}",
	journal = {\apj},
	year = 2007,
	month = apr,
	volume = {658},
	number = {2},
	pages = {1289-1311},
	doi = {10.1086/511668},
}

@ARTICLE{macintosh15,
	author = {{Macintosh}, B. and {Graham}, J.~R. and {Barman}, T. and {De Rosa}, R.~J. and {Konopacky}, Q. and {Marley}, M.~S. and {Marois}, C. and {Nielsen}, E.~L. and {Pueyo}, L. and {Rajan}, A. and {Rameau}, J. and {Saumon}, D. and {Wang}, J.~J. and {Patience}, J. and {Ammons}, M. and {Arriaga}, P. and {Artigau}, E. and {Beckwith}, S. and {Brewster}, J. and {Bruzzone}, S. and {Bulger}, J. and {Burningham}, B. and {Burrows}, A.~S. and {Chen}, C. and {Chiang}, E. and {Chilcote}, J.~K. and {Dawson}, R.~I. and {Dong}, R. and {Doyon}, R. and {Draper}, Z.~H. and {Duch{\^e}ne}, G. and {Esposito}, T.~M. and {Fabrycky}, D. and {Fitzgerald}, M.~P. and {Follette}, K.~B. and {Fortney}, J.~J. and {Gerard}, B. and {Goodsell}, S. and {Greenbaum}, A.~Z. and {Hibon}, P. and {Hinkley}, S. and {Cotten}, T.~H. and {Hung}, L. -W. and {Ingraham}, P. and {Johnson-Groh}, M. and {Kalas}, P. and {Lafreniere}, D. and {Larkin}, J.~E. and {Lee}, J. and {Line}, M. and {Long}, D. and {Maire}, J. and {Marchis}, F. and {Matthews}, B.~C. and {Max}, C.~E. and {Metchev}, S. and {Millar-Blanchaer}, M.~A. and {Mittal}, T. and {Morley}, C.~V. and {Morzinski}, K.~M. and {Murray-Clay}, R. and {Oppenheimer}, R. and {Palmer}, D.~W. and {Patel}, R. and {Perrin}, M.~D. and {Poyneer}, L.~A. and {Rafikov}, R.~R. and {Rantakyr{\"o}}, F.~T. and {Rice}, E.~L. and {Rojo}, P. and {Rudy}, A.~R. and {Ruffio}, J. -B. and {Ruiz}, M.~T. and {Sadakuni}, N. and {Saddlemyer}, L. and {Salama}, M. and {Savransky}, D. and {Schneider}, A.~C. and {Sivaramakrishnan}, A. and {Song}, I. and {Soummer}, R. and {Thomas}, S. and {Vasisht}, G. and {Wallace}, J.~K. and {Ward-Duong}, K. and {Wiktorowicz}, S.~J. and {Wolff}, S.~G. and {Zuckerman}, B.},
	title = "{Discovery and spectroscopy of the young jovian planet 51 Eri b with the Gemini Planet Imager}",
	journal = {Science},
	year = 2015,
	month = oct,
	volume = {350},
	number = {6256},
	pages = {64-67},
	doi = {10.1126/science.aac5891},
}

@ARTICLE{abt95,
	author = {{Abt}, Helmut A. and {Morrell}, Nidia I.},
	title = "{The Relation between Rotational Velocities and Spectral Peculiarities among A-Type Stars}",
	journal = {\apjs},
	year = 1995,
	month = jul,
	volume = {99},
	pages = {135},
	doi = {10.1086/192182},
}

@ARTICLE{esposito20,
	author = {{Esposito}, Thomas M. and {Kalas}, Paul and {Fitzgerald}, Michael P. and {Millar-Blanchaer}, Maxwell A. and {Duch{\^e}ne}, Gaspard and {Patience}, Jennifer and {Hom}, Justin and {Perrin}, Marshall D. and {De Rosa}, Robert J. and {Chiang}, Eugene and {Czekala}, Ian and {Macintosh}, Bruce and {Graham}, James R. and {Ansdell}, Megan and {Arriaga}, Pauline and {Bruzzone}, Sebastian and {Bulger}, Joanna and {Chen}, Christine H. and {Cotten}, Tara and {Dong}, Ruobing and {Draper}, Zachary H. and {Follette}, Katherine B. and {Hung}, Li-Wei and {Lopez}, Ronald and {Matthews}, Brenda C. and {Mazoyer}, Johan and {Metchev}, Stan and {Rameau}, Julien and {Ren}, Bin and {Rice}, Malena and {Song}, Inseok and {Stahl}, Kevin and {Wang}, Jason and {Wolff}, Schuyler and {Zuckerman}, Ben and {Ammons}, S. Mark and {Bailey}, Vanessa P. and {Barman}, Travis and {Chilcote}, Jeffrey and {Doyon}, Rene and {Gerard}, Benjamin L. and {Goodsell}, Stephen J. and {Greenbaum}, Alexandra Z. and {Hibon}, Pascale and {Hinkley}, Sasha and {Ingraham}, Patrick and {Konopacky}, Quinn and {Maire}, J{\'e}r{\^o}me and {Marchis}, Franck and {Marley}, Mark S. and {Marois}, Christian and {Nielsen}, Eric L. and {Oppenheimer}, Rebecca and {Palmer}, David and {Poyneer}, Lisa and {Pueyo}, Laurent and {Rajan}, Abhijith and {Rantakyr{\"o}}, Fredrik T. and {Ruffio}, Jean-Baptiste and {Savransky}, Dmitry and {Schneider}, Adam C. and {Sivaramakrishnan}, Anand and {Soummer}, R{\'e}mi and {Thomas}, Sandrine and {Ward-Duong}, Kimberly},
	title = "{Debris Disk Results from the Gemini Planet Imager Exoplanet Survey's Polarimetric Imaging Campaign}",
	journal = {\aj},
	year = 2020,
	month = jul,
	volume = {160},
	number = {1},
	eid = {24},
	pages = {24},
	doi = {10.3847/1538-3881/ab9199},
}

@ARTICLE{schneider14,
	author = {{Schneider}, Glenn and {Grady}, Carol A. and {Hines}, Dean C. and {Stark}, Christopher C. and {Debes}, John H. and {Carson}, Joe and {Kuchner}, Marc J. and {Perrin}, Marshall D. and {Weinberger}, Alycia J. and {Wisniewski}, John P. and {Silverstone}, Murray D. and {Jang-Condell}, Hannah and {Henning}, Thomas and {Woodgate}, Bruce E. and {Serabyn}, Eugene and {Moro-Martin}, Amaya and {Tamura}, Motohide and {Hinz}, Phillip M. and {Rodigas}, Timothy J.},
	title = "{Probing for Exoplanets Hiding in Dusty Debris Disks: Disk Imaging, Characterization, and Exploration with HST/STIS Multi-roll Coronagraphy}",
	journal = {\aj},
	year = 2014,
	month = oct,
	volume = {148},
	number = {4},
	eid = {59},
	pages = {59},
	doi = {10.1088/0004-6256/148/4/59},
}

@ARTICLE{engler18,
	author = {{Engler}, N. and {Schmid}, H.~M. and {Quanz}, S.~P. and {Avenhaus}, H. and {Bazzon}, A.},
	title = "{Detection of scattered light from the hot dust in HD 172555}",
	journal = {\aap},
	year = 2018,
	month = oct,
	volume = {618},
	eid = {A151},
	pages = {A151},
	doi = {10.1051/0004-6361/201832674},
}

@ARTICLE{engler19,
	author = {{Engler}, N. and {Boccaletti}, A. and {Schmid}, H.~M. and {Milli}, J. and {Augereau}, J. -C. and {Mazoyer}, J. and {Maire}, A. -L. and {Henning}, T. and {Avenhaus}, H. and {Baudoz}, P. and {Feldt}, M. and {Galicher}, R. and {Hinkley}, S. and {Lagrange}, A. -M. and {Mawet}, D. and {Olofsson}, J. and {Pantin}, E. and {Perrot}, C. and {Stapelfeldt}, K.},
	title = "{Investigating the presence of two belts in the HD 15115 system}",
	journal = {\aap},
	year = 2019,
	month = feb,
	volume = {622},
	eid = {A192},
	pages = {A192},
	doi = {10.1051/0004-6361/201833542},
}

@ARTICLE{marton24,
	author = {{Marton}, G. and {Gezer}, I. and {Madar{\'a}sz}, M. and {Dionatos}, O. and {Audard}, M. and {Roquette}, J. and {Hernandez}, D. and {Paladini}, R. and {Altieri}, B.},
	title = "{The new Herschel/PACS Point Source Catalogue}",
	journal = {\aap},
	year = 2024,
	month = aug,
	volume = {688},
	eid = {A203},
	pages = {A203},
	doi = {10.1051/0004-6361/202450032},
}

@ARTICLE{ertel14b,
	author = {{Ertel}, S. and {Marshall}, J.~P. and {Augereau}, J. -C. and {Krivov}, A.~V. and {L{\"o}hne}, T. and {Eiroa}, C. and {Mora}, A. and {del Burgo}, C. and {Montesinos}, B. and {Bryden}, G. and {Danchi}, W. and {Kirchschlager}, F. and {Liseau}, R. and {Maldonado}, J. and {Pilbratt}, G.~L. and {Sch{\"u}ppler}, Ch. and {Th{\'e}bault}, Ph. and {White}, G.~J. and {Wolf}, S.},
	title = "{Potential multi-component structure of the debris disk around HIP 17439 revealed by Herschel/DUNES}",
	journal = {\aap},
	year = 2014,
	month = jan,
	volume = {561},
	eid = {A114},
	pages = {A114},
	doi = {10.1051/0004-6361/201219945},
}

@ARTICLE{han26,
	author = {{Han}, Yinuo and {Wyatt}, Mark and {Su}, Kate Y.~L. and {Sefilian}, Antranik A. and {Lovell}, Joshua B. and {del Burgo}, Carlos and {Marshall}, Jonathan P. and {Marino}, Sebastian and {Wilner}, David J. and {Matthews}, Brenda C. and {Sommer}, Max and {Hughes}, A. Meredith and {Carpenter}, John M. and {MacGregor}, Meredith A. and {Pawellek}, Nicole and {Henning}, Thomas},
	title = "{A Radially Broad Collisional Cascade in the Debris Disk of {\ensuremath{\gamma}} Ophiuchi Observed by JWST}",
	journal = {\apj},
	year = 2026,
	month = jun,
	volume = {1004},
	number = {2},
	eid = {182},
	pages = {182},
	doi = {10.3847/1538-4357/ae6ef7},
}

@ARTICLE{matra25,
	author = {{Matr{\`a}}, L. and {Marino}, S. and {Wilner}, D.~J. and {Kennedy}, G.~M. and {Booth}, M. and {Krivov}, A.~V. and {Williams}, J.~P. and {Hughes}, A.~M. and {del Burgo}, C. and {Carpenter}, J. and {Davies}, C.~L. and {Ertel}, S. and {Kral}, Q. and {Lestrade}, J.-F. and {Marshall}, J.~P. and {Milli}, J. and {{\"O}berg}, K.~I. and {Pawellek}, N. and {Sepulveda}, A.~G. and {Wyatt}, M.~C. and {Matthews}, B.~C. and {MacGregor}, M.},
	title = "{REsolved ALMA and SMA Observations of Nearby Stars (REASONS): A population of 74 resolved planetesimal belts at millimetre wavelengths}",
	journal = {\aap},
	year = 2025,
	month = jan,
	volume = {693},
	eid = {A151},
	pages = {A151},
	doi = {10.1051/0004-6361/202451397},
	archivePrefix = {arXiv},
	eprint = {2501.09058},
	primaryClass = {astro-ph.EP},
	adsurl = {https://ui.adsabs.harvard.edu/abs/2025A&A...693A.151M}
}

@ARTICLE{ochsenbein00,
	author = {{Ochsenbein}, F. and {Bauer}, P. and {Marcout}, J.},
	title = "{The VizieR database of astronomical catalogues}",
	journal = {\aaps},
	year = 2000,
	month = apr,
	volume = {143},
	pages = {23-32},
	doi = {10.1051/aas:2000169},
}

@MISC{cutri06,
	author = {{Cutri}, R. and {Stiening}, R. and {Huchra}, J.},
	title = "{Explanatory Supplement to the 2MASS All Sky Data Release and Extended Mission Products}",
	year = 2006,
	url = {https://irsa.ipac.caltech.edu/data/2MASS/docs/releases/allsky/doc/sec1_1.html},
}
\bibliographystyle{aasjournal}

\end{document}